\documentclass[a4paper,11pt]{article}
\pdfoutput=1 

\usepackage{jcappub} 

\usepackage[T1]{fontenc} 
\usepackage{subcaption}
\usepackage{multirow}

\title{\boldmath Reduced bispectrum seeded by helical primordial magnetic fields }

\author[a,b]{H\'ector Javier Hort\'ua}
\author[b]{Leonardo Casta\~neda,}

\affiliation[a]{Universidad Nacional de Colombia-Bogot\'a, Facultad de Ciencias, Departamento de F\'isica, Carrera 30 Calle 45-03, C.P. 111321 Bogot\'a, Colombia\\}
\affiliation[b]{Grupo de Gravitaci\'on y Cosmolog\'ia, Observatorio Astron\'omico Nacional, Universidad Nacional de Colombia, cra 45 No 26-85, Edificio Uriel Gutierr\'ez, Bogot\'a, D.C., Colombia}

\emailAdd{hjhortuao@unal.edu.co}

\abstract{
In this paper, we investigate the effects of helical primordial magnetic fields (PMFs) on the cosmic microwave background (CMB) reduced bispectrum. We derive the full three-point statistics of helical magnetic fields and numerically calculate the even contribution in the collinear configuration. We then numerically compute the CMB reduced bispectrum induced by passive and compensated PMF modes on large angular scales. There is a negative signal on the bispectrum due to the helical terms of the fields and we also observe that the biggest contribution to the bispectrum comes from the non-zero IR cut-off for causal fields, unlike the two-point correlation case. For negative spectral indices, the reduced bispectrum is enhanced by the passive modes. This gives a lower value of the upper limit for the mean amplitude of the magnetic field on a given characteristic scale. However, high values of IR cut-off in the bispectrum, and the helical terms of the magnetic field relaxes this bound. This demonstrates the importance of the IR cut-off and helicity in the study of the nature of PMFs from CMB observations. 
}

\begin{document}
\maketitle
\flushbottom
\section{Introduction}
Recent observational evidence of intergalactic magnetic fields from  $\gamma$-ray observations of blazars  and  constraints imposed by CMB power spectrum suggest the existence of magnetic fields created in the early Universe \cite{1,2,3,4,5}. Interesting theoretical models have been proposed to explain  generation processes  which gave rise to this likely primordial field. Some of them are originated by causal
requirements via cosmological phase transitions or by  non-linear evolution of primordial density perturbations\cite{4a,4b,4c}; and the other ones could be generated during inflation, with them being the most appealing  models due to the production
of large-scale magnetic fields beyond the horizon scale \cite{6,7,8,9,9a,10,11,12,13,14,15,16,17,17a,17b,17c,17d,18,18l,18l1,18l2}. Moreover, several additional attempts from string theory or extra dimensions have been done to generate the seed magnetic fields needed to be coherent on cosmological scales \cite{19n1,19n2,19n3,19n4}.
One  way to find out  the process which gave rise to this field and determine its main features is by  making theoretical predictions about the signatures in the CMB
from primordial magnetic fields (PMFs) \cite{19,20,20a,21,21a}. Indeed, some authors have shown  that vector modes dominate all the temperature and polarization  anisotropies for higher multipolar numbers  while the scalar mode contribution is larger for lower $l$ \cite{21bn,21cn,23}.
Other  effect of PMFs on the CMB comes from the non-Gaussian (NG) signals  because its  contribution to  energy-momentum tensor  is  quadratic in the fields. The relevant NG signal from PMFs with an amplitude similar to the curvature turns out to be a feature that is important for constraining mean-field amplitude depending on what signal is induced by passive or compensated modes. 
Studies of NG signals via bispectrum on CMB have found upper limits of PMFs around $2-22$ nG  derived from scalar magnetic modes, and  $3.2-10$ nG  derived from vector-tensor magnetic modes smoothed on a scale of  $1$ Mpc \cite{24,25,26,27,28,29,30,31,32}. The Planck Collaboration also reported limits on  the  amplitude of  $B_{1Mpc}< 3$nG for $n_B=-2.9$; $B_{1Mpc}< 0.07$nG for $n_B=-2$; and $B_{1Mpc}<0.04$nG for $n_B=2$ from compensated modes; and $B_{1Mpc}<4.5$nG for $n_B=-2.9$ from the passive-scalar mode  \cite{33}. PMFs have also been constrained by the POLARBEAR experiment, where they reported that the PMF amplitude from the two-point correlation functions is less than 3.9 nG at the 95\% confidence level \cite{33n}.
On the other hand,  distinct signatures on the parity-odd CMB cross correlations would carry valuable clues about a primordial magnetic helicity. 
In fact, helical contribution  in the field (in the perfect conductivity limit) has been  widely studied because it produces  efficient transference of power from smaller to larger scales, and thus  be able to explain the actual observed magnetic fields\cite{4a,33n1}. Further, observational evidence of helical primordial magnetic fields would offer 
a window  for probing physics beyond the standard models of particle physics, particularly  processes of  parity violation in the early Universe\cite{33n2,33n3,33n4}.\\
The study of  helical fields via NG   will give a deeper understanding of the magnetic field generation model and  help us to strengthen the constraints of PMF amplitude.   
Thus, the main goal of this paper is to investigate the effects  of helical PMFs on the CMB bispectrum. Following previous formulation for calculating the bispectrum, we have found signals that arise by considering a minimal cut-off  and we observed that local-type shape contains the biggest contribution to the bispectrum. This paper is organized as follows. Sec. \ref{sec:intro}  describes the statistical properties of PMFs. In Sec. \ref{bissection} we define the magnetic bispectrum and in Sec.\ref{redbisa4} through the numerical computation, we solve exactly the  bispectrum  for a  collinear configuration and observe some important signals. Sec. \ref{redbisa5}   is defined   the reduced bispectrum and in Sec. \ref{redbisa} we present  numerical results of the primary bispectrum sourced by helical PMFs. Sec. \ref{discu} is devoted to further discussion and  conclusions.

\section{ Statistical Aspects of Primordial Magnetic Fields.}
\label{sec:intro}
We consider a stochastic primordial magnetic field (PMF)  generated in the very early Universe which could have been produced during inflation (non-causal field) or after inflation (causal field).
This field acts like a source of fluctuations on the CMB anisotropies under a FLRW background Universe described by the metric
\begin{equation}
ds^2=a^2(t)(-dt^2+\delta_{ij}dx^idx^j)
\end{equation}
with $t$ being the conformal time.  The PMF power spectrum  which is defined as the Fourier transform of the two point correlation  can be written as
\begin{equation}\label{PMFespectro1}
\langle B_i(\mathbf{k})B_j^*(\mathbf{k}^{\prime})\rangle =(2\pi)^3\delta^3(\mathbf{k}-\mathbf{k}^{\prime})\bigg( P_{ij}(k)P_B(k)+i \epsilon_{ijl}\hat{k}^l P_H(k) \bigg),
\end{equation}
where $P_{ij}(k)=\delta_{ij}-\hat{k}_i\hat{k}_j$ is a projector onto the transverse plane\footnote{This projector has the property $P_{ij}\hat{k}^i=0$  with $\hat{\mathbf{k}}^i=\frac{\mathbf{k}^i}{k}$ and we
use the Fourier transform notation  $ B_i(\mathbf{k})=\int d^3 x\exp^{i\mathbf{k}\cdot x} B_i(\mathbf{x})$. }, $\epsilon_{ijk}$ is the 3D Levi-Civita tensor and, $P_B(k)$, $P_H(k)$  are the
symmetric/anti-symmetric parts of the power spectrum which represent the magnetic field energy density and  absolute value of the  kinetic helicity respectively \cite{23}
\begin{eqnarray}\label{PMFespectro}
  \langle B_i(\mathbf{k})B_i^*(\mathbf{k}^{\prime})\rangle& =&2(2\pi)^3\delta^3(\mathbf{k}-\mathbf{k}^{\prime})P_B(k),\\
   -i\langle \epsilon_{ijl}\hat{k}^lB_i(\mathbf{k})B_j^*(\mathbf{k}^{\prime})\rangle& =&2(2\pi)^3\delta^3(\mathbf{k}-\mathbf{k}^{\prime})P_H(k).
\end{eqnarray}
We assume that power spectrum scales as a simple power law \cite{21}
\begin{equation}\label{powerPMF1}
P_B(k)=A_Bk^{n_B}, \quad  \mbox{ with} \quad A_B=\frac{B^2_{\lambda}2\pi^2\lambda^{n_B+3}}{\Gamma(\frac{n_B+3}{2})},
\end{equation}
\begin{equation}\label{powerPMF2}
P_H(k)=A_H k^{n_H}, \quad  \mbox{ with} \quad A_H=\frac{H^2_{\lambda}2\pi^2\lambda^{n_H+3}}{\Gamma(\frac{n_H+4}{2})},
\end{equation}
with $\Gamma$ being  the Gamma function. In order to avoid infrared divergences (when we do not consider an infrared cutoff),  $n_B > -3$, $n_H > -4$. Also, $B_\lambda$, $H_\lambda$ are  the comoving PMF strength and magnetic helicity smoothing over a Gaussian sphere of comoving radius $\lambda$ \cite{34,35}.
The more general case of the power  spectrum for   magnetic fields  can be studied if we assume that it is defined for  $k_{m} \leq k \leq k_{D}$, being $k_{D}$ an ultraviolet cut-off corresponding to  damping scale  where the field is suppressed on small scales\cite{3} as $k_D \sim\mathcal{O}(10)\text{Mpc}^{-1}$ and we also consider a possible dependence on an infrared cut-off, $k_m$.
Given the Schwarz inequality\cite{34}, 
\begin{equation}
\lim\limits_{k^\prime \rightarrow k}\langle \mathbf{B}(k)\cdot \mathbf{B}^*(k^\prime) \rangle \ge |\lim\limits_{k^\prime \rightarrow k}\langle (\hat{\mathbf{k}} \times \mathbf{B}(k))\cdot \mathbf{B}^*(k^\prime) \rangle |,
\end{equation}
 an additional constraint is found for these fields
\begin{equation}\label{cond1a}
 |A_H| \le A_Bk^{n_B-n_H}.
\end{equation}
In the case where   $A_H=A_B$ and $n_B=n_H$ we define the maximal helicity condition. We will also use the procedure in \cite{21} to parametrize the infrared cut-off by a single constant parameter $\alpha$,
\begin{equation}\label{mcut-off}
k_m=\alpha k_D,\quad 0\leq\alpha<1
\end{equation}
which in the case of inflationary scenarios would correspond to the wave mode that exits the horizon at inflation epoch and for causal modes would be important when this scale is larger than the wavenumber of interest (as claimed by Kim et.al. \cite{21d}). Thus, this infrared cut-off would be important in order to constrain PMF parameters and magnetogenesis models\cite{21,21b,21bb,21c,21d}.
Equation (\ref{mcut-off}) gives only an useful mathematical representation   to constrain these cut-off values via cosmological datasets (for this case, the parameter space would be given by $(\alpha, k_D, B_\lambda, H_\lambda, n_H, n_B)$), and therefore we want to point out that latter expression does not state any physical relation between both wave numbers. In \cite{21b,21bb}, they showed constraints on the maximum wave number $k_D$ as a function of $n_B$ via big bang nucleosynthesis (BBN), and they considered  the maximum and minimum wave numbers as  independent parameters. In fact, we have found out that the  integration scheme used for calculating the spectrum and bispectrum of PMFs is exactly the same if we parametrize $k_m$ as seen in  (\ref{mcut-off}), or if we consider $(k_m, k_D, B_\lambda, H_\lambda, n_H, n_B)$ as  independent parameters. Thus the inclusion of $k_m$ is done only for studying at a phenomenological level its effects on the CMB bispectrum.  
On the other hand, the equations  for  the adimensional energy density of magnetic field  and  spatial part of the electromagnetic  energy momentum tensor  respectively  written in Fourier space are given as
\begin{equation}
\rho_B(\mathbf{k1})=\frac{1}{8\pi \rho_{\gamma,0}}\int \frac{d^3p}{(2\pi)^3}B_l(\mathbf{p})B^l(\mathbf{k1}-\mathbf{p}),\nonumber
\end{equation}
\begin{equation}
\Pi_{ij}(\mathbf{k1})=\frac{1}{4\pi \rho_{\gamma,0}}  \int \frac{d^3p}{(2\pi)^3}\left[\frac{\delta_{ij}}{2}B_l(\mathbf{p})B^l(\mathbf{k1}-\mathbf{p})-B_i(\mathbf{p})B_j(\mathbf{k1}-\mathbf{p}) \right],\label{convdensity}
\end{equation}
 where in the last expressions we are  considering  high conductivity so, the electric field is suppressed; the magnetic field evolves as  $B^2\sim a^{-4}(t)$, and therefore we can express each component
 of the energy momentum tensor in terms of photon energy density $\rho_{\gamma}=\rho_{\gamma,0}a^{-4}$, with $\rho_{\gamma,0}$ being its present value.\footnote{The adimensional energy density of magnetic field showed here is written  with different notation in \cite{31}: $\Omega_B\equiv\frac{B^2}{8 \pi a^4\rho_\gamma}$  and   in \cite{30,25}: $\Delta_B\equiv\frac{B^2}{8 \pi a^4\rho_\gamma}$.}

 Given that spatial electromagnetic energy momentum tensor is symmetric, we can decompose this tensor into the two scalar ($\rho_B$, $\Pi^{(S)}$), one vector ($\Pi^{(V)}_i$) and one tensor ($\Pi^{(T)}_{ij}$)  components as
\begin{equation}
\Pi_{ij}=\frac{1}{3}\delta_{ij}\rho_B+(\hat{k}_i\hat{k}_j-\frac{1}{3}\delta_{ij})\Pi^{(S)}+(\hat{k}_i\Pi_j^{(V)}+\hat{k}_j\Pi_i^{(V)})+\Pi_{ij}^{(T)}\label{tensor1}
\end{equation}
which obey to $\hat{k}^i\Pi_i^{(V)}=\hat{k}^i\Pi_{ij}^{(T)}=\Pi_{ii}^{(T)}=0$ \cite{21,36,37}. The components of this tensor are recovered by applying projector operators defined as 
\begin{eqnarray}\label{projecto}
\rho_B  &=&\delta^{ij}\Pi_{ij}\nonumber \\
\Pi^{(S)}&=&(\delta^{ij}-\frac{3}{2}P^{ij})\Pi_{ij}= \mathcal{P}^{ij} \Pi_{ij}  \nonumber \\
\Pi^{(V)}_i&=& \hat{k}^{(j} P^{l)}_i \Pi_{l\,j}=\mathcal{Q}_{i}^{j\,l} \Pi_{l\,j} \nonumber \\
\Pi^{(T)}_{ij}&=& (P^{(a}_iP^{b)}_j-\frac{1}{2}P^{ab}P_{ij})\Pi_{ab}=\mathcal{P}^{ab}_{ij}\Pi_{ab},
\end{eqnarray}
where $(..)$ in the indices denotes symmetrization\cite{24}.
\section{The Magnetic Bispectrum}\label{bissection}
\begin{figure}
\centering
\begin{subfigure}{.5\textwidth}
  \centering
  \includegraphics[width=1\linewidth]{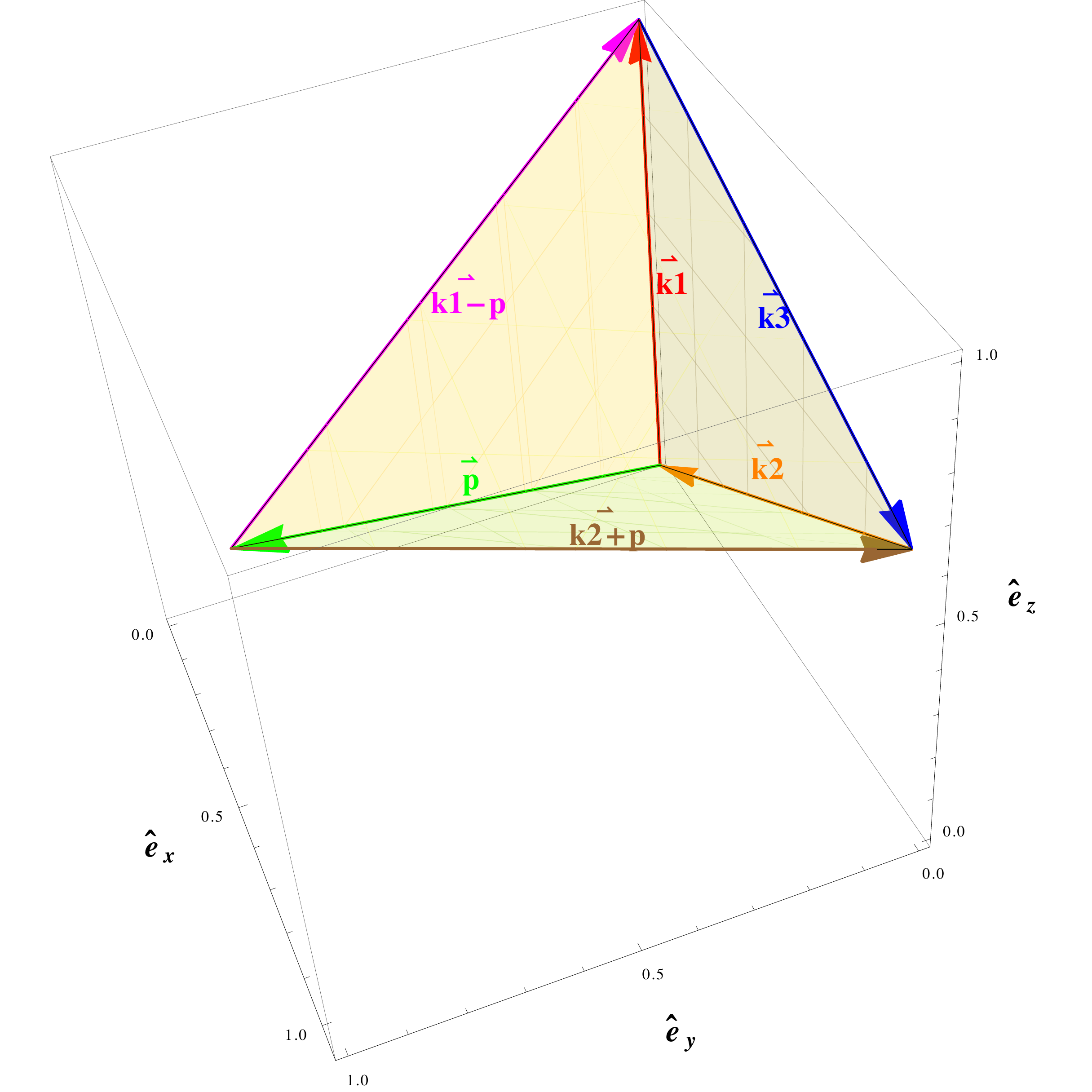}
\end{subfigure}%
\begin{subfigure}{.5\textwidth}
  \centering
  \includegraphics[width=1\linewidth]{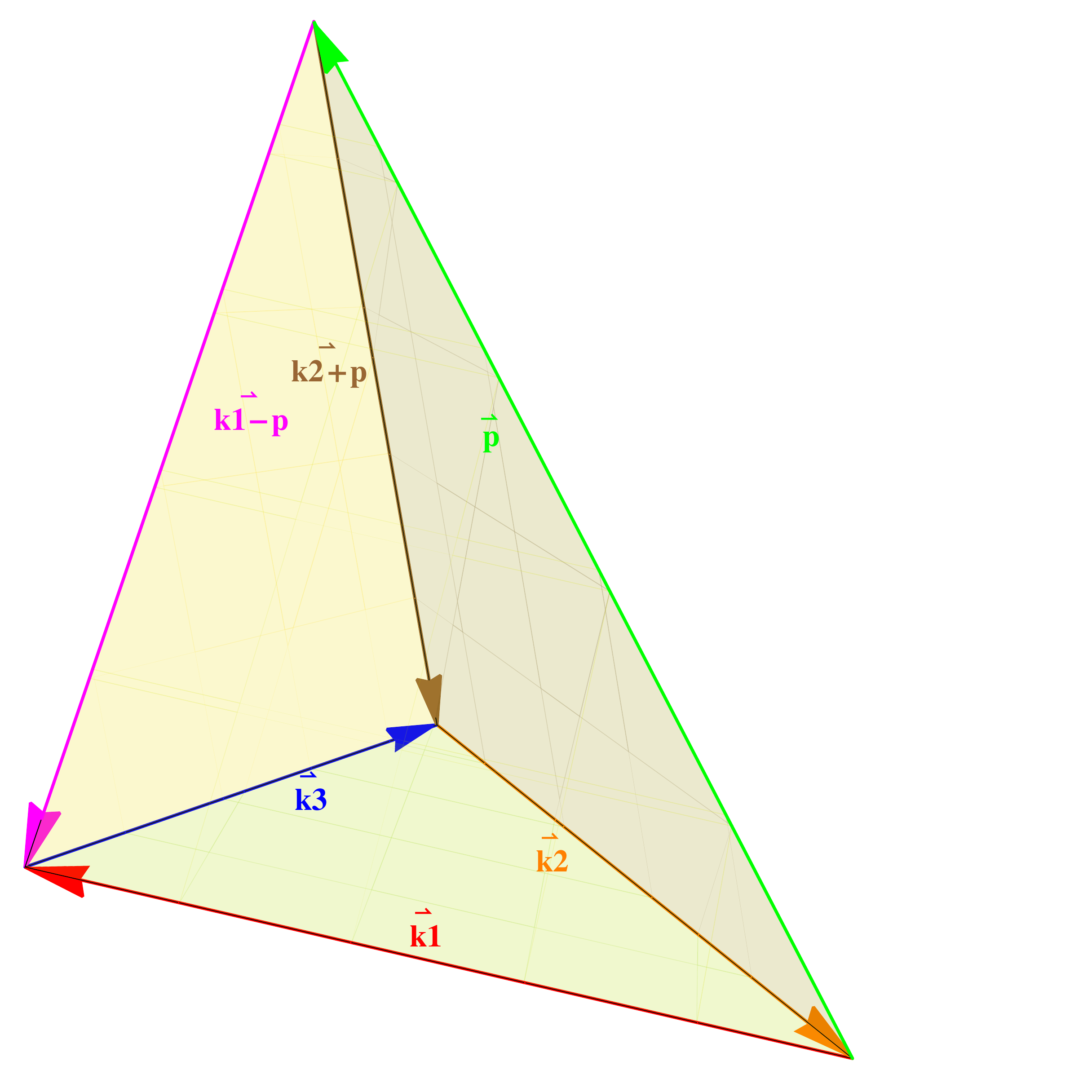}
\end{subfigure}
\caption{ Geometrical configuration for bispectrum. The wavevectors $\mathbf{k1}$, $\mathbf{k2}$ and $\mathbf{k3}$ are free, while $\mathbf{p}$ is the integration mode.}
\label{ref}
\end{figure}

Since the magnetic field is assumed as a Gaussianly-distributed stochastic helical field and the electromagnetic energy momentum tensor is  quadratic in the fields,  the statistics  must be non-Gaussian and the bispectrum is non-zero  as was claimed by \cite{24}. Using eq. (\ref{convdensity}) we have that three-point correlation function is expressed as

\begin{eqnarray}
\langle \Pi_{ij}(\mathbf{k1}) \Pi_{tl}(\mathbf{k2}) \Pi_{mn}(\mathbf{k3})\rangle&=&\frac{-1}{(4\pi \rho_{\gamma,0})^3 }\int \frac{d^3p}{(2\pi)^3}\int \frac{d^3q}{(2\pi)^3} \int \frac{d^3s}{(2\pi)^3} \times \nonumber \\
&&\langle B_i(\mathbf{p})
B_j(\mathbf{k1}-\mathbf{p})B_l(\mathbf{k2}-\mathbf{q})B_n(\mathbf{k3}-\mathbf{s})B_t(\mathbf{q})B_m(\mathbf{s})\rangle \nonumber \\
&-&\frac{\delta_{ij}}{2}\langle ... \rangle-\frac{\delta_{tl}}{2}\langle ... \rangle-\frac{\delta_{mn}}{2}\langle ... \rangle+\frac{\delta_{ij}\delta_{mn}}{4}\langle ... \rangle+\frac{\delta_{ij}\delta_{tl}}{4}\langle ... \rangle \nonumber\\
&+&\frac{\delta_{tl}\delta_{mn}}{4}\langle ... \rangle-\frac{\delta_{ij}\delta_{mn}\delta_{tl}}{8}\langle ... \rangle. \label{TEMbis}
\end{eqnarray}
Where $\langle ... \rangle$ describes  an ensemble average over six stochastic fields. We can use Wicks theorem to decompose the six point correlation function into products of the magnetic field power spectrum expressed in eq. (\ref{PMFespectro}).
Eight of fifteen terms contribute to the bispectrum and they are proportional to $\delta(\mathbf{k1}+\mathbf{k2}+\mathbf{k3})$ due to the homogeneity condition.
In \cite{37} they  point out that expression (\ref{TEMbis}) can be  reduce to just one  contribution if the projection tensor used for extracting each one of the contribution  is  symmetric in ($ij$), ($tl$) and ($mn$). Therefore one can write the bispectrum as follows

\begin{eqnarray}
\langle \Pi_{ij}(\mathbf{k1}) \Pi_{tl}(\mathbf{k2}) \Pi_{mn}(\mathbf{k3})\rangle&=&  \delta(\mathbf{k1}+\mathbf{k2}+\mathbf{k3}) \times \nonumber \\
&&\frac{8}{(4\pi \rho_{\gamma,0})^3 }\int \frac{d^3p}{(2\pi)^3} F_{it}(\mathbf{p})F_{jm}(\mathbf{k1}-\mathbf{p})F_{ln}(\mathbf{k2}+\mathbf{p}), \label{TEMbis1}
\end{eqnarray}
being $F_{ij}(\mathbf{k})=P_{ij}(\mathbf{k})P_{B}(\mathbf{k})+i\epsilon_{ijm}\hat{\mathbf{k}}^m P_H(\mathbf{k})$. Wavevectors that appear  in eq.(\ref{TEMbis1})   generate a tetrahedron configuration (see figure \ref{ref}) such that fifteen angles must be included for calculating the bispectrum. So, in order to make comparison with previos works, we use non only the geometry configuration for  bispectrum but as well the notation of these angles defined in \cite{37} given as
\begin{eqnarray}\label{angles}
&&\beta=\hat{\mathbf{p}}\cdot\widehat{\mathbf{k1}-\mathbf{p}},\quad \gamma=\hat{\mathbf{p}}\cdot\widehat{\mathbf{k2}+\mathbf{p}}, \quad \mu=\widehat{\mathbf{k1}-\mathbf{p}}\cdot\widehat{\mathbf{k2}+\mathbf{p}}, \quad \theta_{kp}=\hat{\mathbf{k1}}\cdot\hat{\mathbf{k2}} \nonumber \\
&&\theta_{kq}=\hat{\mathbf{k1}}\cdot\hat{\mathbf{k3}},\quad \theta_{pq}=\hat{\mathbf{k2}}\cdot\hat{\mathbf{k3}}, \quad \alpha_{k}=\hat{\mathbf{k1}}\cdot \hat{\mathbf{p}},\quad \alpha_{p}=\hat{\mathbf{k2}}\cdot \hat{\mathbf{p}},\quad \alpha_{q}=\hat{\mathbf{k3}}\cdot \hat{\mathbf{p}}\nonumber \\
&& \beta_{k}=\hat{\mathbf{k1}}\cdot \widehat{\mathbf{k1}-\mathbf{p}}, \beta_{p}=\hat{\mathbf{k2}}\cdot \widehat{\mathbf{k1}-\mathbf{p}}, \quad,\beta_{q}=\hat{\mathbf{k3}}\cdot \widehat{\mathbf{k1}-\mathbf{p}}, \quad \gamma_{k}=\hat{\mathbf{k1}}\cdot \widehat{\mathbf{k2}+\mathbf{p}}\nonumber\\
&&\gamma_{p}=\hat{\mathbf{k2}}\cdot \widehat{\mathbf{k2}+\mathbf{p}}, \quad \gamma_{q}=\hat{\mathbf{k3}}\cdot \widehat{\mathbf{k2}+\mathbf{p}}.
\end{eqnarray}
As we will see, the bispectrum has two main contributions, the first contribution contains terms proportional to $A_B^3$ or $A_BA^2_H$ and this is called the even contribution denoted here with $B^{(S)}$. The second contribution is proportional to terms like $A_H^3$ or $A_B^2A_H$ and it is called the odd contribution denoted as  $\mathcal{B}^{(A)}$. Hence, we can define the three-point  correlation for the scalar modes  as
\begin{equation}
\langle Z_1(\mathbf{k1}) Z_2(\mathbf{k2}) Z_3(\mathbf{k3})\rangle \equiv \delta\left(\sum_{i=1}^3\mathbf{ki}\right)\left(B_{Z_1Z_2Z_3}^{(S)}-i\epsilon_{ljk}\mathcal{B}_{Z_1Z_2Z_3}^{(A\,)\,ljk}\right),\label{rrr}
\end{equation}
where $Z_{\{1,2,3\}}=\{\rho_B,\Pi^{(S)}_B\}$.  We begin calculating  the bispectrum of the magnetic energy density. To do so, we shall apply the projector defined in eq.(\ref{projecto}) three times $\delta_{ij}\delta_{tl}\delta_{mn}$ on  eq.( \ref{TEMbis}) to obtain the following 
\begin{equation}
\langle \Pi_{ij}(\mathbf{k1}) \Pi_{tl}(\mathbf{k2}) \Pi_{mn}(\mathbf{k3})\rangle\delta_{ij}\delta_{tl}\delta_{mn}=\langle \rho_B(\mathbf{k1}) \rho_B(\mathbf{k2}) \rho_B(\mathbf{k3})\rangle,
\end{equation}
using eq.(\ref{rrr}), a straightforward calculation gives the following expression
\begin{eqnarray}\label{passiveee1}
B_{\rho_B\rho_B\rho_B}^{(S)}&=&\frac{8}{(2\pi)^3(4\pi \rho_{\gamma,0})^3 }\int d^3p \left(P_B(p)P_B(\left| \mathbf{k1}-\mathbf{p}\right|)P_B(\left|\mathbf{p}+\mathbf{k2}\right|)F_{\rho \rho \rho}^{1} \nonumber \right.\\
&+&P_H(\left| \mathbf{p}+\mathbf{k2}\right|)(-P_B(p)P_H(\left|\mathbf{k1}-\mathbf{p}\right|)F_{\rho \rho \rho}^{2}+P_B(\left|\mathbf{k1}-\mathbf{p}\right|)P_H(p)F_{\rho \rho \rho}^{3} )\nonumber \\
&-& \left. P_B(\left|\mathbf{p}+\mathbf{k2}\right|) P_H(p)    P_H(\left| \mathbf{k1}-\mathbf{p}\right|)F_{\rho \rho \rho}^{4} \right),
\end{eqnarray}
for the even contribution. The values of $F_{\rho\rho\rho}$ are shown  along with  the odd contribution  in appendix \ref{A1}. In orden to find the three-point cross-correlation between scalar anisotropic stress and magnetic energy density, we will apply the three projections $\delta_{ij}\delta_{tl}\mathcal{P}_{mn}$ defined in eq.(\ref{projecto}) on  eq.(\ref{TEMbis}) which gives us
\begin{equation}
\langle \Pi_{ij}(\mathbf{k1}) \Pi_{tl}(\mathbf{k2}) \Pi_{mn}(\mathbf{k3})\rangle\delta_{ij}\delta_{tl}\mathcal{P}_{mn}=\langle \rho_B(\mathbf{k1}) \rho_B(\mathbf{k2}) \Pi_B^{(S)}(\mathbf{k3})\rangle,
\end{equation}
and using the  three-point correlation eq.(\ref{rrr}),
 the even contribution yields
\begin{eqnarray}
B_{\rho_B\rho_B\Pi_B^{(S)}}^{(S)}&=&\frac{8}{2(2\pi)^3(4\pi \rho_{\gamma,0})^3 }\int d^3p \left(-P_B(p)P_B(\left| \mathbf{k1}-\mathbf{p}\right|)P_B(\left|\mathbf{p}+\mathbf{k2}\right|)F_{\rho \rho \Pi^{(S)}}^{1} \right. \nonumber \\
&-&P_B(p)P_H(\left|\mathbf{k1}-\mathbf{p}\right|)P_H(\left|\mathbf{p}+\mathbf{k2}\right|)F_{\rho \rho \Pi^{(S)}}^{2}\nonumber \\
&+&P_H(p)\left(-P_H(\left| \mathbf{p}+\mathbf{k2}\right|)P_B(\left|\mathbf{k1}-\mathbf{p}\right|)F_{\rho \rho \Pi^{(S)}}^{3} \right. \nonumber \\
&+&\left.\left. P_B(\left| \mathbf{p}+\mathbf{k2}\right|)P_H(\left|\mathbf{k1}-\mathbf{p}\right|)F_{\rho \rho \Pi^{(S)}}^{4} \right) \right).
\end{eqnarray}
 Other cross-bispectra is obtained by applying $\delta_{ij}\mathcal{P}_{tl}\mathcal{P}_{mn}$ on  eq.(\ref{TEMbis}) and this gives
\begin{equation}
\langle \Pi_{ij}(\mathbf{k1}) \Pi_{tl}(\mathbf{k2}) \Pi_{mn}(\mathbf{k3})\rangle\delta_{ij}\mathcal{P}_{tl}\mathcal{P}_{mn}=\langle \rho_B(\mathbf{k1}) \Pi_B^{(S)}(\mathbf{k2}) \Pi_B^{(S)}(\mathbf{k3})\rangle,
\end{equation}
as result we found the following expression
\begin{eqnarray}
B_{\rho_B\Pi_B^{(S)}\Pi_B^{(S)}}^{(S)}&=&\frac{8}{4(2\pi)^3(4\pi \rho_{\gamma,0})^3}\int d^3p \left(P_B(p)P_B(\left| \mathbf{k1}-\mathbf{p}\right|)P_B(\left|\mathbf{p}+\mathbf{k2}\right|)F_{\rho \Pi^{(S)}\Pi^{(S)}}^{1} \right. \nonumber \\
&+&P_B(p)P_H(\left| \mathbf{p}+\mathbf{k2}\right|)P_H(\left|\mathbf{k1}-\mathbf{p}\right|)F_{\rho \Pi^{(S)}\Pi^{(S)}}^{2} \nonumber \\
&-&P_H(p)P_H(\left| \mathbf{p}+\mathbf{k2}\right|)P_B(\left|\mathbf{k1}-\mathbf{p}\right|)F_{\rho \Pi^{(S)}\Pi^{(S)}}^{3}\nonumber \\
&-&\left. P_H(p)P_H(\left|\mathbf{k1}-\mathbf{p}\right|)P_B(\left|\mathbf{p}+\mathbf{k2}\right|)F_{\rho \Pi^{(S)}\Pi^{(S)}}^{4} \right).
\end{eqnarray}
 Finally the three-point correlation of scalar anisotropic stress is obtained by applying $\mathcal{P}_{ij}\mathcal{P}_{tl}\mathcal{P}_{mn}$ on  eq.(\ref{TEMbis}) finding that
\begin{equation}
\langle \Pi_{ij}(\mathbf{k1}) \Pi_{tl}(\mathbf{k2}) \Pi_{mn}(\mathbf{k3})\rangle\mathcal{P}_{ij}\mathcal{P}_{tl}\mathcal{P}_{mn}=\langle \Pi_B^{(S)}(\mathbf{k1}) \Pi_B^{(S)}(\mathbf{k2}) \Pi_B^{(S)}(\mathbf{k3})\rangle,
\end{equation}
thus, the expression for the bispectrum for that contribution is
\begin{eqnarray}\label{passiveee}
B_{\Pi_B^{(S)}\Pi_B^{(S)}\Pi_B^{(S)}}^{(S)}&=&\frac{1}{(2\pi)^3(4\pi \rho_{\gamma,0})^3}\int d^3p \left(-P_B(p)P_B(\left| \mathbf{k1}-\mathbf{p}\right|)P_B(\left|\mathbf{p}+\mathbf{k2}\right|)F_{\Pi^{(S)} \Pi^{(S)}\Pi^{(S)}}^{1} \right. \nonumber \\
&-& P_B(p)P_H(\left| \mathbf{p}+\mathbf{k2}\right|)P_H(\left|\mathbf{k1}-\mathbf{p}\right|)F_{\Pi^{(S)} \Pi^{(S)}\Pi^{(S)}}^{2} \nonumber\\
&+&  P_H(p)P_B(\left| \mathbf{p}+\mathbf{k2}\right|)P_H(\left|\mathbf{k1}-\mathbf{p}\right|) F_{\Pi^{(S)} \Pi^{(S)}\Pi^{(S)}}^{3} \nonumber \\
&+& \left. P_H(p)P_B(\left|\mathbf{k1}-\mathbf{p}\right|)P_H(\left|\mathbf{p}+\mathbf{k2}\right|) F_{\Pi^{(S)} \Pi^{(S)}\Pi^{(S)}}^{4} \right).  
\end{eqnarray}
Again, the  $F_{\Pi\Pi\Pi}$'s values   can be checked  along with   the odd contribution in appendix \ref{A1}.
In the case where the helicity of the field is not considered $(A_H=0)$, the only contribution that remains is sourced from $A^3_B$. The results of this contribution were  reported in \cite{37} and we have found the same expressions. Therefore, we have generalized those previous results to  even and odd contributions of the PMFs bispectrum and thus these findings are  the first results of the paper.
\section{Full Evaluation}\label{redbisa4}
\begin{figure}
\centering
\includegraphics[width=.8\linewidth]{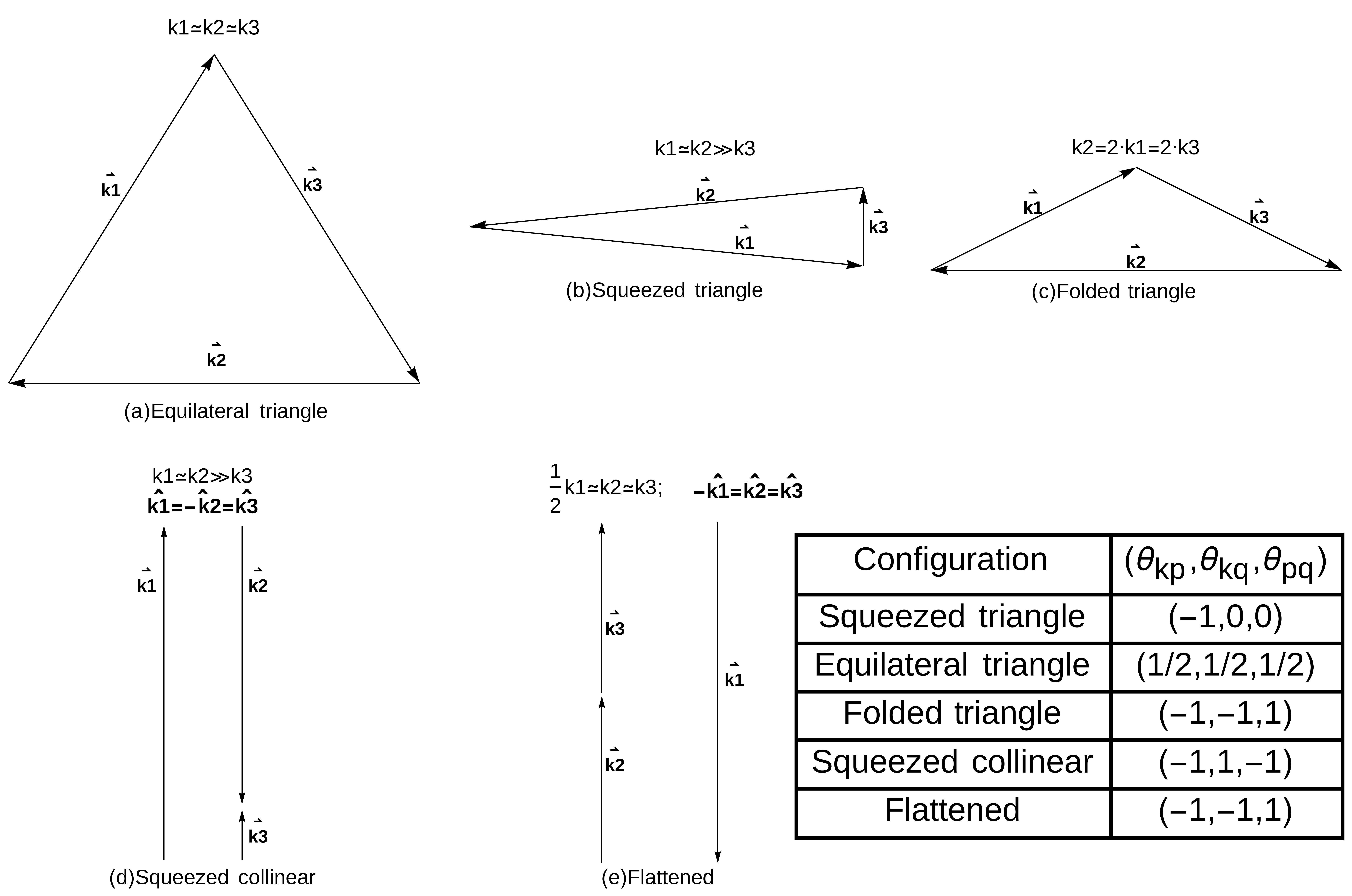}
\caption{Geometrical representations for the bispectrum. The figure shows a  visual representation of the triangles and the collinear configuration of the bispectrum shape. The table in the bottom-right panel describes the values
of the $\mathbf{p}$-independent terms for each configuration.}
\label{fig2}
\end{figure}

With the derivation of the angular part of the three-point correlation for each component of the magnetic tensor, we proceed to make the evaluation of the above integrals. Due to the complexity of the calculation, 
we are going to follow the exact methodology proposed in \cite{25,26,27}. They consider two cases for finding solution of the correlator. In the first case, the terms dependent on the integration vector $\mathbf{p}$ 
are not considered in the evaluation. For the second case, the squeezed collinear configuration is used to make  predictions. We will use five  representative shapes of the  bispectrum which are shown in the Figure \ref{fig2}.
The odd signal arising from $\mathcal{B}^{(A\,)}$ will not be considered here  but will be deferred  for later work.
\subsection{$\mathbf{p}$-independent}\label{pindep}
For this case, the only terms which appear in the evaluation are those angles given in eq.(\ref{angles}) independent of $\mathbf{p}$; they are ($\theta_{kp}$, $\theta_{kq}$, $\theta_{pq}$). The values of these angles for each configuration are shown in figure \ref{fig2}. The $F$'s functions  defined above take the following values under this approximation
\begin{equation}
F_{\rho \rho\rho}^{1}=\mu^2, \quad F_{\rho \rho\rho}^{2}=\mu, \quad F_{\rho \rho\rho}^{3}=F_{\rho \rho\rho}^{4}=0,
\end{equation}
\begin{equation}
F_{\rho \rho\Pi}^{1}=\mu^2-3, \quad F_{\rho \rho\Pi}^{2}=-\mu, \quad F_{\rho \rho\Pi}^{3}=F_{\rho \rho\Pi}^{4}=0,
\end{equation}
\begin{equation}
F_{\rho \Pi\Pi}^{1}=\mu^2-6+9\theta_{pq}^2, \quad F_{\rho \Pi\Pi}^{2}=(-7+9\theta_{pq}^2)\mu, \quad F_{\rho \Pi\Pi}^{3}=F_{\rho \Pi\Pi}^{4}=0,
\end{equation}
\begin{eqnarray}
F_{\Pi \Pi\Pi}^{1}&=&-9+9\theta_{kp}^2-27\theta_{kp}\theta_{kq}\theta_{pq}+9\theta_{pq}^2+\mu^2+9\theta_{kq}^2,  \quad F_{\Pi \Pi\Pi}^{3}=0, \nonumber \\
F_{\Pi \Pi\Pi}^{2}&=&(-13+18\theta_{kp}^2+9\theta_{kq}^2-27\theta_{kp}\theta_{kq}\theta_{pq}+9\theta_{pq}^2)\mu, \quad F_{\Pi \Pi\Pi}^{4}=0,
\end{eqnarray}
where the result  given for  $F_{\Pi \Pi\Pi}^{1}$  is in agreement with the reported  in \cite{25}\footnote{There is a difference with a minus sign because we are taking a different signature in the metric.}.
The  values of $F$ for each geometrical representation of the bispectrum are shown in table \ref{table1} and $\mu=0$ for all configuration except to squeezed configuration where it takes $\mu \sim -1$.
\begin{table}[!hbt]
\begin{center}
\begin{tabular}[b]{| l | c|c|c|c|c|c|c| r |}
\hline
\textbf{Configuration} & $F_{\rho \rho\rho}^{1}$ & $F_{\rho \rho\rho}^{2}$ & $F_{\rho\rho\Pi}^{1}$ & $F_{\rho \rho\Pi}^{2}$ & $F_{\rho \Pi\Pi}^{1}$ & $F_{\rho \Pi\Pi}^{2}$ & $F_{\Pi \Pi\Pi}^{1}$ & $F_{\Pi \Pi\Pi}^{2}$ \\
\hline
Squeezed triangle & 0 & 0 & -3 & 0 & -6 & 0 & 0 & 0 \\
\hline
Equilateral triangle & 0 & 0& -3 & 0 & -15/4 & 0 & -45/8 & 0 \\
\hline
Folded triangle & 0 & 0 & -3 & 0 & 3 & 0 & -9 & 0 \\
\hline
Squeezed collinear & 1 & -1 & -2 & 1 & 4 & -2 & -8 & 4 \\
\hline
Flattened & 0 & 0 & -3 & 0 & 3 & 0 & -9 & 0 \\
\hline
\end{tabular}
\caption{Values of $F$ for different geometrical configurations in the $\mathbf{p}$-independent case.}\label{table1}
\end{center}
\end{table}
\subsection{Squeezed Collinear Configuration}
\begin{figure}[h!]
    \centering
    \begin{subfigure}[b]{0.3\textwidth}
        \includegraphics[width=\textwidth]{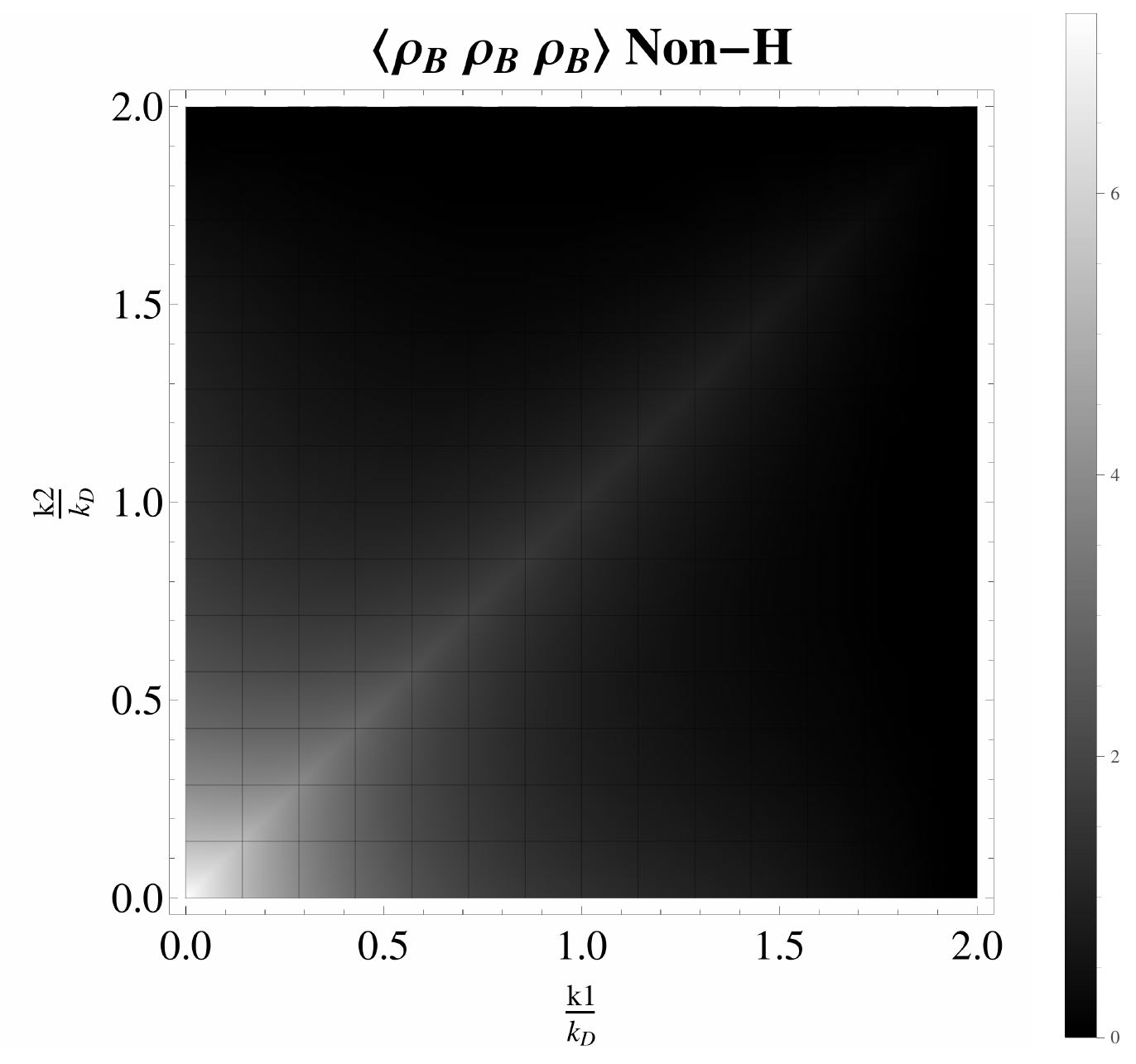}
        \caption{{\footnotesize Three-point correlation of $\langle \rho_B \rho_B \rho_B \rangle$ in units of $\frac{8}{(2\pi)^3(4\pi \rho_{\gamma,0})^3}$ only for $A_B^3$.}} 
        \label{fig:gull}
    \end{subfigure}
    ~ 
    \begin{subfigure}[b]{0.3\textwidth}
        \includegraphics[width=\textwidth]{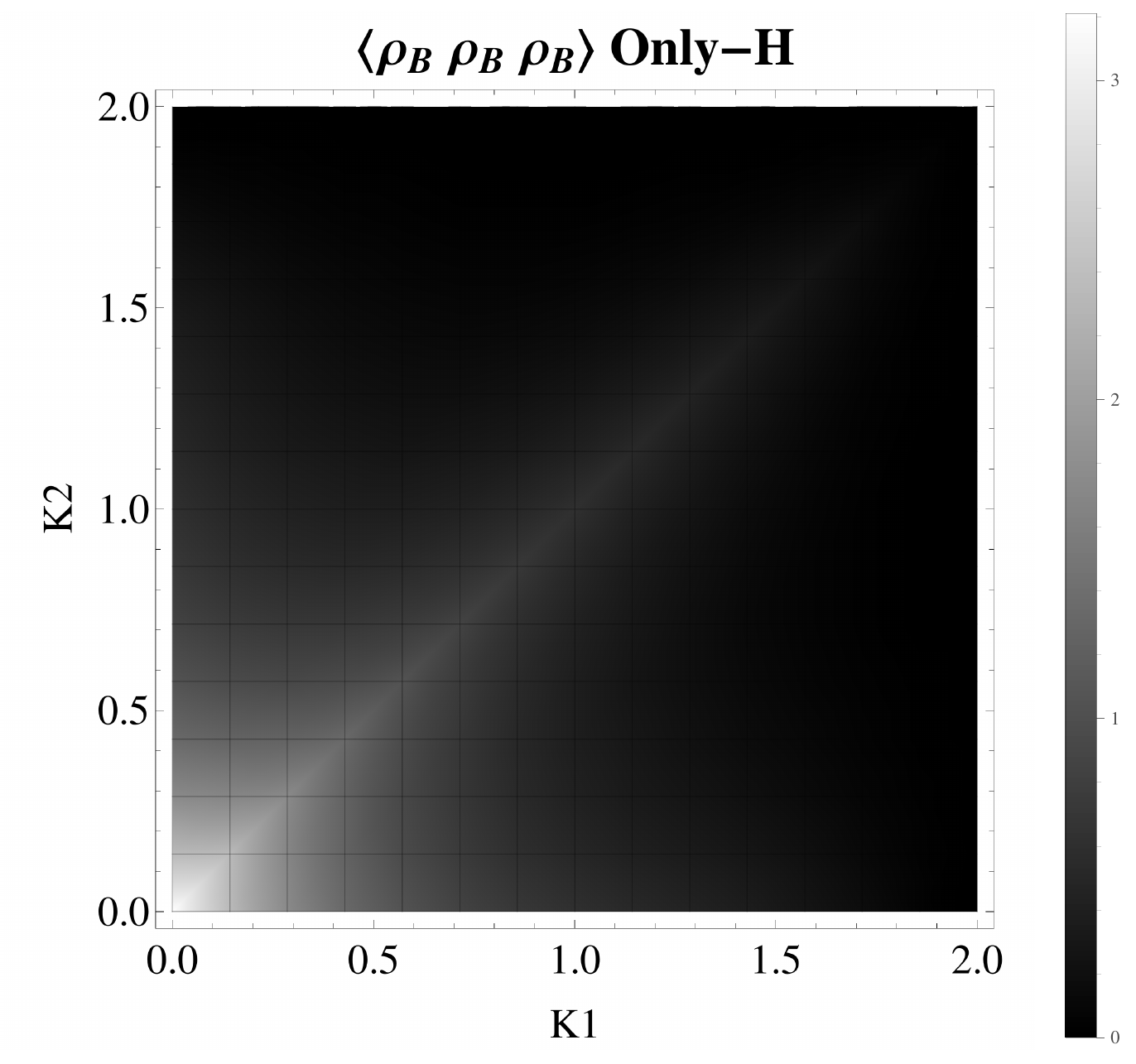}
        \caption{{\footnotesize Three-point correlation of $\langle \rho_B \rho_B \rho_B \rangle$ in units of $\frac{8}{(2\pi)^3(4\pi \rho_{\gamma,0})^3}$ only with $A_BA_H^2$.}}
        \label{fig:tiger}
    \end{subfigure}
    ~ 
    \begin{subfigure}[b]{0.3\textwidth}
        \includegraphics[width=\textwidth]{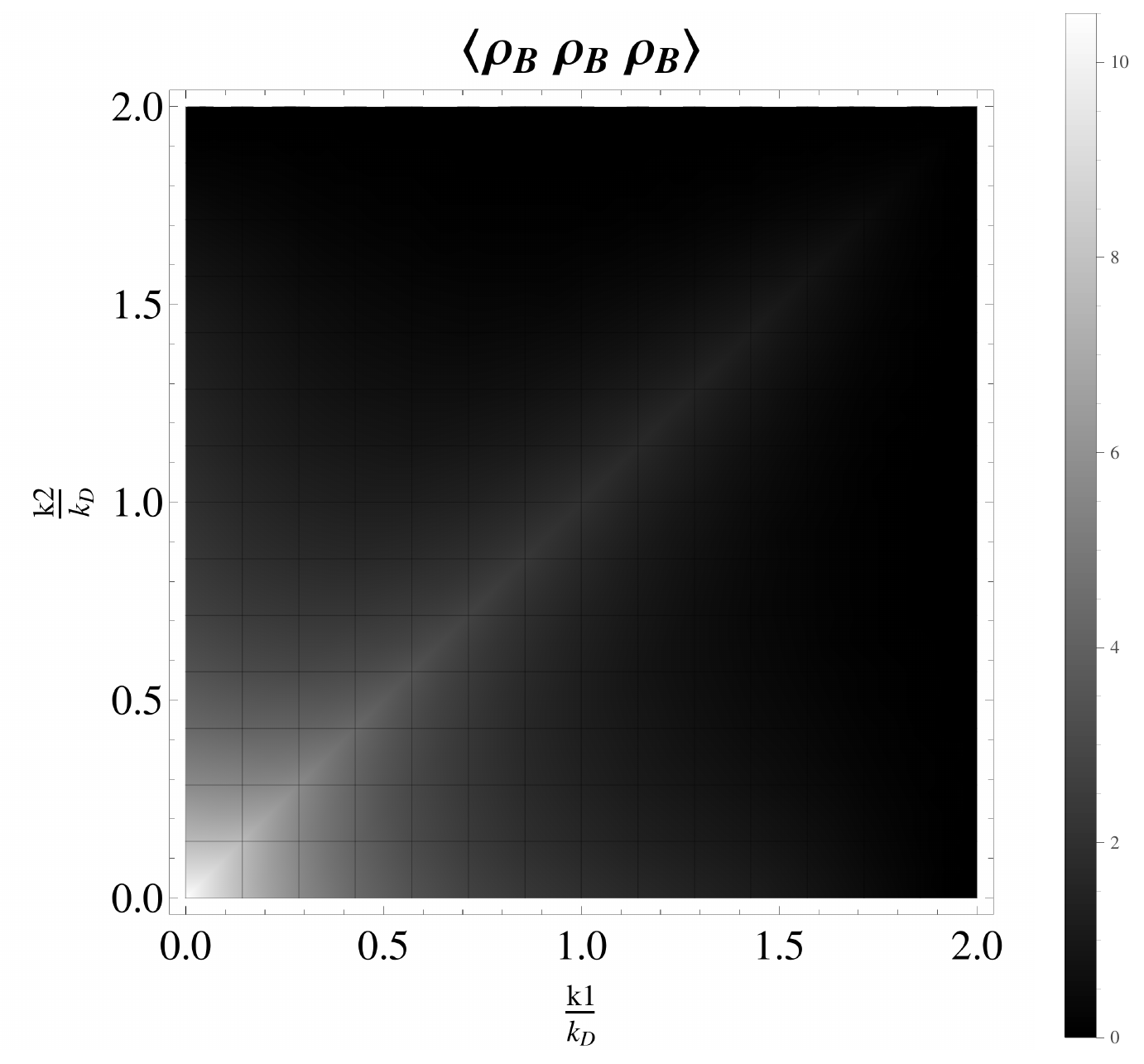}
        \caption{{\footnotesize Even contribution of  three-point correlation of $\langle \rho_B \rho_B \rho_B \rangle$ in units of $\frac{8}{(2\pi)^3(4\pi \rho_{\gamma,0})^3}$.}}
        \label{fig:mouse}
    \end{subfigure}
    \begin{subfigure}[b]{0.3\textwidth}
        \includegraphics[width=\textwidth]{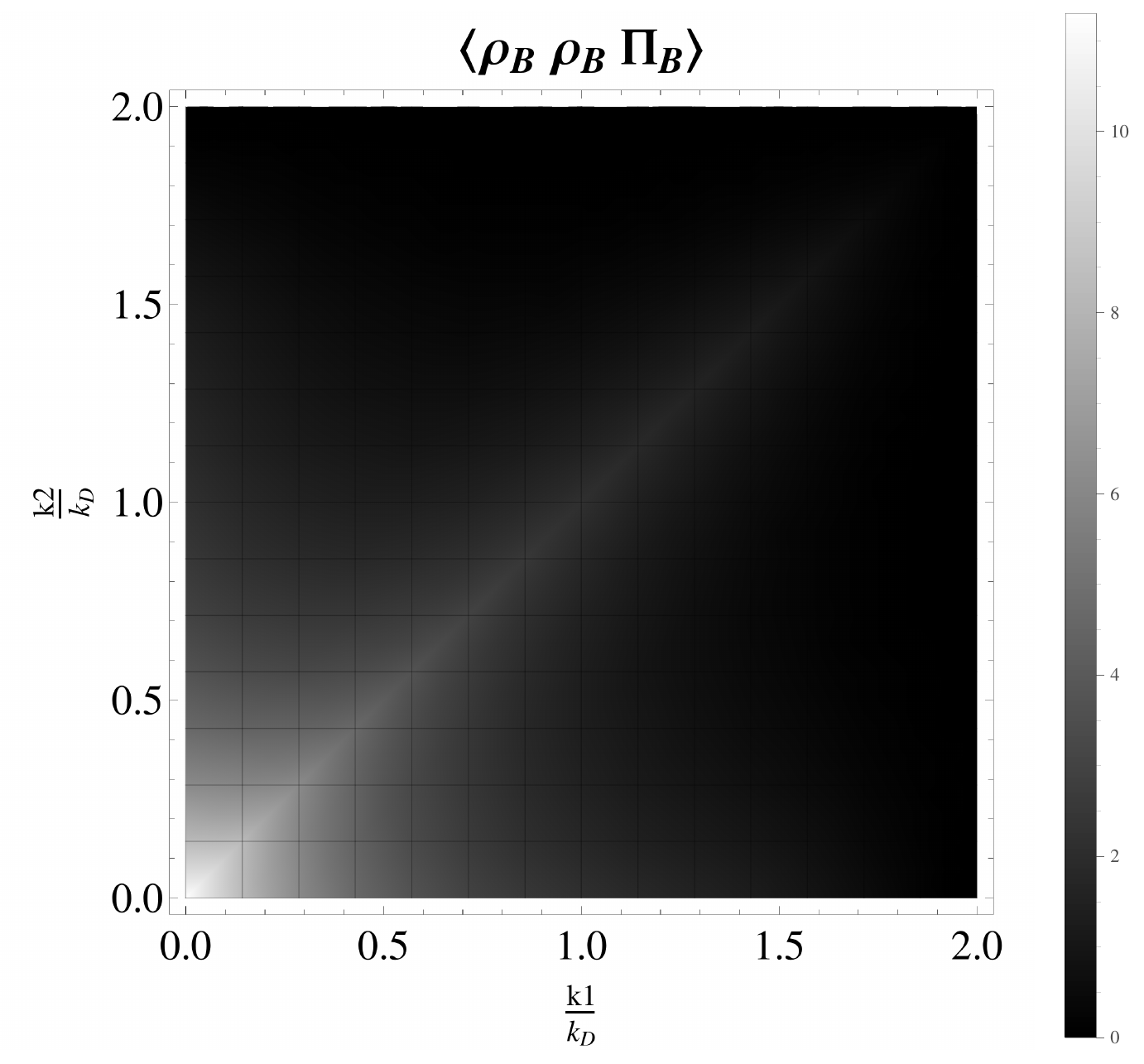}
        \caption{{\footnotesize Even contribution of  three-point correlation of $\langle \rho_B \rho_B \Pi_B \rangle$ in units of $\frac{4}{(2\pi)^3(4\pi \rho_{\gamma,0})^3}$.}}
        \label{fig1aa}
    \end{subfigure}
    ~ 
    \begin{subfigure}[b]{0.3\textwidth}
         \includegraphics[width=\textwidth]{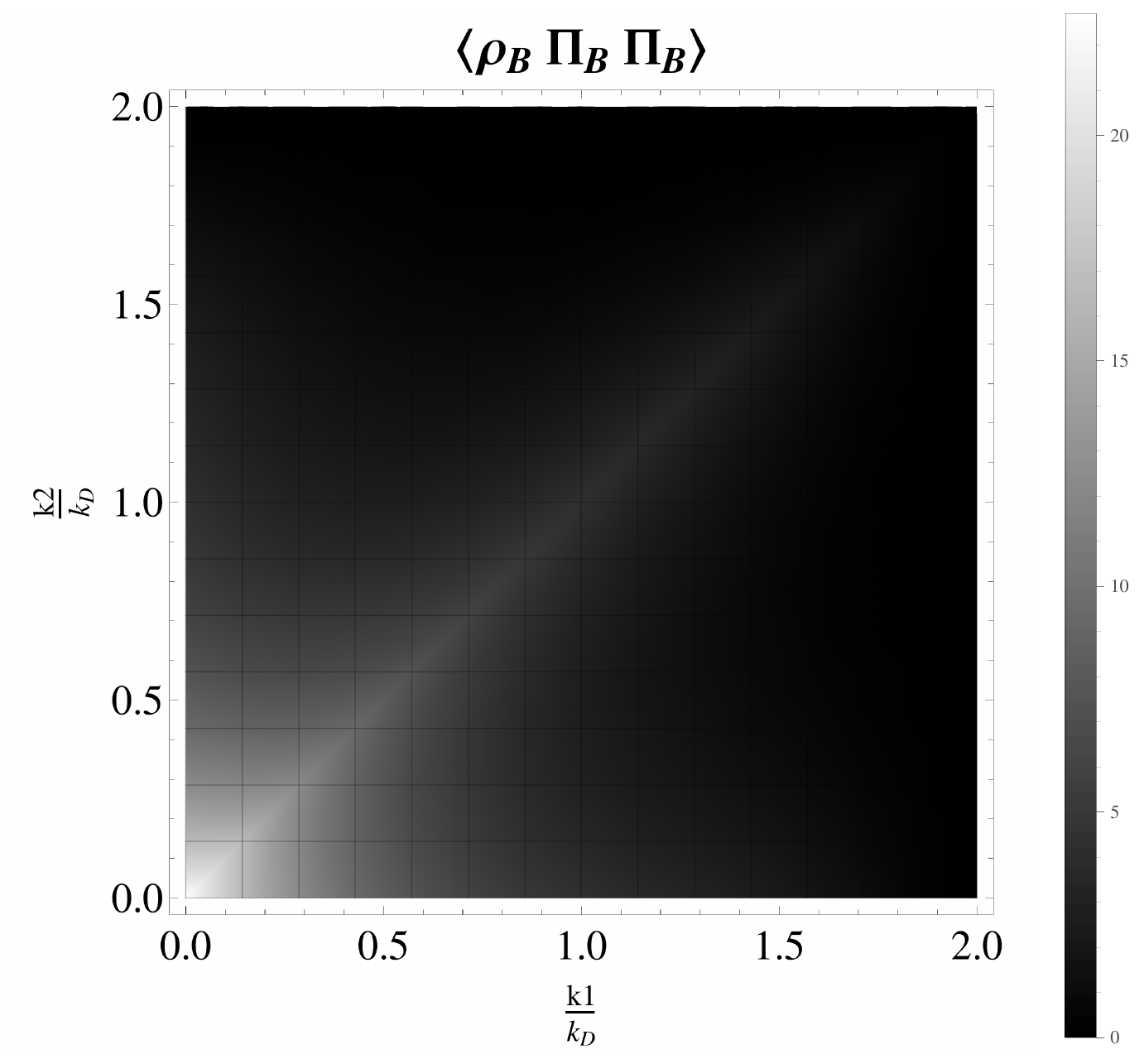}
        \caption{{\footnotesize Even contribution of  three-point correlation of $\langle \rho_B \Pi_B  \Pi_B \rangle$ in units of $\frac{2}{(2\pi)^3(4\pi \rho_{\gamma,0})^3}$.}}
        \label{fig2aa}
    \end{subfigure}
    ~ 
    \begin{subfigure}[b]{0.3\textwidth}
        \includegraphics[width=\textwidth]{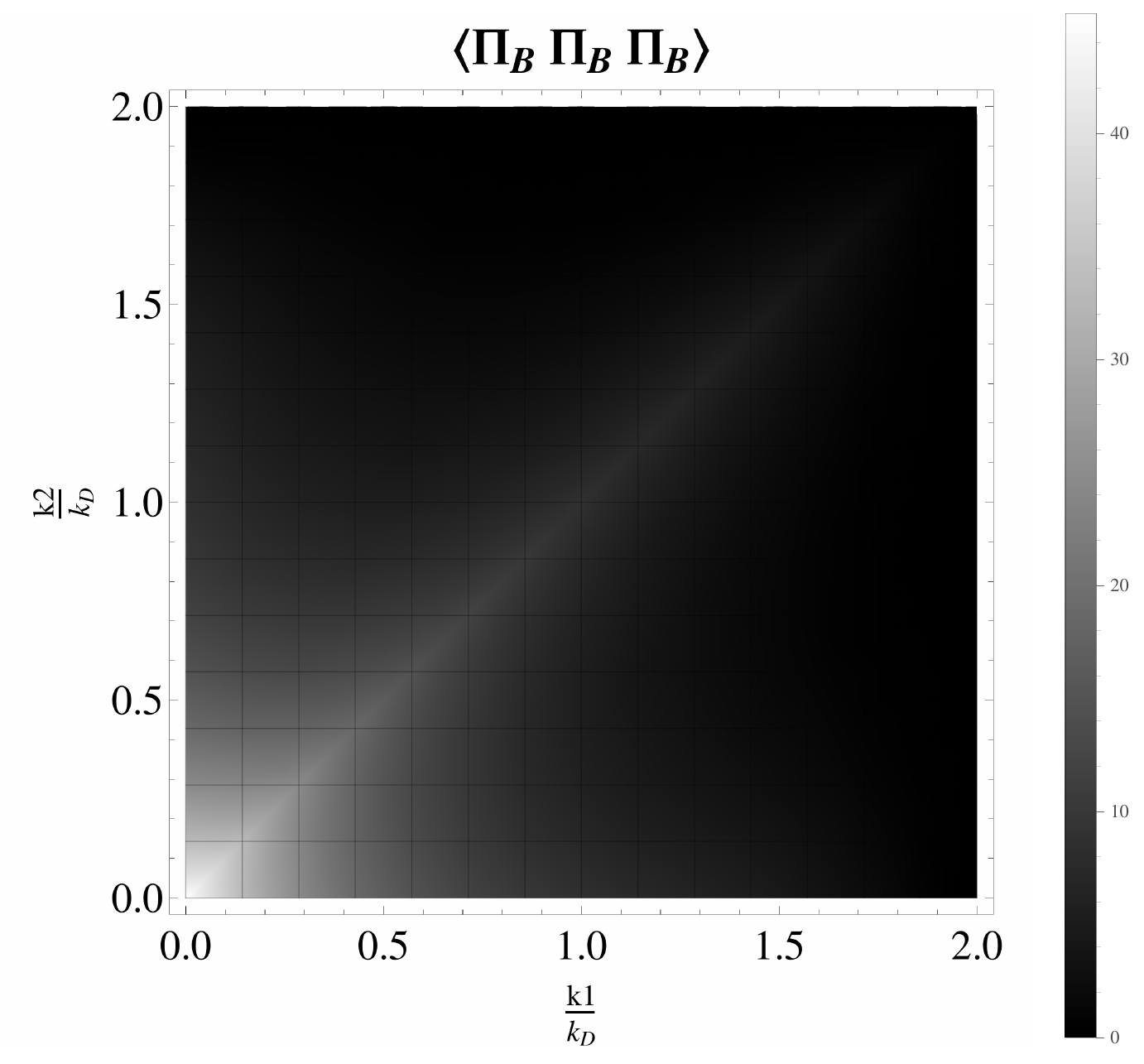}
       \caption{{\footnotesize Even contribution of  three-point correlation of $\langle \Pi_B \Pi_B  \Pi_B \rangle$ in units of $\frac{1}{(2\pi)^3(4\pi \rho_{\gamma,0})^3}$.}}
        \label{fig:mouse}
    \end{subfigure}
    
       \begin{subfigure}[b]{0.3\textwidth}
        \includegraphics[width=\textwidth]{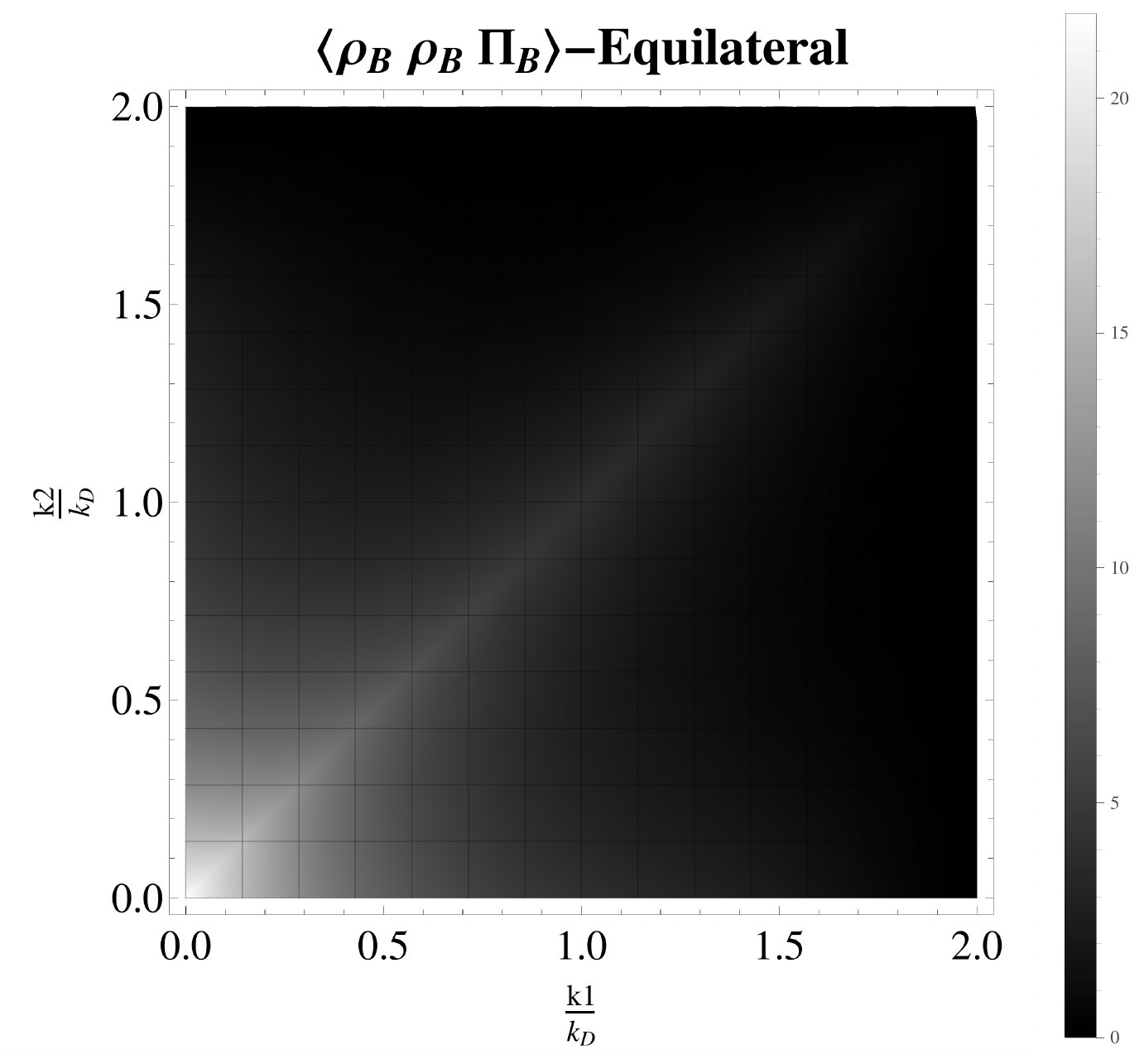}
        \caption{{\footnotesize Even contribution of  three-point correlation of $\langle \rho_B \rho_B \Pi_B \rangle$ in units of $\frac{4}{(2\pi)^3(4\pi \rho_{\gamma,0})^3}$ in the  equilateral configuration.}}
        \label{fig1aa}
    \end{subfigure}
    ~ 
    \begin{subfigure}[b]{0.3\textwidth}
         \includegraphics[width=\textwidth]{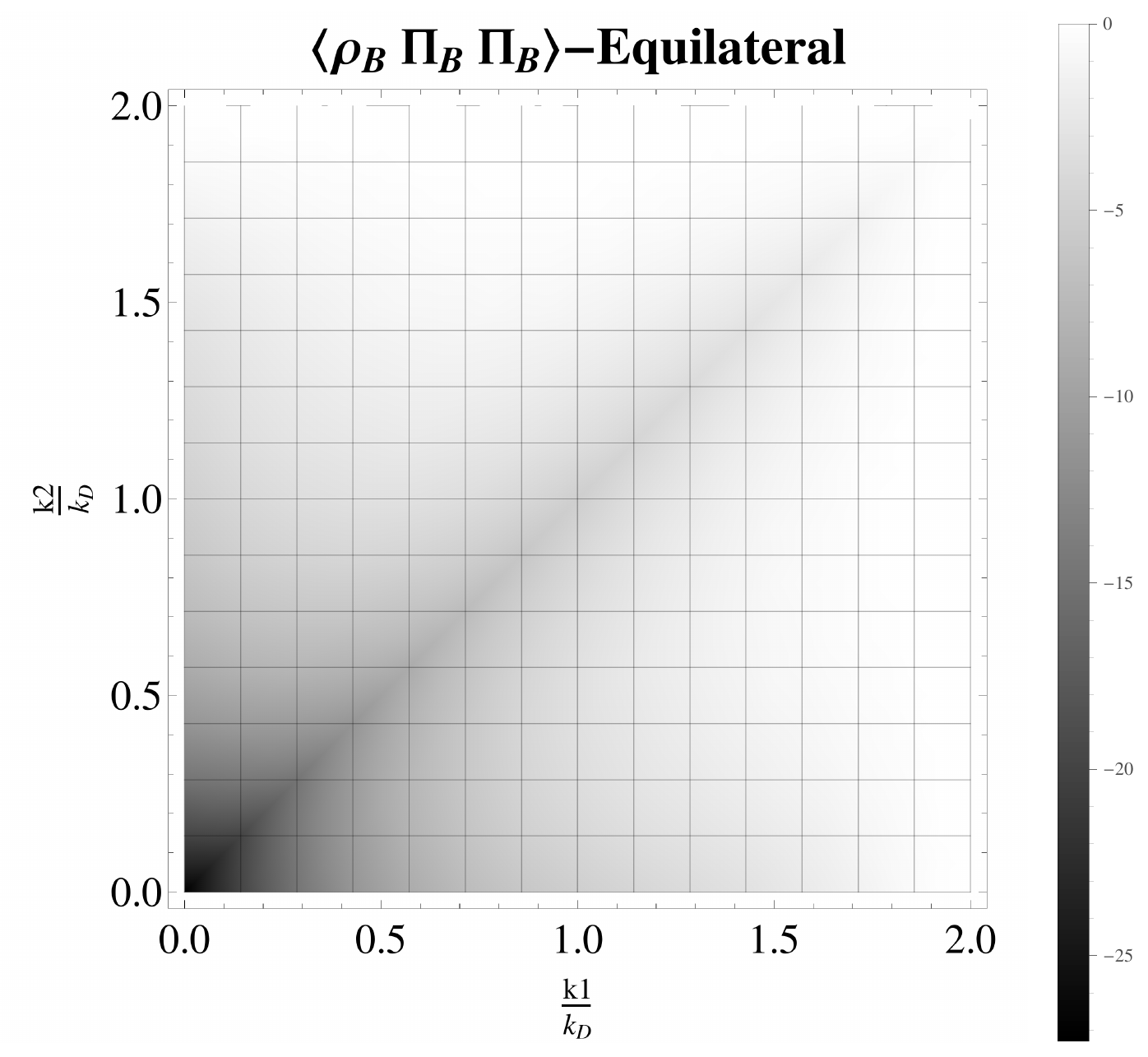}
        \caption{{\footnotesize Even contribution of  three-point correlation of $\langle \rho_B \Pi_B  \Pi_B \rangle$ in units of $\frac{2}{(2\pi)^3(4\pi \rho_{\gamma,0})^3}$ in the  equilateral configuration.}}
        \label{fig2aa}
    \end{subfigure}
    ~ 
    \begin{subfigure}[b]{0.3\textwidth}
        \includegraphics[width=\textwidth]{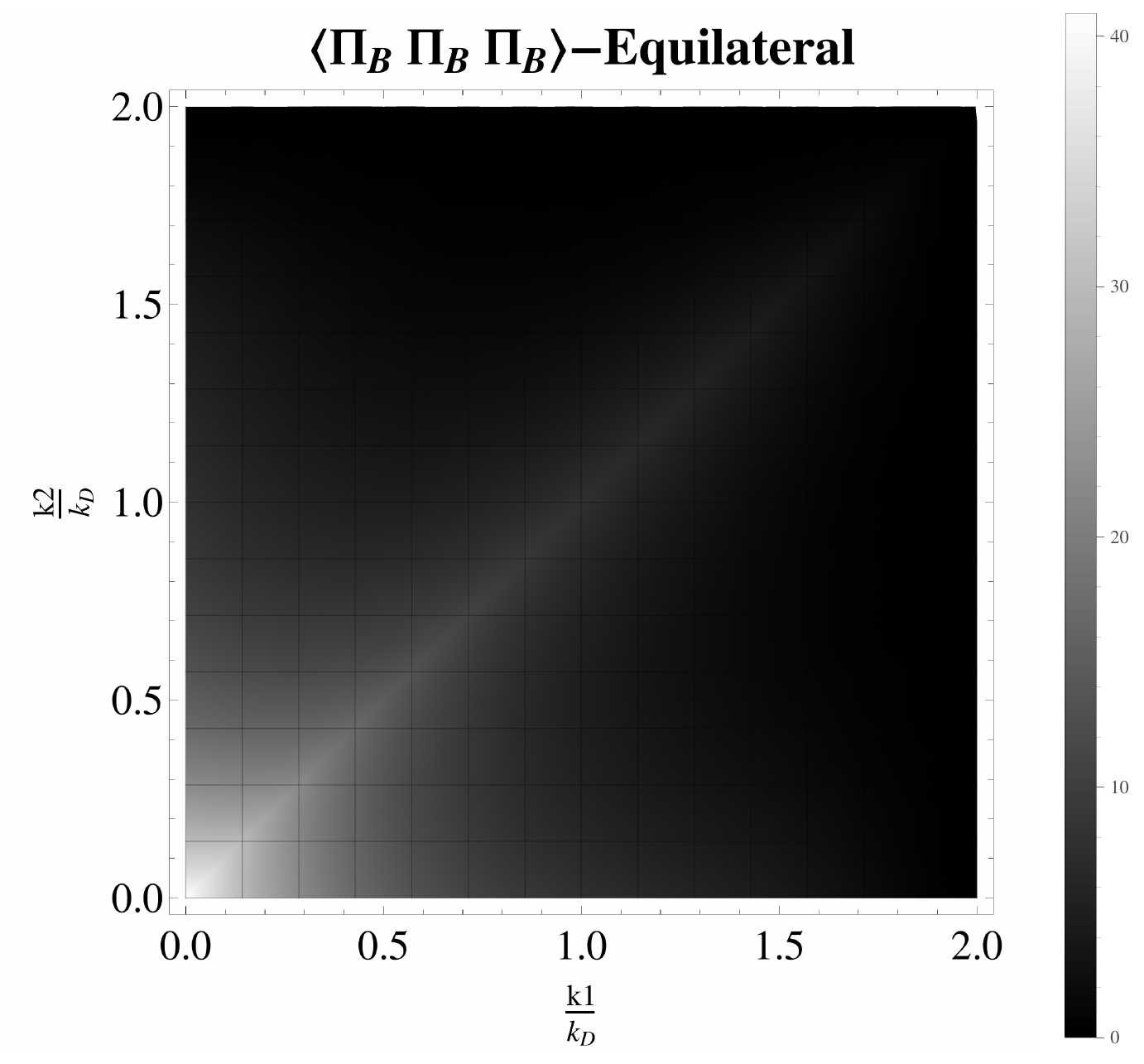}
       \caption{{\footnotesize Even contribution of  three-point correlation of $\langle \Pi_B \Pi_B  \Pi_B \rangle$ in units of $\frac{1}{(2\pi)^3(4\pi \rho_{\gamma,0})^3}$ in the  equilateral configuration.}}
        \label{fig:mouse}
    \end{subfigure}
    \caption{{\footnotesize
Total contribution of  three-point correlation of all scalar modes described in the text using the $\textbf{p}$-independent approximation. The figures (a), (b) (c) show the three-point correlation of the energy density of the magnetic field without, with $A_BA_H^2$ and full contribution respectively. The figures (d), (e) and (f) show the cross three-point correlation of the field. Finally, the  figures (g), (h) (i) show the cross three-point correlation field in the equilateral configuration, where the  $\langle \rho_B \Pi_B  \Pi_B \rangle$ has a total negative contribution. We can see that largest contribution to the bispectrum is obtained when $k1\sim k2$. We can also see that   biggest contributions to the scalar modes is given by $\langle \Pi_B \Pi_B  \Pi_B \rangle$ when we consider a squeezed configuration. }}\label{figparte1}
\end{figure}

n this approximation, the magnitude of one wave vector  ($\hat{\mathbf{k3}}$) is small while the others have equal magnitudes but have opposing directions ($\hat{\mathbf{k1}}=-\hat{\mathbf{k2}}$) as shown in figure \ref{fig2}. 
With this assumption the angles can be reduced to
\begin{eqnarray}
\beta&=&\hat{\mathbf{p}}\cdot\widehat{\mathbf{k1}-\mathbf{p}}\sim -\hat{\mathbf{p}}\cdot\widehat{\mathbf{k1}-\mathbf{p}}\sim-\gamma, \quad \mu\sim-\widehat{\mathbf{k1}-\mathbf{p}}\cdot\widehat{\mathbf{k1}-\mathbf{p}}\sim -1, \nonumber\\
\alpha_k&\sim& -\alpha_p\sim \alpha_q, \quad \beta_k \sim -\beta_p\sim \beta_q\sim -\gamma_k\sim \gamma_p\sim-\gamma_q.
\end{eqnarray}
By using this approximation the  values of the $F$'s are simplified to
\begin{equation}
F_{\rho \rho\rho}^{1}=1+\beta^2, \quad F_{\rho \rho\rho}^{2}=-(1+\beta^2),  \quad F_{\rho \rho\rho}^{3}=-2\beta,\quad F_{\rho \rho\rho}^{4}=2\beta,
\end{equation}
\begin{eqnarray}
F_{\rho \rho\Pi}^{1}&=&-2+3\alpha_k^2+\beta^2-6\alpha_k\beta\beta_k+3\beta_k^2+3\beta^2\beta_k^2, \quad F_{\rho \rho\Pi}^{2}=1-3\alpha_k^2-2\beta^2+6\alpha_k\beta\beta_k, \nonumber \\
F_{\rho \rho\Pi}^{3}&=&-\beta(-1+3\beta_k^2), \quad F_{\rho \rho\Pi}^{4}=\beta(-1+3\beta_k^2),
\end{eqnarray}
\begin{eqnarray}
F_{\rho \Pi\Pi}^{1}&=&-9\alpha_k\beta\beta_k^3+(2-3\beta_k^2)^2+\beta^2(1+3\beta_k^2)+\alpha_k^2(-3+9\beta_k^2), \quad F_{\rho \Pi\Pi}^{3}=3\alpha_k\beta_k-\beta(4-3\beta_k^2), \nonumber\\
F_{\rho \Pi\Pi}^{2}&=&-2-2\beta^2+3\alpha_k\beta\beta_k+3\beta_k^2,\quad F_{\rho \Pi\Pi}^{4}=-6\alpha_k\beta_k+9\alpha_k\beta_k^3-\beta(-5+6\beta_k^2),
\end{eqnarray}
\begin{eqnarray}
F_{\Pi \Pi\Pi}^{1}&=&-8+\beta^2+18\beta_k^2+3\beta^2\beta_k^2-9\beta_k^4+6\alpha_k\beta\beta_k(1-3\beta_k^2)+9\alpha_k^2(1-3\beta_k^2+3\beta_k^4),  \nonumber \\
F_{\Pi \Pi\Pi}^{2}&=& 4-2\beta^2-3\beta_k^2+\alpha_k^2(-6+9\beta_k^2), \quad  F_{\Pi \Pi\Pi}^{3}=F_{\Pi \Pi\Pi}^{4}=(2\beta-3\alpha_k\beta_k)(1-3\beta_k^2).
\end{eqnarray}
Same results have been obtained in \cite{25} for $F_{\Pi \Pi\Pi}^{1}$ (case where there is not helicity).  The angular part of the integrals must be written in spherical coordinates $d^3p=p^2 dp d\alpha_kd\theta$, where $\theta$ is  the azimuthal angle. Since we consider an upper cut-off $k_D$,
we must introduce the ($k1,k2$)-dependence on the angular integration domain; this implies that we should split the integral domain
in different regions such as  $0 <k1,k2< 2k_D$.
The  integration domains we use for calculating the integrals are shown in appendix \ref{apenb}. By using the power spectrum expression for the magnetic fields
eqs.(\ref{powerPMF1})-(\ref{powerPMF2}) and the F's values for the $\mathbf{p}-$independent case given above, we get the  causal magnetic bispectrum ($n_B=n_H=2$) which is shown in the Figure \ref{figparte1}. We see that the most contribution for the bispectrum occurs when $k1 \sim k2$  and $\langle \Pi_B \Pi_B  \Pi_B \rangle$ generates the largest value for  scalar mode. Hence, we conclude that the shape of the non-Gaussian associated with the PMF  can be classified into the local-type configuration as was previously reported in \cite{28} for a scale invariant shape.  We also observe that effects from $A_BA_H^2$  contribution are smaller with respect to  $A_B^3$.  
The figures \ref{fig1aa}, \ref{fig2aa} show the cross-correlation of the bispectrum obtaining the same behavior and a large contribution (with respect to the energy density bispectrum).

\begin{figure}[h!]
    \centering
    \begin{subfigure}[b]{0.3\textwidth}
         \includegraphics[width=\textwidth]{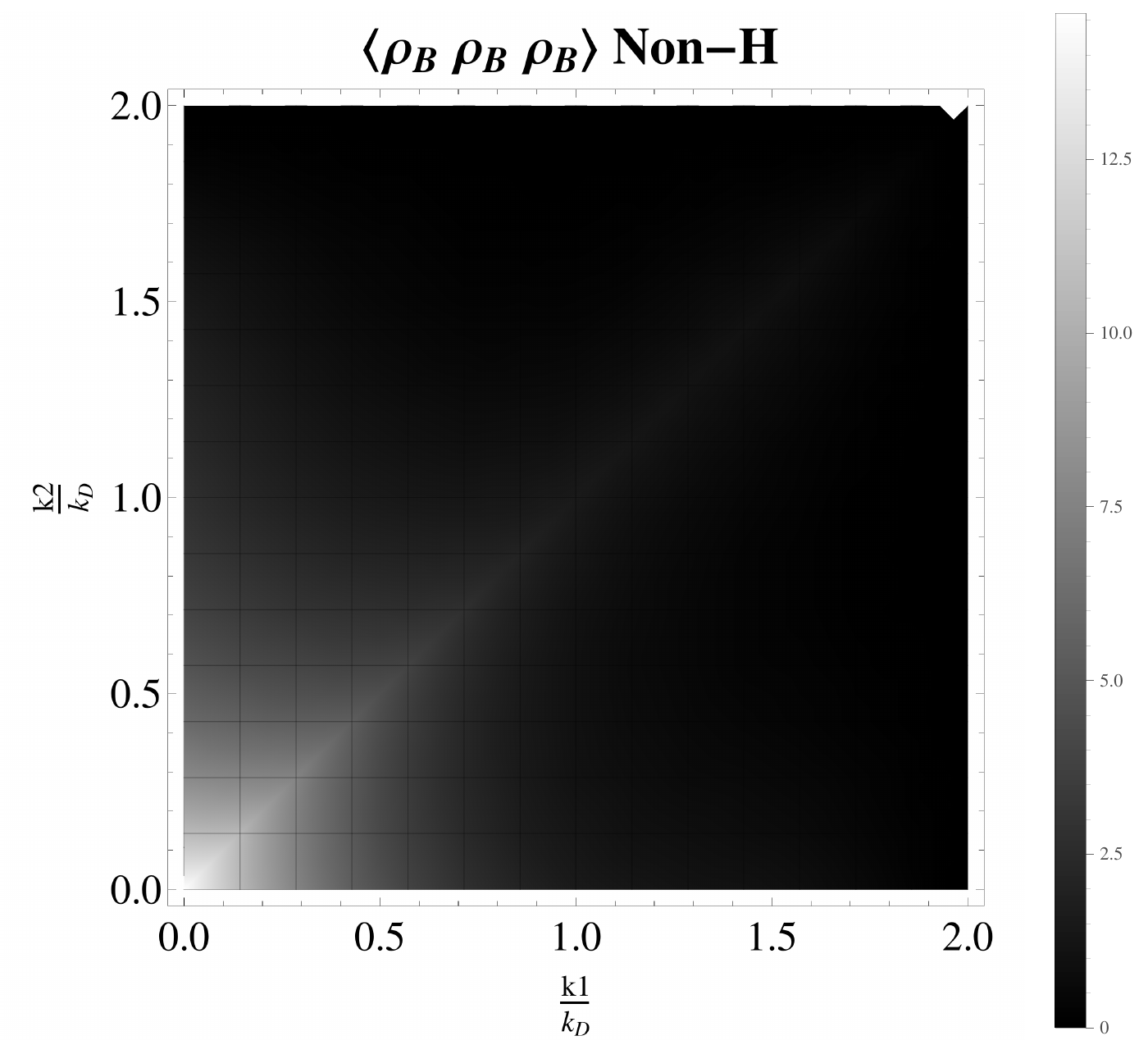}
        \caption{{\footnotesize Three-point correlation of $\langle \rho_B \rho_B \rho_B \rangle$ in units of $\frac{8}{(2\pi)^3(4\pi \rho_{\gamma,0})^3}$ without $A_BA_H^2$.}} 
        \label{fig:gull}
    \end{subfigure}
    ~ 
    \begin{subfigure}[b]{0.3\textwidth}
     \includegraphics[width=\textwidth]{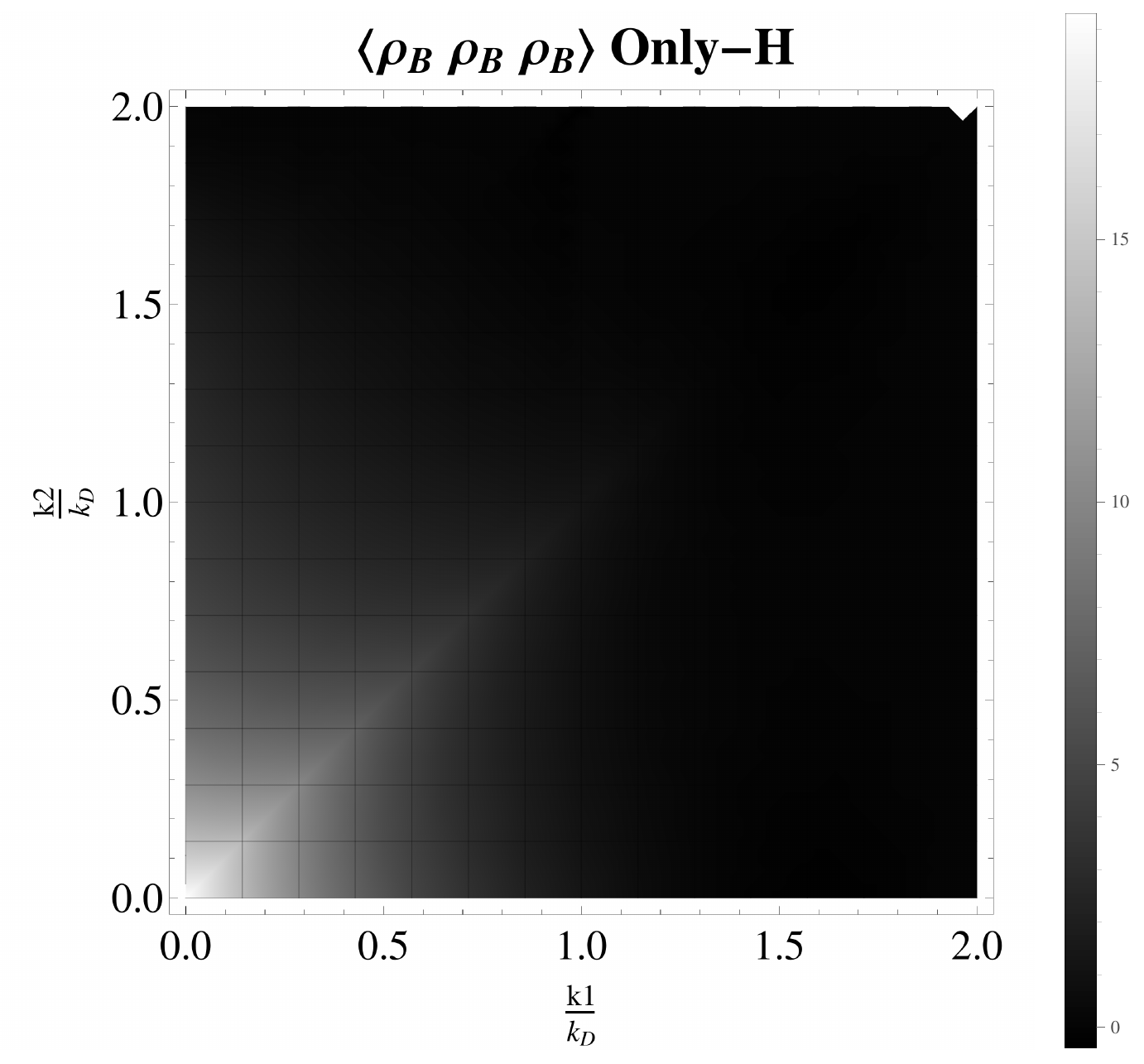}
        \caption{{\footnotesize Three-point correlation of $\langle \rho_B \rho_B \rho_B \rangle$ in units of $\frac{8}{(2\pi)^3(4\pi \rho_{\gamma,0})^3}$ only with $A_BA_H^2$.}}
        \label{fig:tiger}
    \end{subfigure}
    ~ 
    \begin{subfigure}[b]{0.3\textwidth}
         \includegraphics[width=\textwidth]{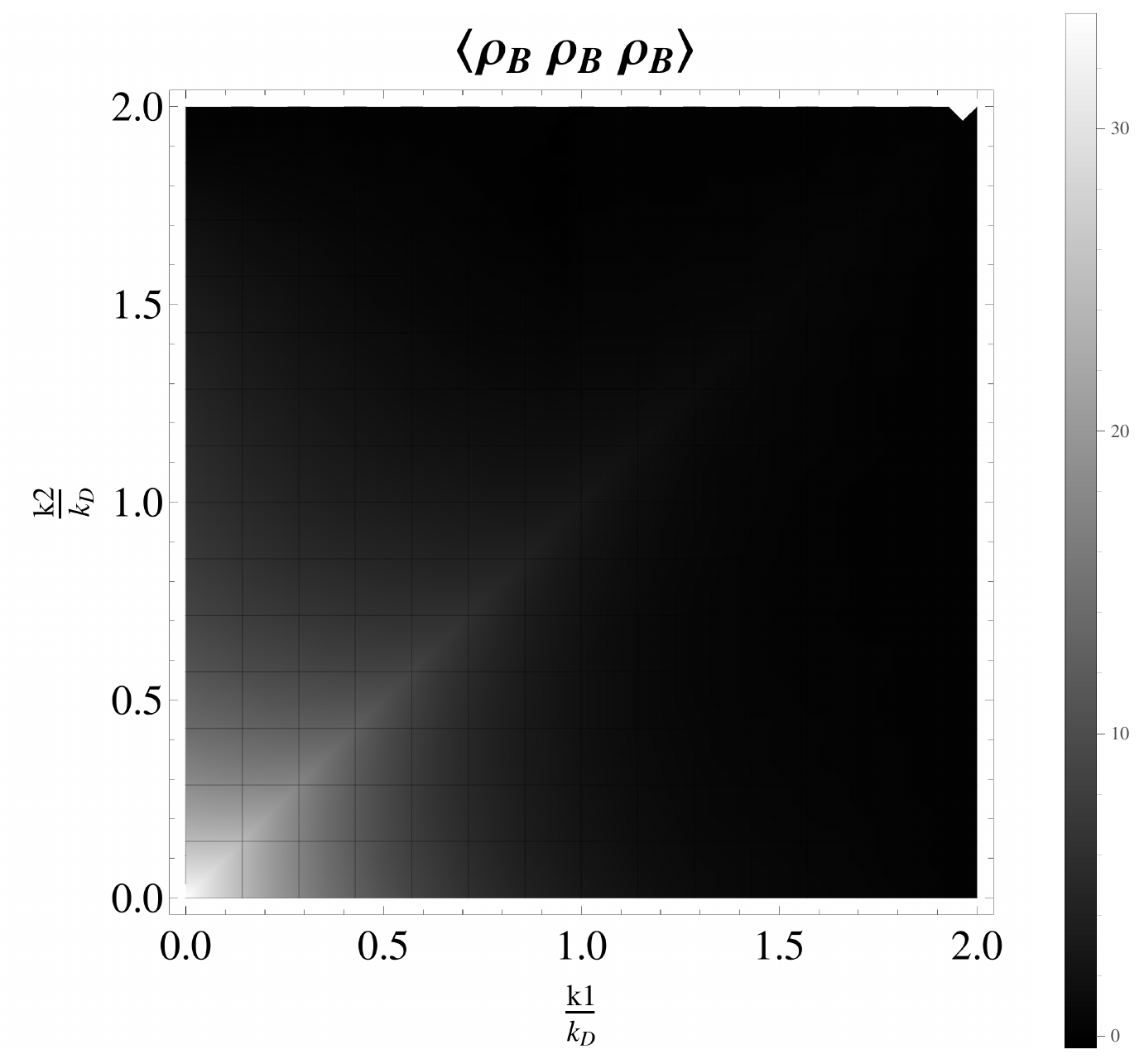}
        \caption{{\footnotesize Even contribution of  three-point correlation of $\langle \rho_B \rho_B \rho_B \rangle$ in units of $\frac{8}{(2\pi)^3(4\pi \rho_{\gamma,0})^3}$.}}
        \label{fig:mouse}
    \end{subfigure}
    \begin{subfigure}[b]{0.3\textwidth}
        \includegraphics[width=\textwidth]{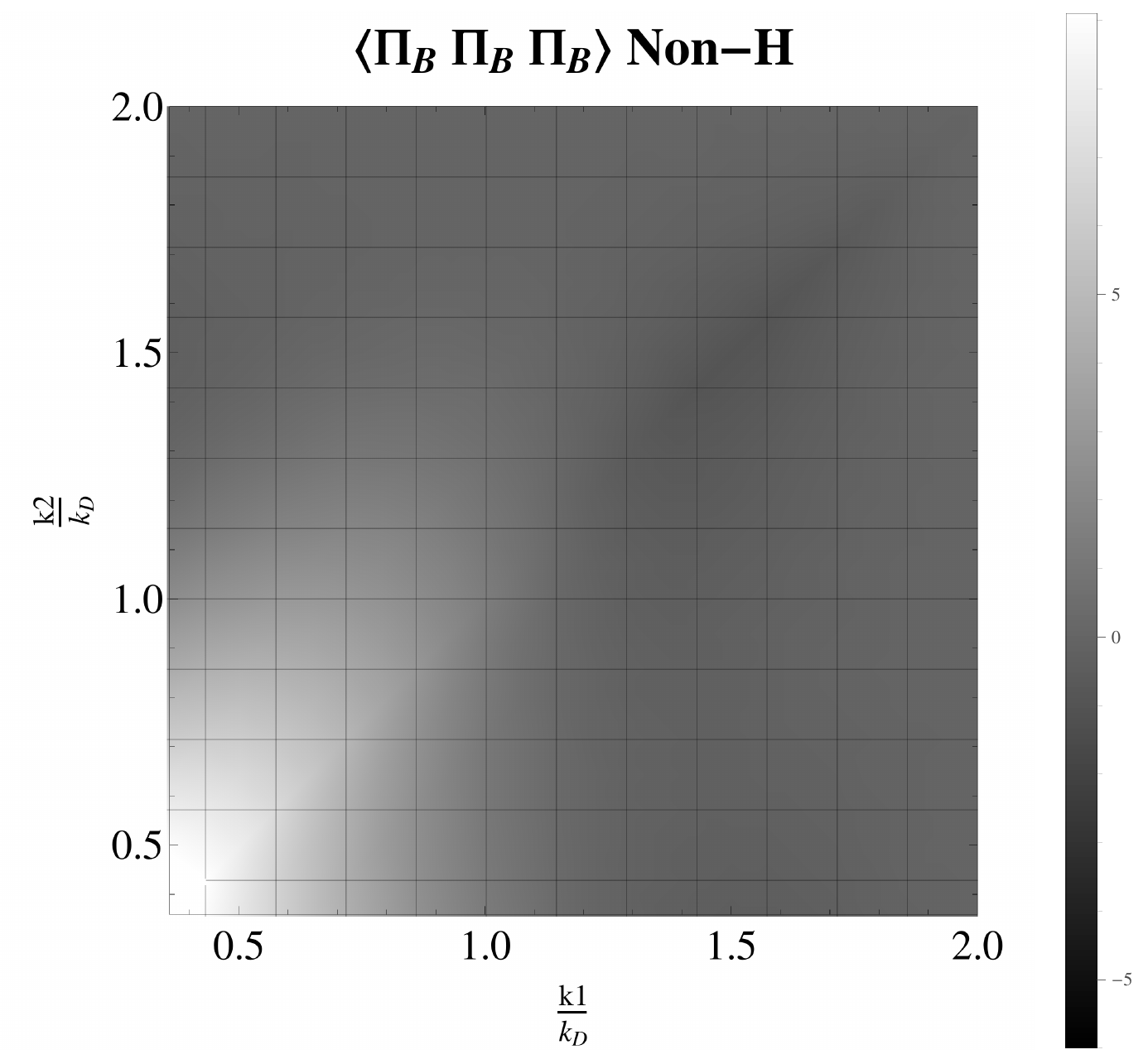}
        \caption{{\footnotesize Three-point correlation of $\langle \Pi_B \Pi_B  \Pi_B \rangle$ in units of $\frac{1}{(2\pi)^3(4\pi \rho_{\gamma,0})^3}$ without $A_BA_H^2$.}}
        \label{fig:gull}
    \end{subfigure}
    ~ 
    \begin{subfigure}[b]{0.3\textwidth}
        \includegraphics[width=\textwidth]{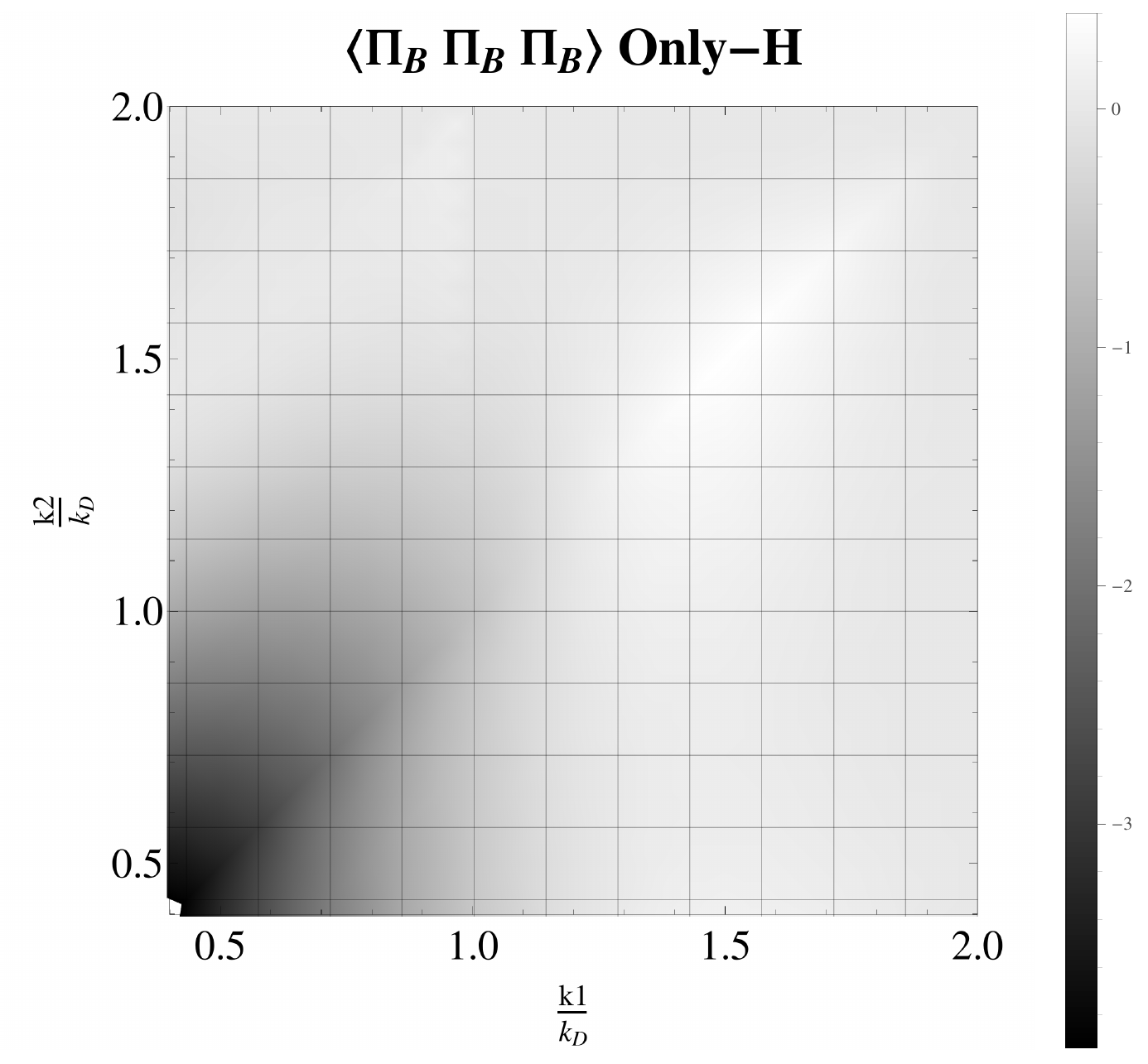}
        \caption{{\footnotesize Three-point correlation of $\langle \Pi_B \Pi_B  \Pi_B \rangle$ in units of $\frac{1}{(2\pi)^3(4\pi \rho_{\gamma,0})^3}$ only with $A_BA_H^2$.}}
        \label{fig:tiger}
    \end{subfigure}
    ~ 
    \begin{subfigure}[b]{0.3\textwidth}
        \includegraphics[width=\textwidth]{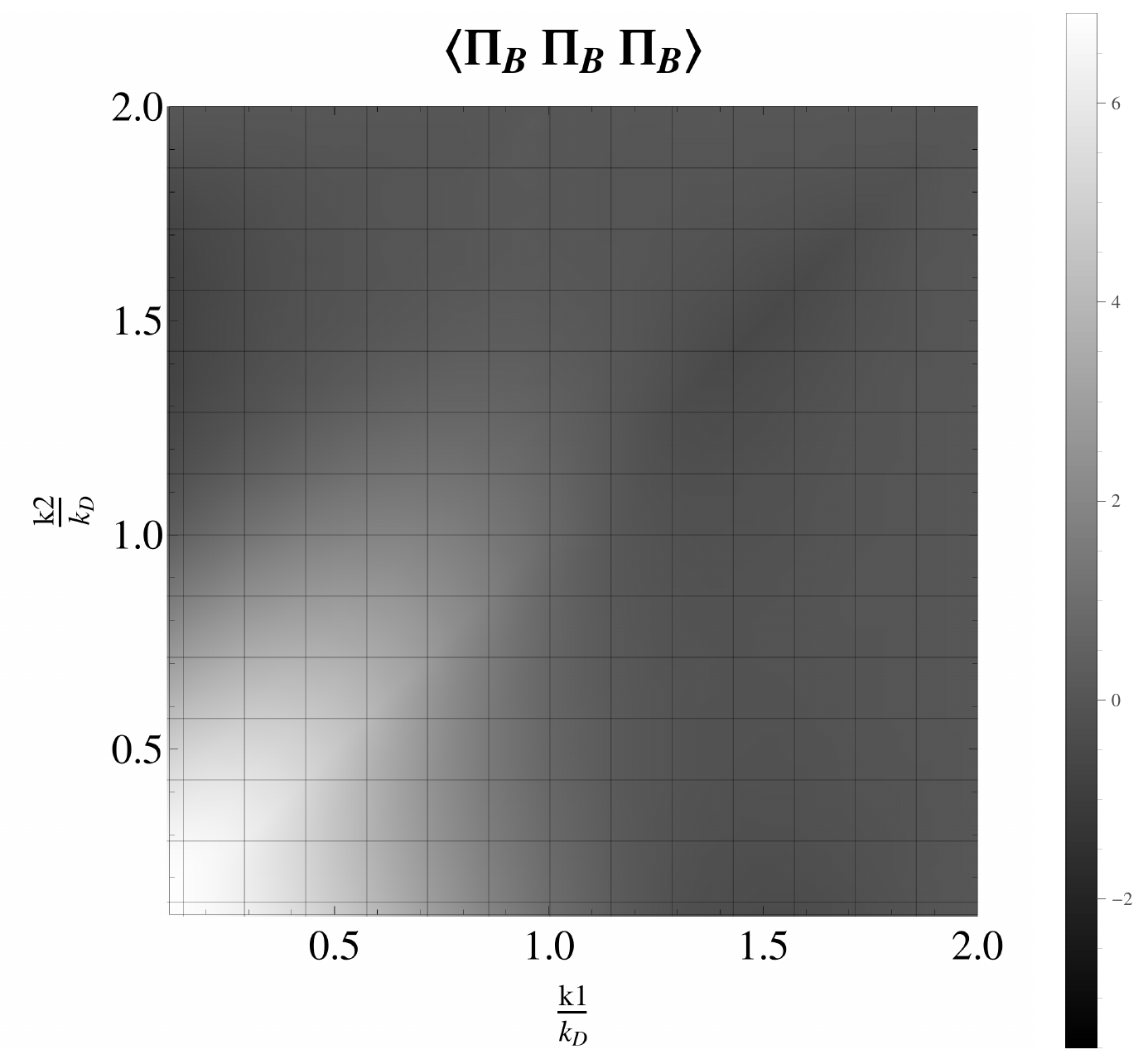}
        \caption{{\footnotesize Even contribution of  three-point correlation of $\langle \Pi_B \Pi_B  \Pi_B \rangle$ in units of $\frac{1}{(2\pi)^3(4\pi \rho_{\gamma,0})^3}$.}}
        \label{fig:mouse}
    \end{subfigure}
     \begin{subfigure}[b]{0.3\textwidth}
        \includegraphics[width=\textwidth]{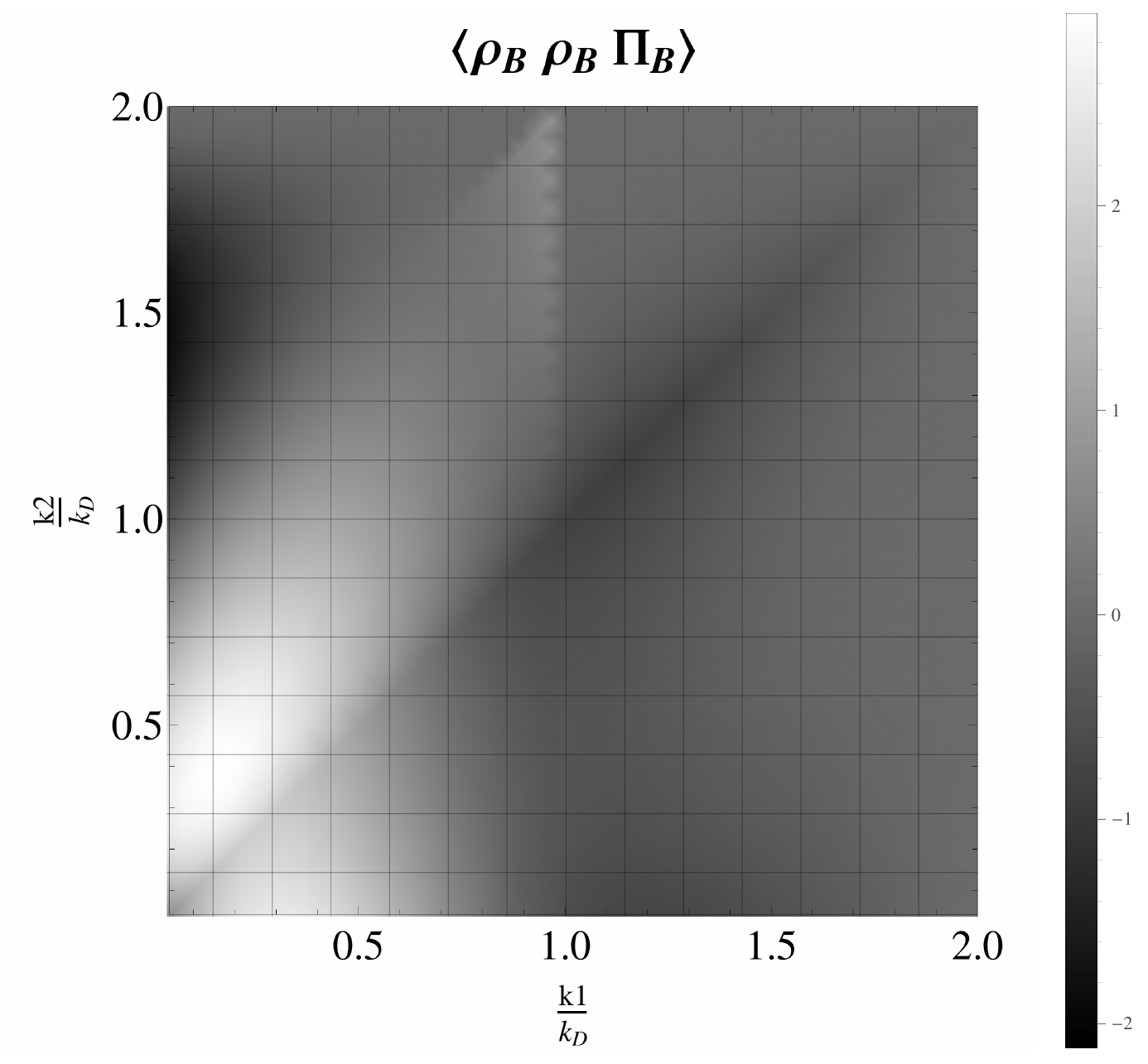}
        \caption{{\footnotesize Three-point correlation of $\langle  \rho_B  \rho_B   \Pi_B \rangle$ in units of $\frac{4}{(2\pi)^3(4\pi \rho_{\gamma,0})^3}$.}}
        \label{fig:gull}
    \end{subfigure}
    ~ 
    \begin{subfigure}[b]{0.3\textwidth}
        \includegraphics[width=\textwidth]{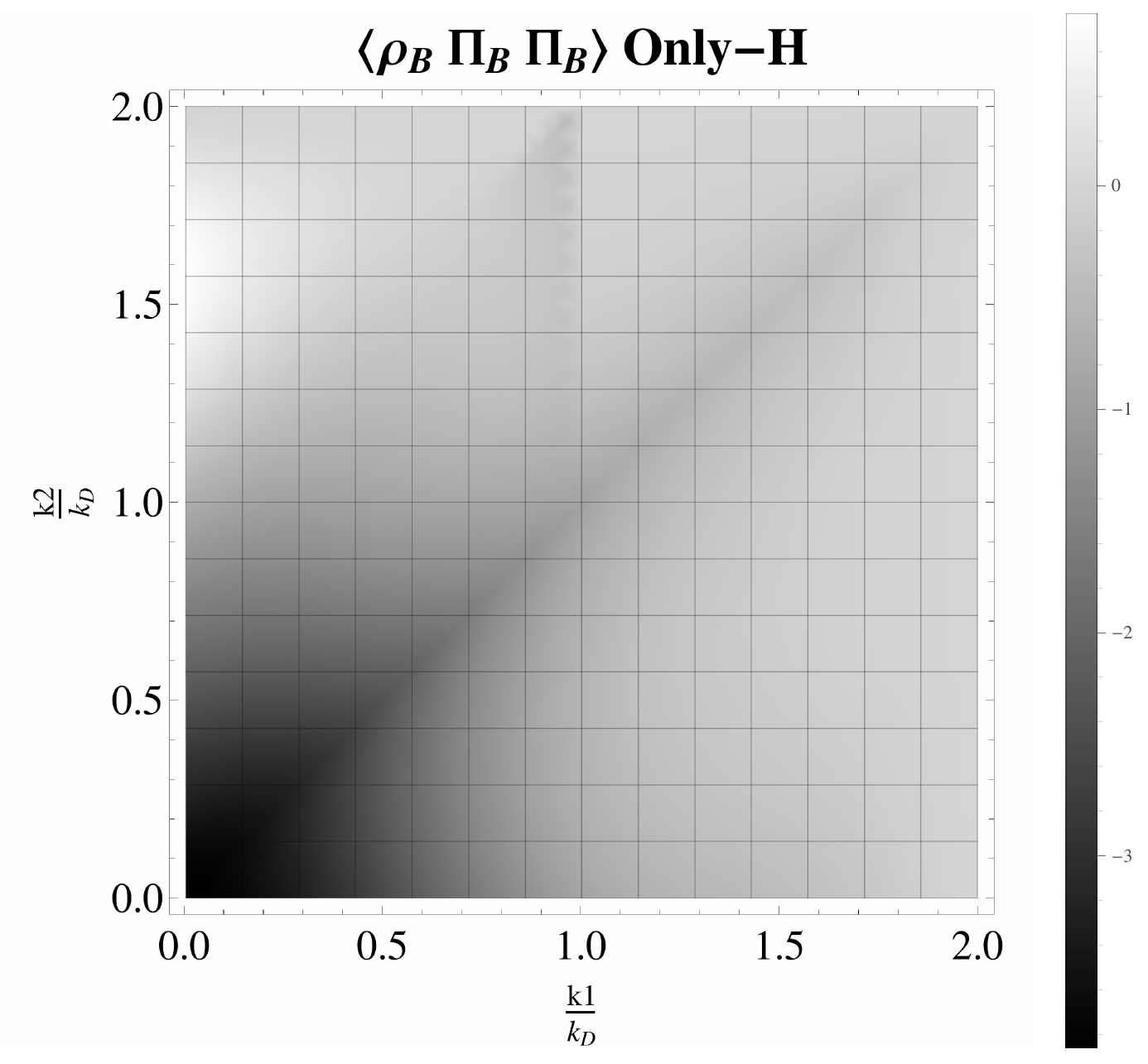}
        \caption{{\footnotesize Three-point correlation of $\langle  \rho_B  \Pi_B  \Pi_B \rangle$ in units of $\frac{2}{(2\pi)^3(4\pi \rho_{\gamma,0})^3}$ only with $A_BA_H^2$.}}
        \label{fig:tiger}
    \end{subfigure}
    ~ 
    \begin{subfigure}[b]{0.3\textwidth}
        \includegraphics[width=\textwidth]{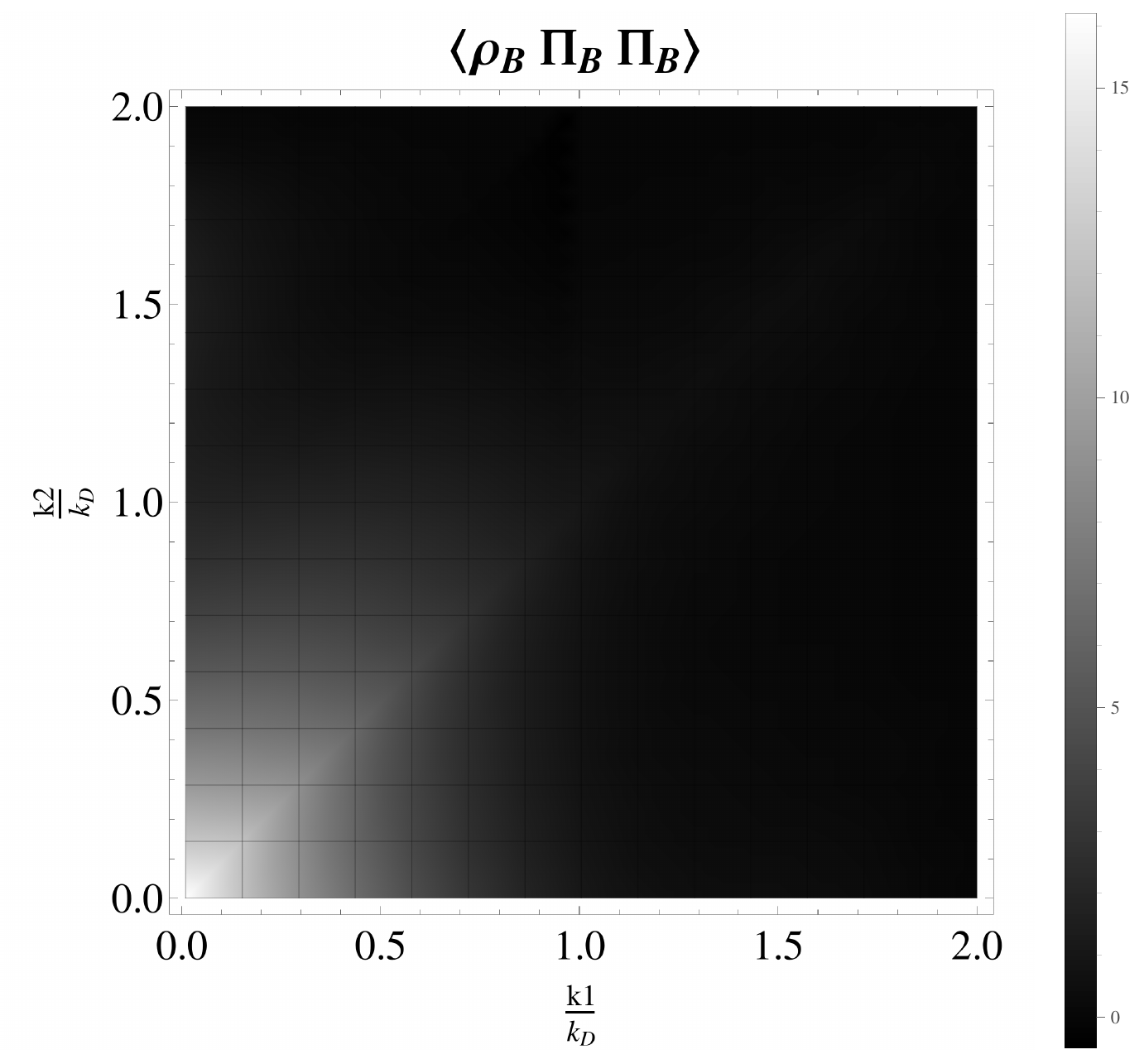}
        \caption{{\footnotesize Even contribution of  three-point correlation of $\langle  \rho_B  \Pi_B  \Pi_B \rangle$ in units of $\frac{2}{(2\pi)^3(4\pi \rho_{\gamma,0})^3}$.}}
        \label{fig:mouse}
    \end{subfigure}
    \caption{{\footnotesize Total contribution of  three-point correlation of non-crossing scalar modes described in the text using the squeezed collinear configuration. The figures (a), (b) (c) show the three-point correlation of the energy density of the magnetic field without, with $A_BA_H^2$ and full contribution respectively; while  figures (d), (e) and (f) show  the three-point correlation of the anisotropic stress of the magnetic field without, with $A_BA_H^2$ and full contribution respectively. On the other hand, the figures  (g), (h) and (i) show the cross three-point correlation of the field in this configuration.  We can see that $k1\sim k2$ has the biggest contribution to the bispectrum. One important feature of the anisotropic stress mode  is  the negative contribution to the total bispectrum for wavevenumbers larger than $K_D/2$. 
      }}\label{figparte2}
\end{figure}
On the other hand, in  figure  \ref{figparte2} we present the results under the squeezed collinear configuration driven also by causal fields. Here, we see that the bispectrum for this configuration is less than the $\mathbf{p}-$independent case. We also see that magnetic anisotropic stress change the sign under this configuration for wavenumbers larger than $k_D/2$, while the $A_BA_H^2$ contribution is practically negative.  Note that all the palettes shown here are sequential colour palettes, so they are very well suited to show the amplitude of the bispectrum.
Finally, for negative spectral indices, an approximate solution for the bispectrum can be found using the formula (5.13) in \cite{31} (see also equation (\ref{apenb1}) in appendix \ref{apenb}). Assuming  ($n_B=n_H=n$) along with  $k2<k1<k_D$ , the  expression for bispectrum becomes
\begin{eqnarray}
B_{\rho_B^{(S)}\rho_B^{(S)}\rho_B^{(S)}}^{(S)}&\sim&\frac{16}{(2\pi)^3(4\pi \rho_{\gamma,0})^3}\left(\frac{nk1^nk2^{2n+3}}{(n+3)(2n+3)}+\frac{nk1^{3n+3}}{(2n+3)(3n+3)}+\frac{k_D^{3n+3}}{(3n+3)}\right) \nonumber\\
&\times&\left(A_s^3 F_{\rho \rho\rho}^{1}+A_H^2A_s\left(F_{\rho\rho\rho}^{3}-F_{\rho\rho\rho}^{4}-F_{\rho\rho\rho}^{2}\right)\right),
\end{eqnarray}
\begin{eqnarray}
B_{\Pi_B^{(S)}\Pi_B^{(S)}\Pi_B^{(S)}}^{(S)}&\sim&\frac{2}{(2\pi)^3(4\pi \rho_{\gamma,0})^3}\left(\frac{nk1^nk2^{2n+3}}{(n+3)(2n+3)}+\frac{nk1^{3n+3}}{(2n+3)(3n+3)}+\frac{k_D^{3n+3}}{(3n+3)}\right) \nonumber\\
&\times&\left(A_s^3 F_{\Pi \Pi\Pi}^{1}+A_H^2A_s\left(F_{\Pi \Pi\Pi}^{3}+F_{\Pi \Pi\Pi}^{4}-F_{\Pi \Pi\Pi}^{2}\right)\right).
\end{eqnarray}
\subsection{Infrared  Cut-off}
We now analyze the effect of an infrared cut-off parametrized by $\alpha$ on the magnetic bispectrum (see equation (\ref{mcut-off})).  We saw that the non-Gaussianity peaks at  $k1 \sim k2$  under a squeezed configuration, so we  compute the magnetic bispectrum using the strategy adopted in \cite{21} and in appendix \ref{apenc}. In figure \ref{figparte22}, the effect of this IR cut-off for causal fields is illustrated. 
\begin{figure}[h!]
    \centering
    \begin{subfigure}[b]{0.45\textwidth}
        \includegraphics[width=\textwidth]{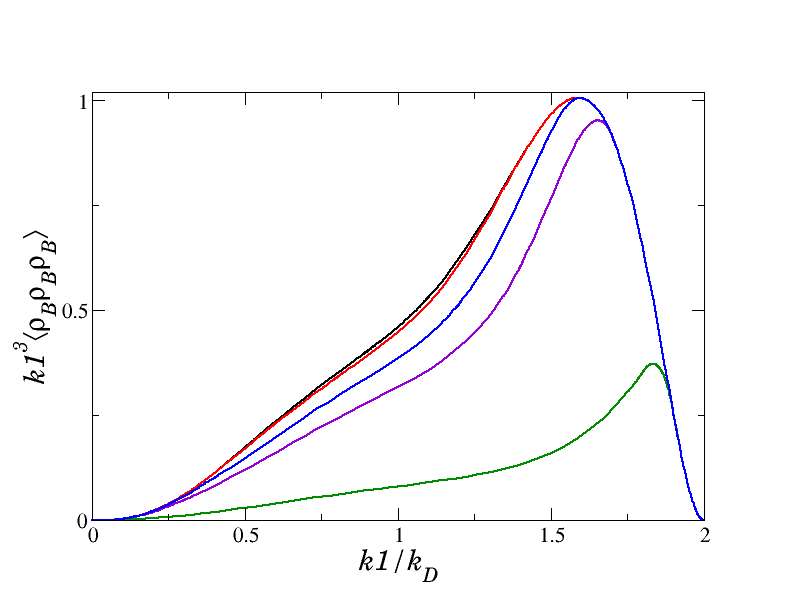}
        \caption{{\footnotesize Change of  $k1^3\langle \rho_B \rho_B \rho_B \rangle$ respect to  the infrared cut-off without $A_BA_H^2$.}} 
        \label{fig6a}
    \end{subfigure}
    ~ 
    \begin{subfigure}[b]{0.45\textwidth}
        \includegraphics[width=\textwidth]{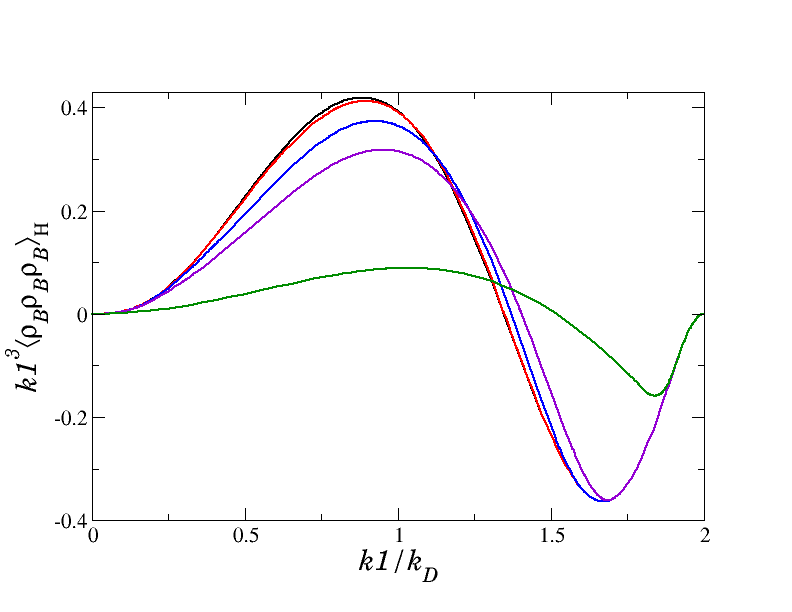}
        \caption{{\footnotesize \footnotesize Change of  $k1^3\langle \rho_B \rho_B \rho_B \rangle$ respect to  the infrared cut-off only with $A_BA_H^2$.}}
        \label{fig6b}
    \end{subfigure}
    ~ 
    \begin{subfigure}[b]{0.45\textwidth}
        \includegraphics[width=\textwidth]{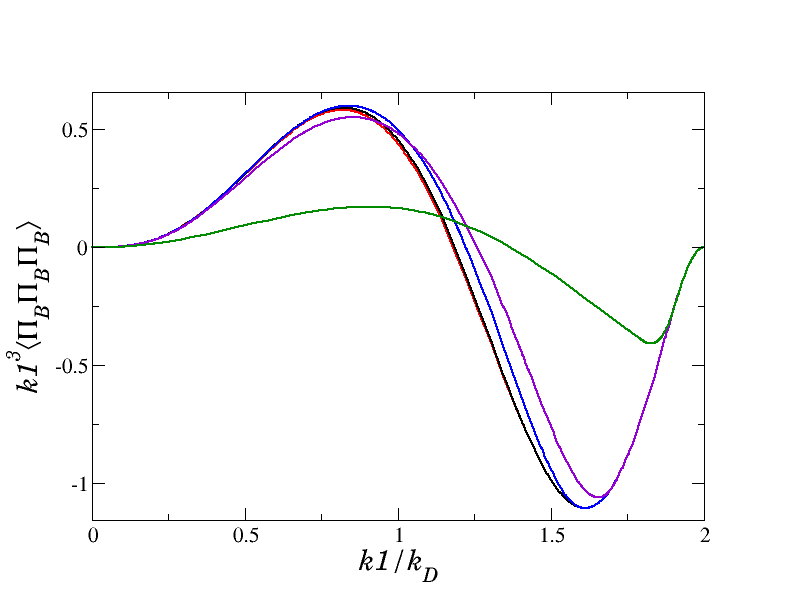}
        \caption{{\footnotesize Change of $k1^3\langle \Pi_B \Pi_B \Pi_B \rangle$ respect to  the infrared cut-off without $A_BA_H^2$.}}
        \label{fig6c}
    \end{subfigure}
    \begin{subfigure}[b]{0.45\textwidth}
        \includegraphics[width=\textwidth]{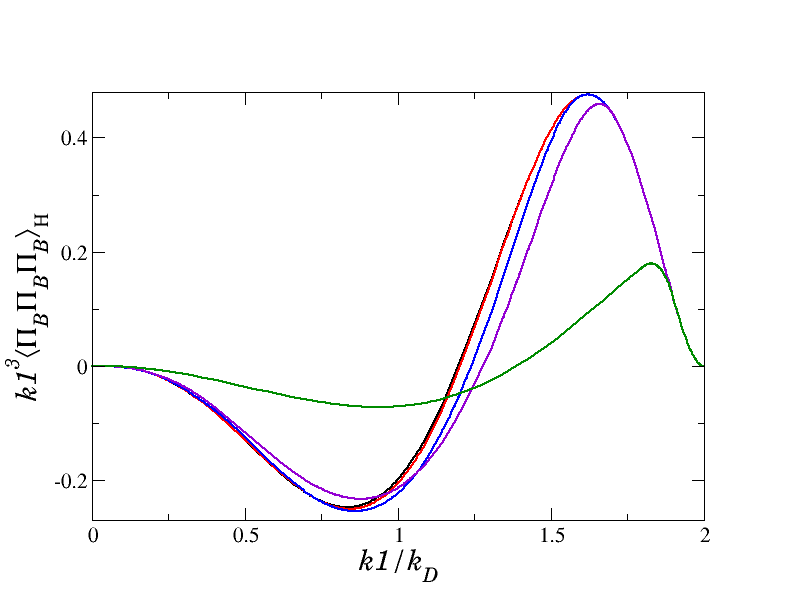}
        \caption{{\footnotesize Change of $k1^3\langle \Pi_B \Pi_B  \Pi_B \rangle$ respect to  the infrared cut-off only with $A_BA_H^2$.}}
        \label{fig6d}
    \end{subfigure}
    \caption{{\footnotesize Effects of a lower cut-off at the  three-point correlation of non-crossing scalar modes described in the text using the squeezed collinear configuration. The figures (a), (b) show the three-point correlation of the energy density of the magnetic field without and only with $A_BA_H^2$ respectively, while  figures (c) and (d) show  the three-point correlation of the anisotropic stress of the magnetic field without and only with $A_BA_H^2$  respectively.    The black, red, blue, violet and green  lines refer to lower cut-off for $\alpha=0.01$, $\alpha=0.4$, $\alpha=0.6$, $\alpha=0.7$, $\alpha=0.9$ respectively.  Here the units are normalized respect to  the values in figure (a) and we use $n_B=n_H=2$.
      }}\label{figparte22}
\end{figure}
\begin{figure}[h!]
    \centering
    \begin{subfigure}[b]{0.45\textwidth}
        \includegraphics[width=\textwidth]{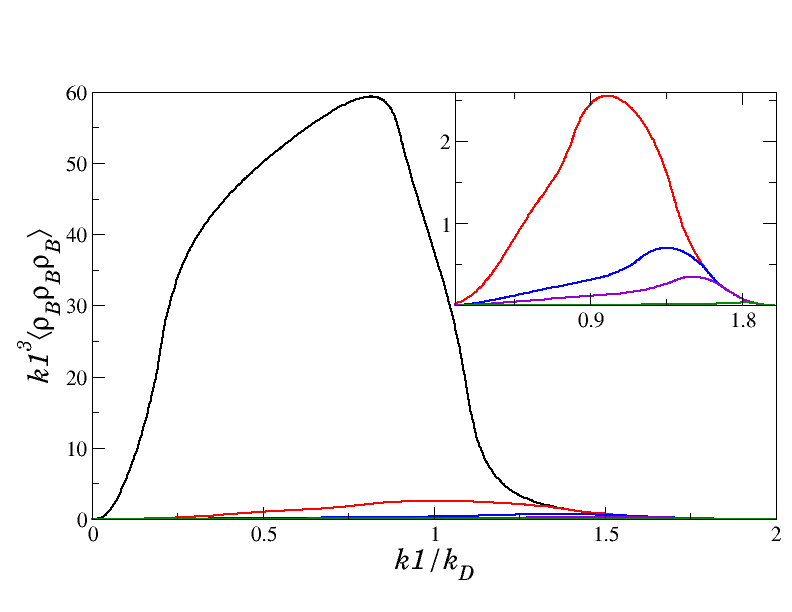}
        \caption{{\footnotesize Change of  $k1^3\langle \rho_B \rho_B \rho_B \rangle$ respect to  the infrared cut-off without $A_BA_H^2$.}} 
        \label{fig6aa}
    \end{subfigure}
    ~ 
    \begin{subfigure}[b]{0.45\textwidth}
        \includegraphics[width=\textwidth]{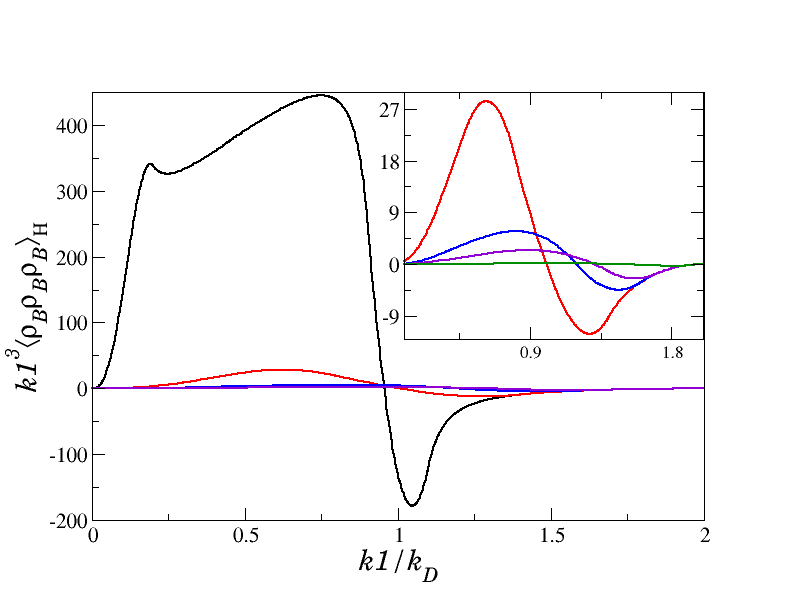}
        \caption{{\footnotesize \footnotesize Change of  $k1^3\langle \rho_B \rho_B \rho_B \rangle$ respect to  the infrared cut-off only with $A_BA_H^2$.}}
        \label{fig6bb}
    \end{subfigure}
    ~ 
    \begin{subfigure}[b]{0.45\textwidth}
        \includegraphics[width=\textwidth]{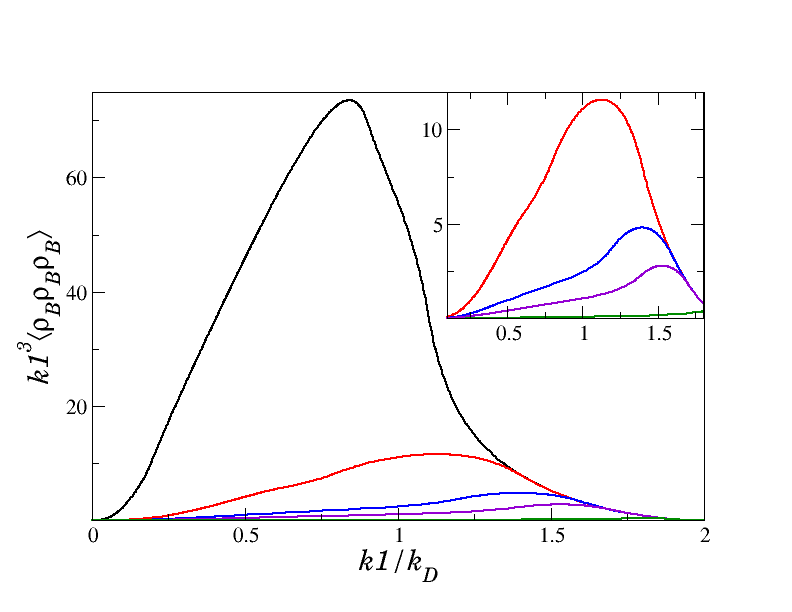}
        \caption{{\footnotesize Change of $k1^3\langle \rho_B \rho_B \rho_B \rangle$ respect to  the infrared cut-off without $A_BA_H^2$.}}
        \label{fig6cc}
    \end{subfigure}
    \begin{subfigure}[b]{0.45\textwidth}
        \includegraphics[width=\textwidth]{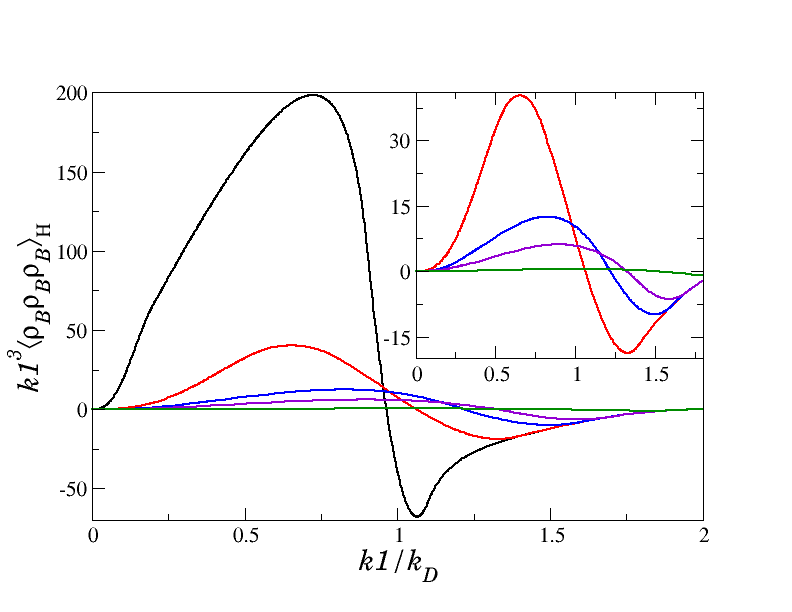}
        \caption{{\footnotesize Change of $k1^3\langle \rho_B \rho_B  \rho_B \rangle$ respect to  the infrared cut-off only with $A_BA_H^2$.}}
        \label{fig6dd}
    \end{subfigure}
    \caption{{\footnotesize Effects of a lower cut-off at the  three-point correlation of non-crossing scalar modes described in the text using the squeezed collinear configuration. The figures (a), (b) show the three-point correlation of the energy density of the magnetic field  for $n_B=n_H=-5/2$ without and only with $A_BA_H^2$ respectively, while  figures (c) and (d) show  the three-point correlation of the energy density of the magnetic field for  $n_B=n_H=-1.9$ without and only with $A_BA_H^2$  respectively. The black, red, blue, violet and green  lines refer to lower cut-off for $\alpha=0.01$, $\alpha=0.4$, $\alpha=0.6$, $\alpha=0.7$, $\alpha=0.9$ respectively.  Here the units are normalized respect to  the values in figure (\ref{fig6a}).
      }}\label{figparte22a}
\end{figure}
The top figures (\ref{fig6a}) and (\ref{fig6b}) show the effect of $\alpha$ on  $k1^3\langle \rho_B \rho_B \rho_B \rangle$ where the lines  refer to different values of the cut-off, while the bottom figures (\ref{fig6c}) and (\ref{fig6d}) show the  effect of $\alpha$ on  $k1^3\langle \Pi_B \Pi_B \Pi_B \rangle$.  What we read off these figures is how the peak of the bispectrum moves to high wavenumbers when we increase the value of $k_m$  in the same way that magnetic power spectrum (and its amplitude decreases also due to reduction of the wavenumber space) and how the effects of the $A_BA_H^2$ contributions PMF are tiny compared with the non-helical case. On the other hand, the figure (\ref{figparte22a}) shows the effect of the infrared cut-off when we are considering non-causal fields. The top panel shows   the three-point correlation of the energy density of the magnetic field for $n_H=n_B=-5/2$ while the bottom panel shows the  three-point correlation for $n_H=n_B=-1.9$. Here we can see that the contribution driven by $A_BA_H^2$ is bigger than $A_B^3$, meaning that for negative spectral indices the effect of helicity becomes  relevant for our studies.

\section{Reduced Bispectrum from PMF} \label{redbisa5}
In this section, we  estimate  the reduced bispectrum  and  give a careful review  the  results  of \cite{25},\cite{38},\cite{39}.
The CMB temperature perturbation at a direction of photon momentum $\hat{\mathbf{n}}$ can be expanded into spherical harmonics
\begin{equation}
\frac{\Delta T^{(Z)}}{T}(\hat{\mathbf{n}})=\sum_{lm}a^{(Z)}_{lm}Y_{lm}(\hat{\mathbf{n}}),
\end{equation}
where $Z=S,V,T$ refers to the contribution given by scalar , vector or tensor perturbations. The coefficient $a^{(Z)}_{lm}$ is written as \cite{30} 
\begin{eqnarray}\label{alm}
a^{Z}_{lm}&=&4\pi(-i)^{l}\int\frac{d^3\mathbf{k}}{(2\pi)^3}\Delta^{Z}_l(k)\sum_\lambda[sgn(\lambda)]^\lambda\xi_{lm}^{(\lambda)}(k), \nonumber\\
\xi_{lm}^{(\lambda)}(k)&=&\int d^2\hat{\mathbf{k}}\xi^\lambda(k) _{-\lambda}\!Y^*_{lm}(\hat{\mathbf{k}}),
\end{eqnarray}
where $\lambda=0,\pm1,\pm2$ describes the helicity of the scalar, vector, tensor mode; $_{-\lambda}\!Y^*_{lm}$ is the spin-weight spherical harmonics; $\xi^\lambda(k)$ is the primordial perturbation and $\Delta^{Z}_l(k)$ is the
transfer function. Let us define the CMB angular bispectrum as
\begin{equation}\label{bis1}
B_{l_1\,l_2\,l_3}^{m_1\,m_2\,m_3}=\bigg\langle \prod_{n=1}^3 a_{l_n\,m_n}^{(Z)} \bigg\rangle,
\end{equation}
where only scalar perturbations (Z=S) will be considered in the paper. Now, by substituting the equation (\ref{alm}) into equation (\ref{bis1}) we can find
\begin{equation}\label{bisalm}
\bigg\langle \prod_{n=1}^3 a_{l_n\,m_n} \bigg\rangle=\left[\prod_{n=1}^3 4\pi(-i)^{l_n}\int \frac{d^3\mathbf{kn}}{(2\pi)^3}\Delta_{l_n}(kn)Y^*_{l_n\,m_n}(\hat{\mathbf{n}})\right]\times\bigg\langle \prod_{n=1}^3 \xi(kn) \bigg\rangle.
\end{equation}
We also consider a rough approximation for the transfer function that works quite well at large angular scales and for primordial adiabatic perturbations given by $\Delta_l(k)=\frac{1}{3}j_l(k(\eta_0-\eta_*))$, where $j_l(x)$ the spherical Bessel function, $\eta_0=14.38$ Gpc the conformal time at present and $\eta_*=284.85$ Mpc the  conformal time at the recombination epoch \cite{39}. This is the large scale Sachs Wolfe effect.  
Since we want to evaluate the contribution of the bispectrum by PMFs, the three-point correlator for the primordial perturbation must satisfy the relation
\begin{equation}\label{pmf1}
\bigg\langle \prod_{n=1}^3 \xi(kn) \bigg\rangle=A_P  \delta(\mathbf{k1}+\mathbf{k2}+\mathbf{k3})B_{\Pi_B\Pi_B\Pi_B}^{(S)}.
\end{equation}
Here, $A_P$ is a constant that depends on the type of perturbation (passive or compensated magnetic  mode) and $B_{\Pi_B\Pi_B\Pi_B}^{(S)}$ is the magnetic bispectrum computed above. Using the fo\-llowing relations \cite{38}
\begin{equation}
\delta(\mathbf{k1}+\mathbf{k2}+\mathbf{k3})=\frac{1}{(2\pi)^3}\int_{-\infty}^{\infty} \exp^{i(\mathbf{k1}+\mathbf{k2}+\mathbf{k3})\cdot \mathbf{x}}d^3x,
\end{equation}
\begin{equation}
\exp^{i\mathbf{k}\cdot \mathbf{x}}=4\pi\sum_li^lj_l(kx)\sum_m Y_{lm}(\hat{\mathbf{k}})Y_{lm}^*(\hat{\mathbf{x}}),
\end{equation}
with the Gaunt integral $\mathcal{G}_{l_1l_2l_3}^{m_1m_2m_3}$  defined by  
\begin{eqnarray}
\mathcal{G}_{l_1l_2l_3}^{m_1m_2m_3}&\equiv& \int d^2 \hat{\mathbf{n}}Y_{l_1m_1}(\hat{\mathbf{n}})Y_{l_2m_2}(\hat{\mathbf{n}})Y_{l_3m_3}(\hat{\mathbf{n}})\nonumber\\
&=&\sqrt{\frac{(2l_1+1)(2l_2+1)(2l_3+1)}{4\pi}}\bigg(\begin{array}{ccc}
l_1 & l_2 & l_3 \\
0 & 0 & 0  \end{array}  \bigg)\bigg(\begin{array}{ccc}
l_1 & l_2 & l_3 \\
m_1 & m_2 & m_3  \end{array}  \bigg),
\end{eqnarray}
and along with the equation (\ref{pmf1}), the equation (\ref{bisalm})  takes the form
\begin{eqnarray}
\bigg\langle \prod_{n=1}^3 a_{l_n\,m_n} \bigg\rangle&=&A_P\sqrt{\frac{(2l_1+1)(2l_2+1)(2l_3+1)}{4\pi}}\bigg(\begin{array}{ccc}
l_1 & l_2 & l_3 \\
0 & 0 & 0  \end{array}  \bigg)\bigg(\begin{array}{ccc}
l_1 & l_2 & l_3 \\
m_1 & m_2 & m_3  \end{array}  \bigg)\times \nonumber \\
&\times&\left[\prod_{n=1}^3 \frac{1}{3\pi^2}\int kn^2\int j_{l_n}(kn\,x)j_{l_n}(kn(\eta_0-\eta_*))dkn\right]B_{\Pi_B\Pi_B\Pi_B}^{(S)}x^2dx.
\end{eqnarray}
Here  the matrix is the Wigner-3j symbol and it vanishes unless the selection rules are satisfied\footnote{The Wigner-3j symbols satisfy that: $|m_1|\leq l_1$, $|m_2|\leq l_2$, $|m_3|\leq l_3$, $m_1+m_2+m_3=0$, $l_1+l_2+l_3=\mathbb{Z}$ and $|l_1-l_2|\leq l_3 \leq l_1+l_2$ \cite{30}. }. 
Given the rotational invariance of the Universe, Komatsu-Spergel \cite{38} defined a real symmetric function of $l_i$   called the reduced bispectrum  $b_{l_1l_2l_3}$  
\begin{equation}
\bigg\langle \prod_{n=1}^3 a_{l_n\,m_n} \bigg\rangle \equiv \mathcal{G}_{l_1l_2l_3}^{m_1m_2m_3} b_{l_1l_2l_3}.
\end{equation}
Checking the last two equations,  the properties of the bispectrum generated by PMFs can be expressed via the reduced bispectrum as
\begin{eqnarray}\label{redbis}
b_{l_1l_2l_3}=A_P\left[\prod_{n=1}^3 \frac{1}{3\pi^2}\int kn^2\int j_{l_n}(kn\,x)j_{l_n}(kn(\eta_0-\eta_*))dkn\right]B_{\Pi_B\Pi_B\Pi_B}^{(S)}x^2dx.
\end{eqnarray}
In order to calculate $A_P$, we must clarify the  sources of primordial perturbations. Prior to neutrino decoupling ($\tau_\nu=1$MeV$^{-1}$), the  Universe is dominated by radiation and it is tightly coupled to baryons such that they cannot have any anisotropic
stress contribution. Since we are also considering magnetic fields, they would be the only ones that develop anisotropic stress and therefore at superhorizon scales the curvature perturbation depends on the  primordial magnetic source\cite{30}. But after neutrino decoupling, neutrinos generated  anisotropic stress which compensates the  one coming
from   PMF  finishing the growth of the perturbations. Shaw-Lewis \cite{40} showed that curvature perturbation is given by
\begin{equation}
\xi(k)\sim -\frac{1}{3}R_\gamma \ln\left(\frac{\tau_\nu}{\tau_B}\right)\Pi_B^{(S)}(k),
\end{equation}
commonly known as passive mode, where $R_\gamma=\frac{\rho_\gamma}{\rho}\sim 0.6$ and  $\tau_B$ is the  epoch of magnetic field generation. Another contribution comes from the density-sourced
mode with unperturbed anisotropic stresses, the  magnetic compensated scalar mode, this is  proportional to the amplitude of the  perturbed magnetic density just as the magnetic Sachs Wolfe effect \cite{40}.
So, if the primordial perturbation is associated with the initial gravitational potential, in  the limit on large-angular scales the compensated modes is expressed as \cite{25,42}
\begin{equation}
\xi(k)\sim \frac{1}{4}R_\gamma \rho_B(k).
\end{equation}
Therefore if we use the passive mode contribution, $B_{\Pi_B\Pi_B\Pi_B}^{(S)}$ is given by $B_{\Pi_B^{(S)}\Pi_B^{(S)}\Pi_B^{(S)}}^{(S)}$ (see equation (\ref{passiveee})) with $A_P=\left(-\frac{1}{3}R_\gamma \ln\left(\frac{\tau_\nu}{\tau_B}\right)\right)^3$, whilst  compensated mode the primordial three-point correlation is described by $B_{\rho_B\rho_B\rho_B}^{(S)}$  (see equation (\ref{passiveee1})) with $A_P=\left(\frac{1}{4}R_\gamma\right)^3$. 
Since the magnetic bispectrum only depends on $(k1;k2)$,  the $k3$ integral in the equation (\ref{redbis}) gives $\frac{\pi}{2x^2} \delta(x-(\eta_0-\eta_*))$  due to the closure relation \cite{25}, and integrating out the delta function one  finally obtains  
\begin{eqnarray}\label{redbis1}
b_{l_1l_2l_3}=A_P\frac{\pi}{2}\left[\prod_{n=1}^2 \frac{1}{3\pi^2}\int kn^2 j_{l_n}(kn(\eta_0-\eta_*))^2 dkn\right]B_{\Pi_B\Pi_B\Pi_B}^{(S)}.
\end{eqnarray}
This is the master formula that we shall use in the following section in order to calculate the CMB reduced bispectrum.
\section{Analysis}\label{redbisa} 
In this section  we show the numerical results of the CMB reduced bispectrum produced by helical PMFs. In order to numerically solve eq.(\ref{redbis1}), we use the  adaptive strategy implemented in Mathematica   called Levin-type rule  which estimates the integral of an oscillatory function with a good accuracy\cite{mathem}.
\subsection{Causal Fields}
\begin{figure}[h!]
    \centering
    \begin{subfigure}[b]{0.45\textwidth}
        \includegraphics[width=\textwidth]{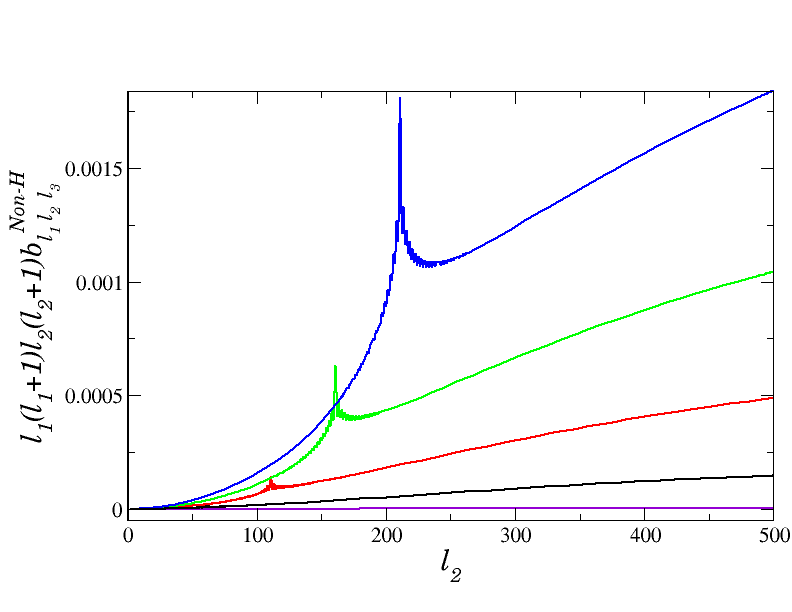}
        \caption{{\footnotesize Reduced bispectrum given by $A_B^3$  contribution of compensated PMFs.}} 
        \label{fig7a}
    \end{subfigure}
    ~ 
    \begin{subfigure}[b]{0.45\textwidth}
        \includegraphics[width=\textwidth]{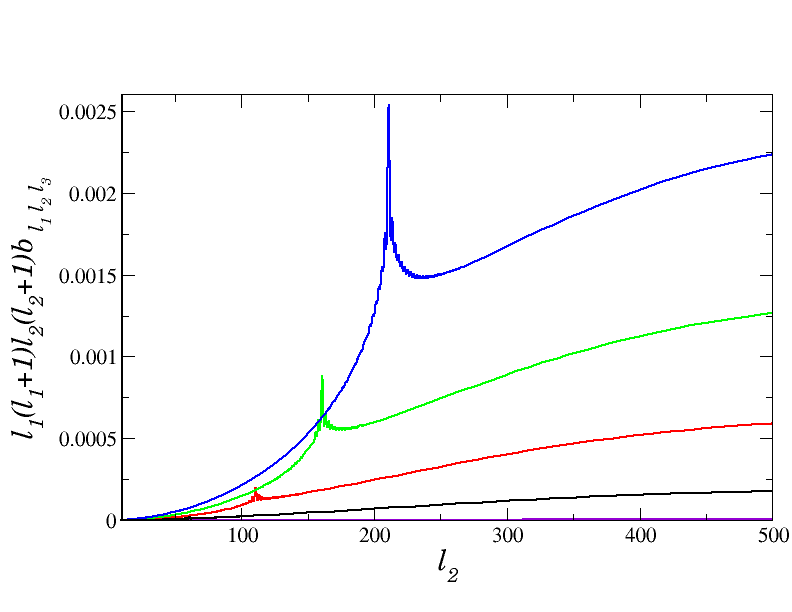}
        \caption{{\footnotesize \footnotesize Reduced bispectrum given by total  contribution of compensated PMFs.}}
        \label{fig7b}
    \end{subfigure}
    ~ 
    \begin{subfigure}[b]{0.45\textwidth}
        \includegraphics[width=\textwidth]{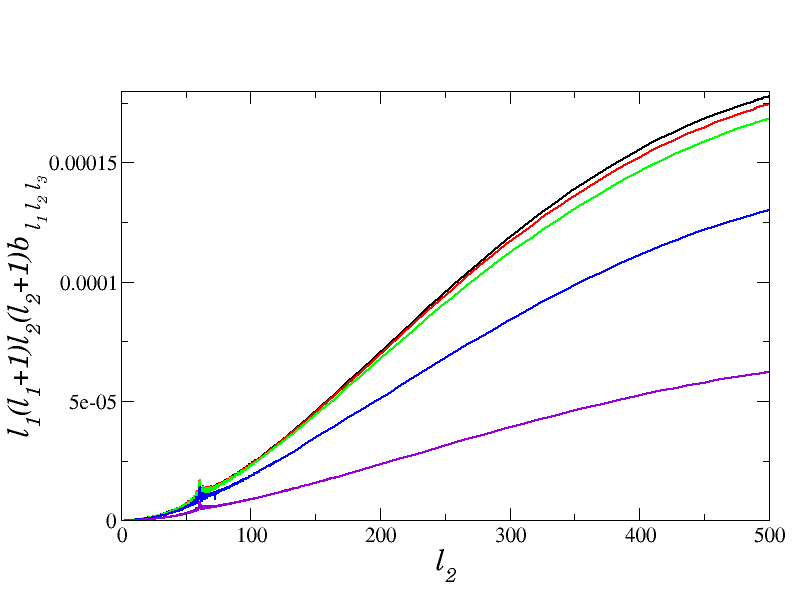}
        \caption{{\footnotesize  Effects of infrared cut-off on the reduced bispectrum with $l_1=61$.}}
        \label{fig7c}
    \end{subfigure}
    \begin{subfigure}[b]{0.45\textwidth}
        \includegraphics[width=\textwidth]{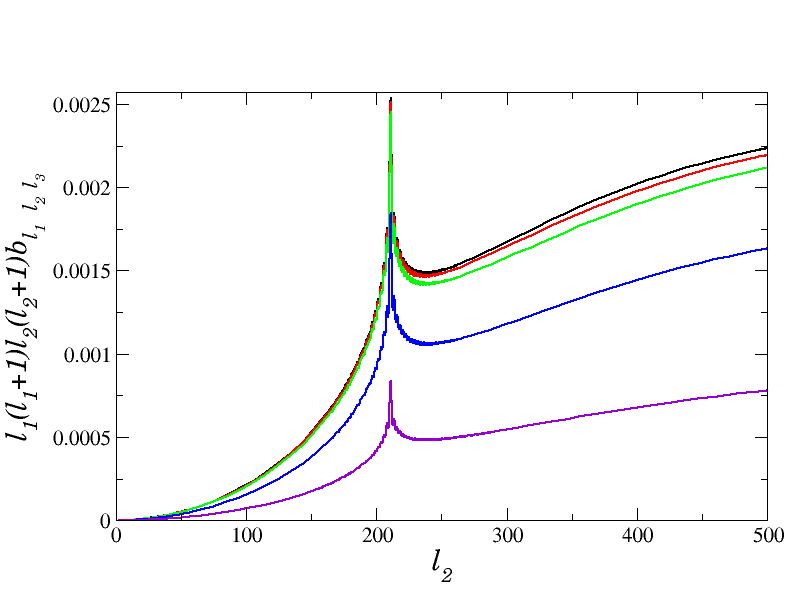}
        \caption{{\footnotesize  Effects of infrared cut-off on the reduced bispectrum with $l_1=210$.}}
        \label{fig7d}
    \end{subfigure}
    \caption{{\footnotesize Reduced bispectrum seeded by  compensated PMFs with $n=2$ using the squeezed collinear configuration. The figures (a)  shows the reduced bispectrum of the magnetic field   with only $A_B^3$, while  figure (b) shows the total contribution of the compensated mode; here the lines refers to different values of $l_1$, violet($l_1=11$), black($l_1=61$), red($l_1=110$), green($l_1=161$) and blue line($l_1=210$). The figures (c), (d)  show the effects of an infrared cut-off on the reduced bispectrum for different values of multipolar numbers $l_1$.
    Black; red; green; blue;  and violet lines refer to lower cut-off of $\alpha=0.001$, $\alpha=0.4$, $\alpha=0.5$, $\alpha=0.7$, $\alpha=0.8$ respectively.
    The reduced bispectrum is in units of $4\pi10^{-8}A_P/(8\pi^2 \rho_{\gamma,0})^3$.
      }}\label{fig7s}
\end{figure}

Figure  \ref{fig7s}  presents the CMB reduced bispectrum generated by compensated PMFs modes under collinear configuration.
In this figure we observe the signal produced by only the $A_B^3$ contribution (\ref{fig7a}),  as welll as the signal by the whole(\ref{fig7b}). We found that $A_BA_H^2$ contribution (helical) is smaller than non-helical part $A_B^3$. Here we plot the change of the reduced bispectrum with respect to $l_1$, finding a large contribution for larges values of $l_1$. We also see that helical contribution reaches a maximum around $l_2 \sim 400$ whilst non-helical contribution tends to increase at least until $l_2 \sim 500$. 
In figures (\ref{fig7c}) and (\ref{fig7d}),  the effect of an IR cut-off on the reduced bispectrum are shown. Each of these plots show the signal for different values of $l_1$. We see that signal is biggest for small $\alpha$ values (being the biggest contribution for spectrum without IR cut-off) similar to the found with the power spectrum case\cite{21}. 

\begin{figure}[h!]
    \centering
    \begin{subfigure}[b]{0.45\textwidth}
        \includegraphics[width=\textwidth]{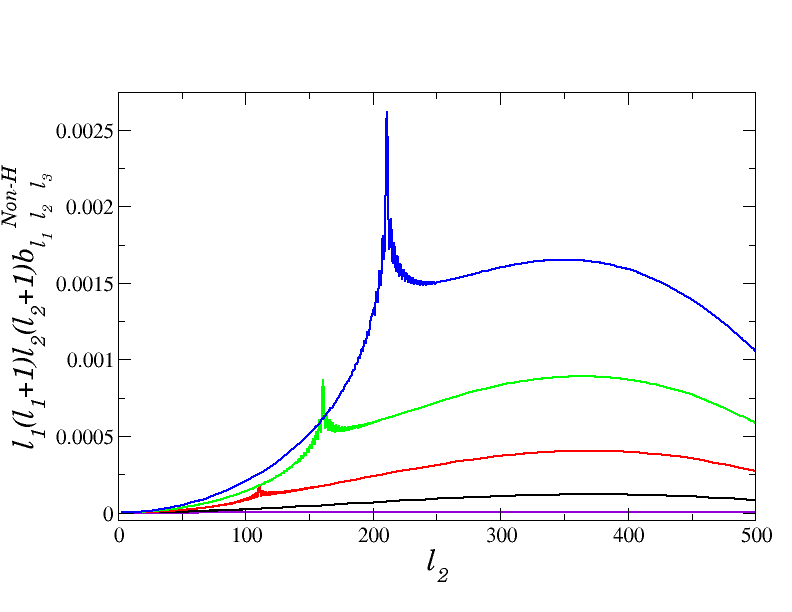}
        \caption{{\footnotesize Reduced bispectrum given by only $A_B^3$  contribution of passive PMFs.}} 
        \label{fig8a}
    \end{subfigure}
    ~ 
    ~ 
    \begin{subfigure}[b]{0.45\textwidth}
        \includegraphics[width=\textwidth]{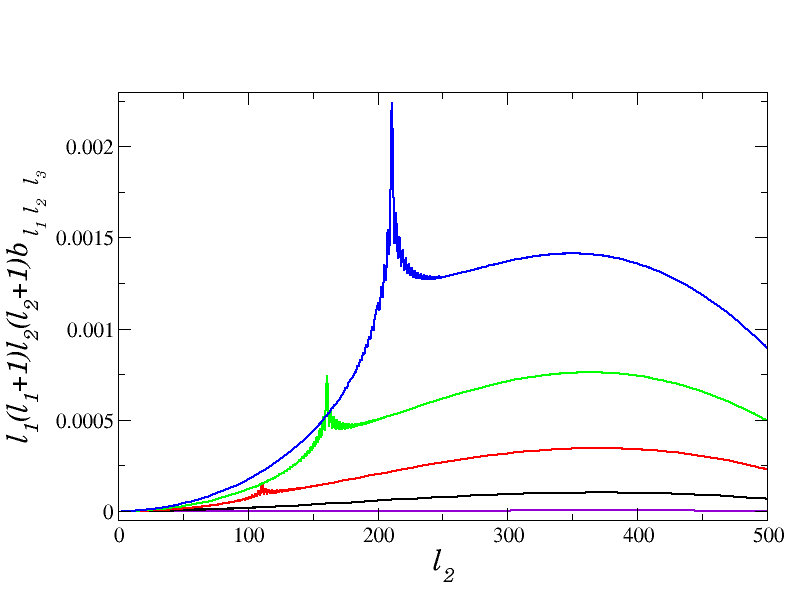}
        \caption{{\footnotesize Reduced bispectrum given by total  contribution of passive PMFs. }}
        \label{fig8b}
    \end{subfigure}
    \begin{subfigure}[b]{0.45\textwidth}
        \includegraphics[width=\textwidth]{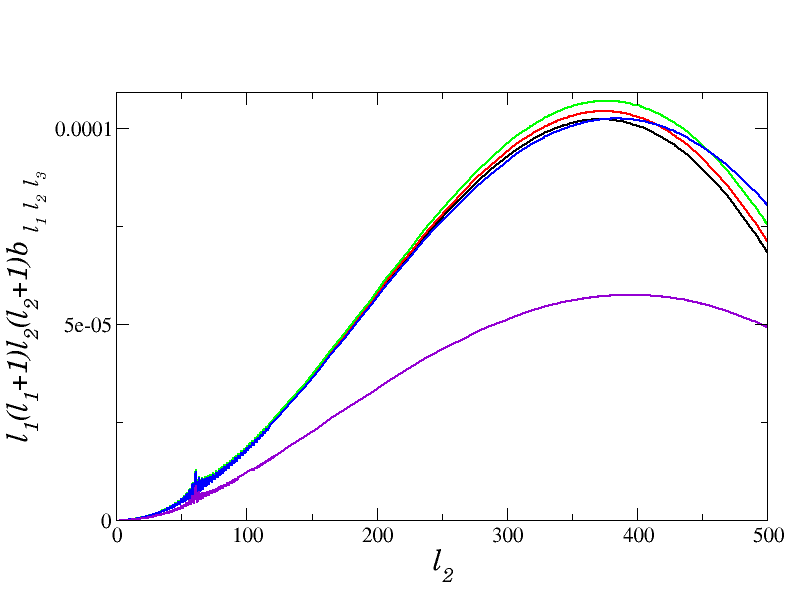}
        \caption{{\footnotesize Effects of infrared cut-off on the reduced bispectrum with $l_1=61$.}}
        \label{fig8c}
    \end{subfigure}
    \begin{subfigure}[b]{0.45\textwidth}
        \includegraphics[width=\textwidth]{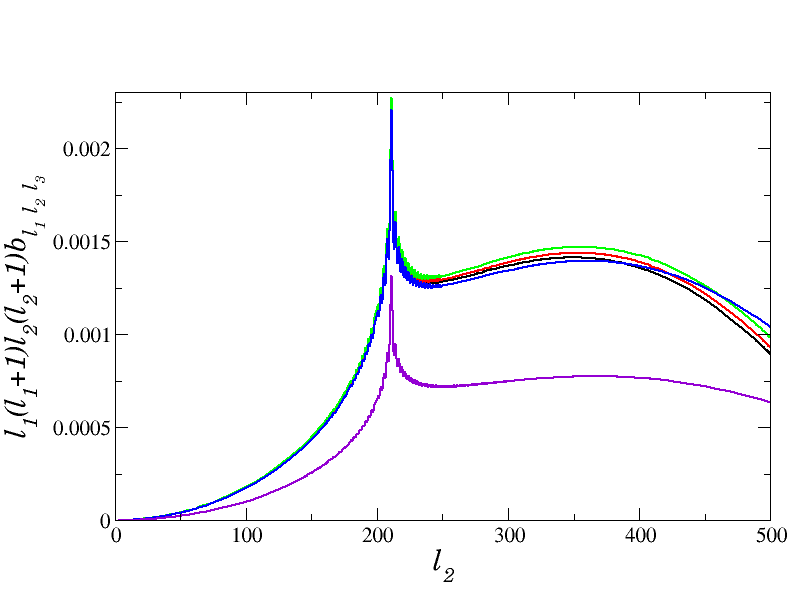}
        \caption{{\footnotesize  Effects of infrared cut-off on the reduced bispectrum with $l_1=210$.}}
        \label{fig8d}
    \end{subfigure}
   
    \caption{{\footnotesize Reduced bispectrum seeded by  passive PMFs with $n_B=n_H=n=2$ using the squeezed collinear configuration. The figure (a) shows reduced bispectrum of the magnetic field   with only $A_B^3$, while  figure (b) shows the total contribution of the passive mode, here the lines refer to different values of $l_1$, violet($l_1=11$), black($l_1=61$), red($l_1=110$), green($l_1=161$) and blue line($l_1=210$). The figures (c) and (d) show the effects of an infrared cut-off on the reduced bispectrum for difference values of multipolar numbers $l_1$.
    Black; red; green; blue;  and violet lines refer to lower cut-off of $\alpha=0.001$, $\alpha=0.4$, $\alpha=0.5$, $\alpha=0.7$, $\alpha=0.8$ respectively. The reduced bispectrum is in units of $4\pi10^{-8}A_P/2(8\pi^2 \rho_{\gamma,0})^3$.
      }}\label{fig8}
\end{figure}
\begin{figure}[h!]
    \centering
    \begin{subfigure}[b]{0.45\textwidth}
        \includegraphics[width=\textwidth]{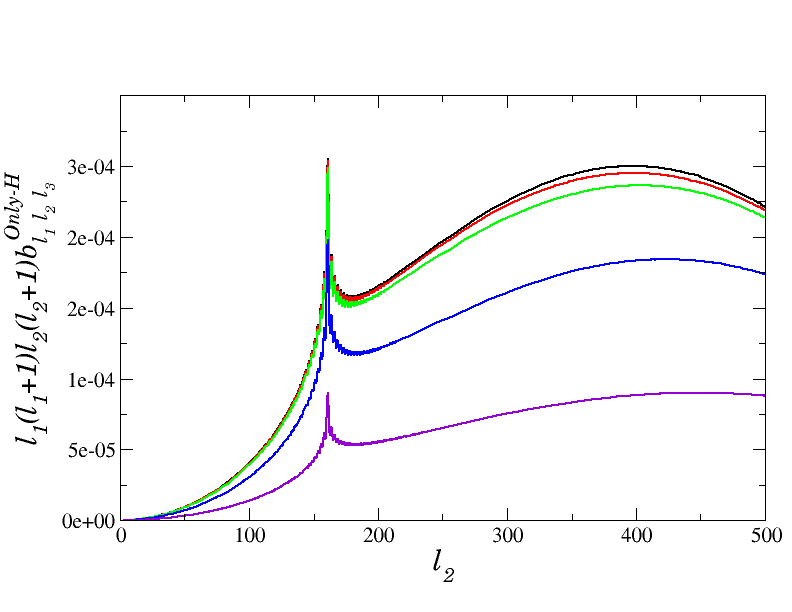}
        \caption{{\footnotesize Effects of infrared cut-off on the reduced bispectrum with $l_1=161$ seeded by total  contribution of compensated helical PMFs.}} 
        \label{fig9a}
    \end{subfigure}
    ~ 
    \begin{subfigure}[b]{0.45\textwidth}
        \includegraphics[width=\textwidth]{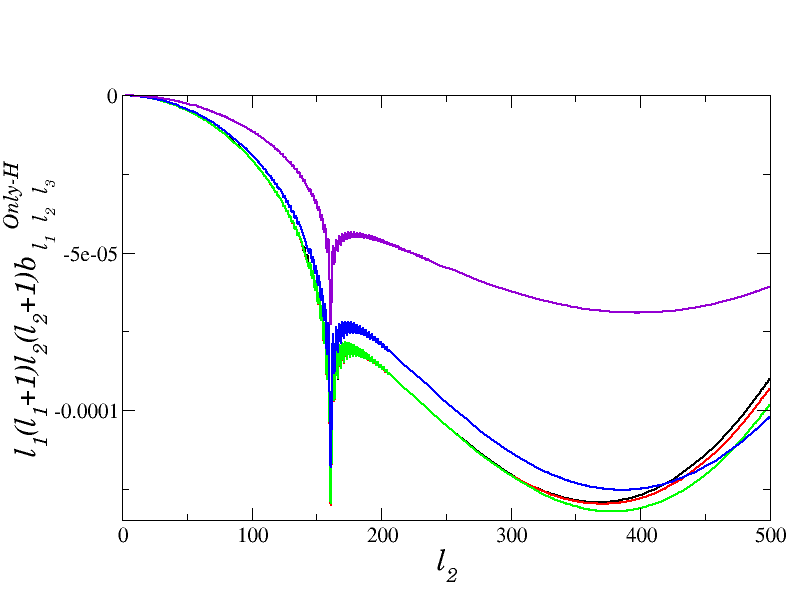}
        \caption{{\footnotesize \footnotesize Effects of infrared cut-off on the reduced bispectrum with $l_1=161$ seeded by total  contribution of passive helical PMFs.}}
        \label{fig9b}
    \end{subfigure}
    ~ 
    \caption{{\footnotesize Reduced bispectrum seeded by  compensated (a) and passive (b) helical PMFs with only $A_BA_H^2$ contribution using the squeezed collinear configuration.
    Black; red; green; blue;  and violet lines refer to lower cut-off of $\alpha=0.001$, $\alpha=0.4$, $\alpha=0.5$, $\alpha=0.7$, $\alpha=0.8$ respectively.  The reduced bispectrum is in units of $4\pi10^{-8}A_P/2(8\pi^2 \rho_{\gamma,0})^3$.
      }}\label{fighel}
\end{figure}
\begin{figure}[h!]
    \centering
    \begin{subfigure}[b]{0.45\textwidth}
        \includegraphics[width=\textwidth]{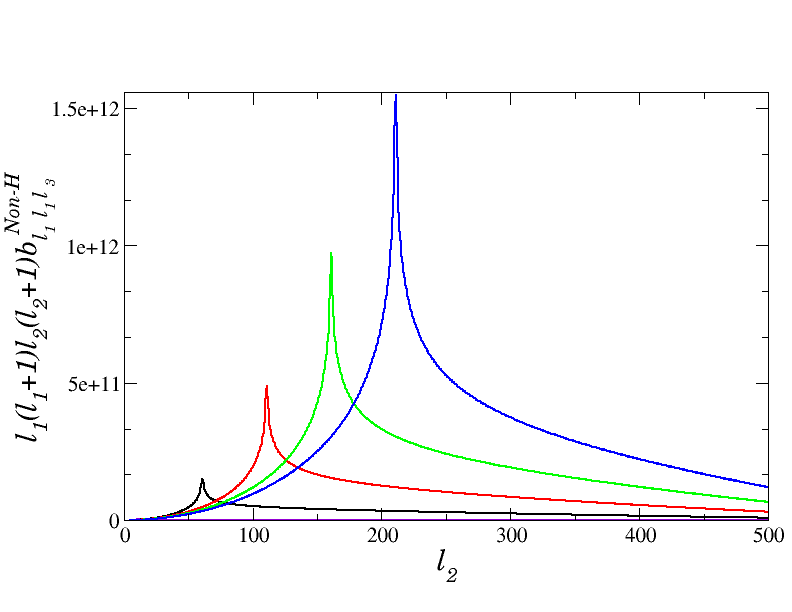}
        \caption{{\footnotesize Reduced bispectrum given by $A_B^3$  contribution of compensated PMFs.}} 
        \label{fig10a}
    \end{subfigure}
    ~ 
    ~ 
    \begin{subfigure}[b]{0.45\textwidth}
        \includegraphics[width=\textwidth]{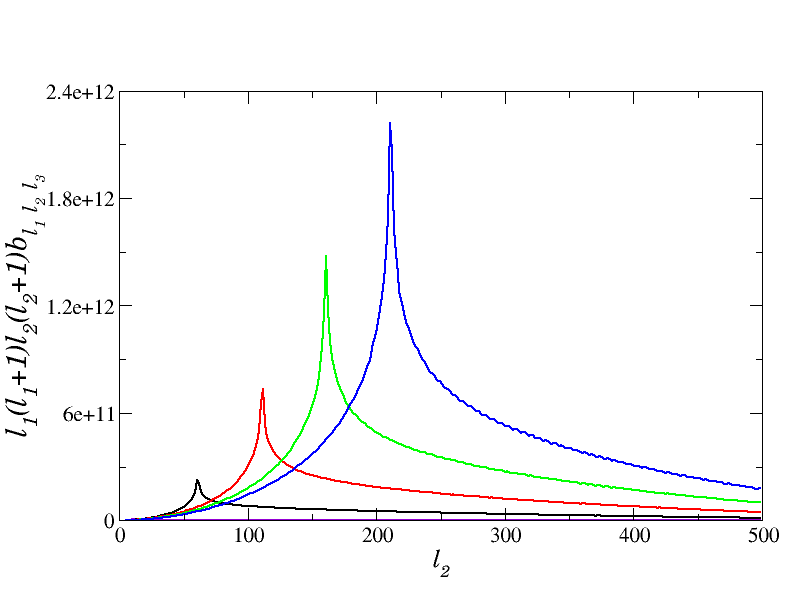}
        \caption{{\footnotesize Reduced bispectrum given by total  contribution of compensated PMFs.}}
        \label{fig10b}
    \end{subfigure}
    \begin{subfigure}[b]{0.45\textwidth}
        \includegraphics[width=\textwidth]{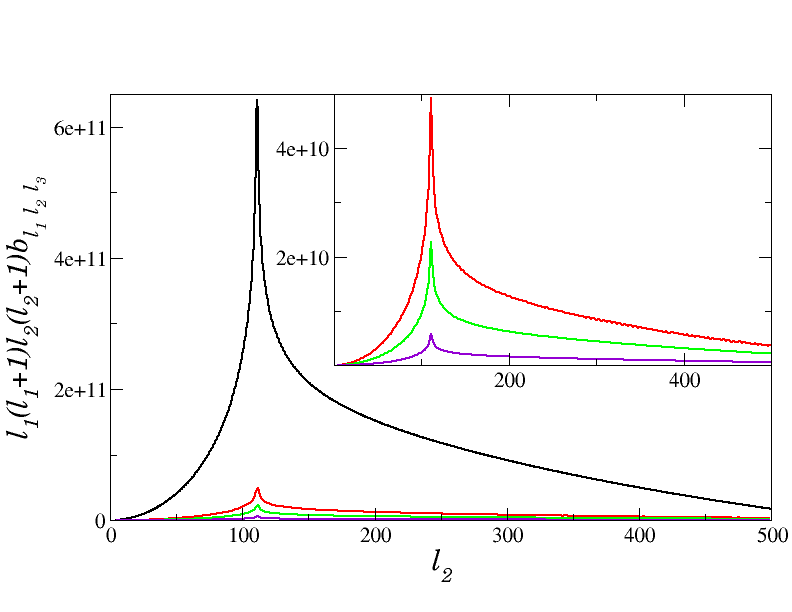}
        \caption{{\footnotesize  Effects of infrared cut-off on the reduced bispectrum with $l_1=111$.}}
        \label{fig10c}
    \end{subfigure}
      \begin{subfigure}[b]{0.45\textwidth}
        \includegraphics[width=\textwidth]{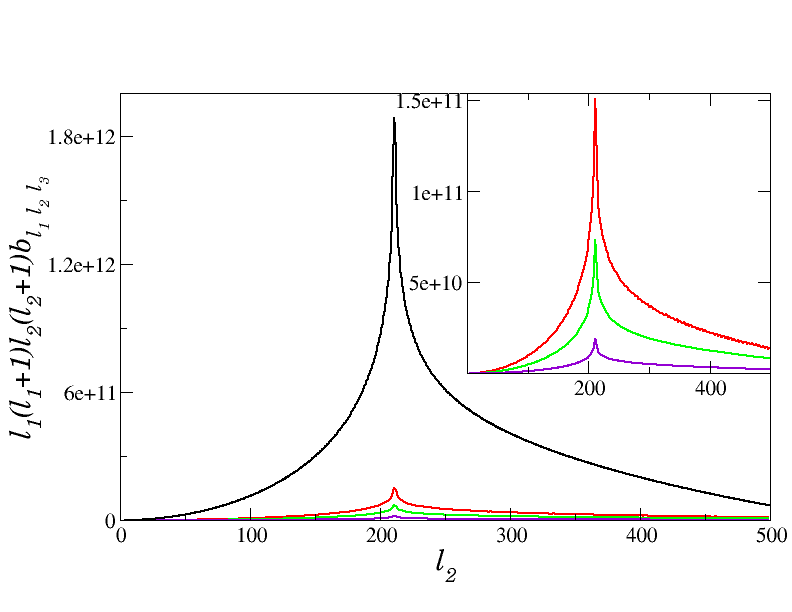}
        \caption{{\footnotesize  Effects of infrared cut-off on the reduced bispectrum with $l_1=210$.}}
        \label{fig10d}
    \end{subfigure}
    \caption{{\footnotesize Absolute value of  reduced bispectrum seeded by  compensated PMFs with $n=-5/2$ using the squeezed collinear configuration. The figure (a) shows reduced bispectrum of the magnetic field with only $A_B^3$ contribution, while  figure (b) shows the total contribution of the compensated mode; here the lines refers to different values of $l_1$, violet($l_1=11$), black($l_1=61$), red($l_1=110$), green($l_1=161$) and blue line($l_1=210$). The figures (c) and (d)  show the effects of an infrared cut-off on the reduced bispectrum for difference values of multipolar numbers $l_1$.
    Black; red; green;   and violet lines refer to lower cut-off of $\alpha=0.001$, $\alpha=0.4$, $\alpha=0.6$, $\alpha=0.8$ respectively. The reduced bispectrum is in units of $4\pi10^{16}A_P/(8\pi^2 \rho_{\gamma,0})^3$.
      }}\label{fig7s1}
\end{figure}
\begin{figure}[h!]
    \centering
    \begin{subfigure}[b]{0.45\textwidth}
        \includegraphics[width=\textwidth]{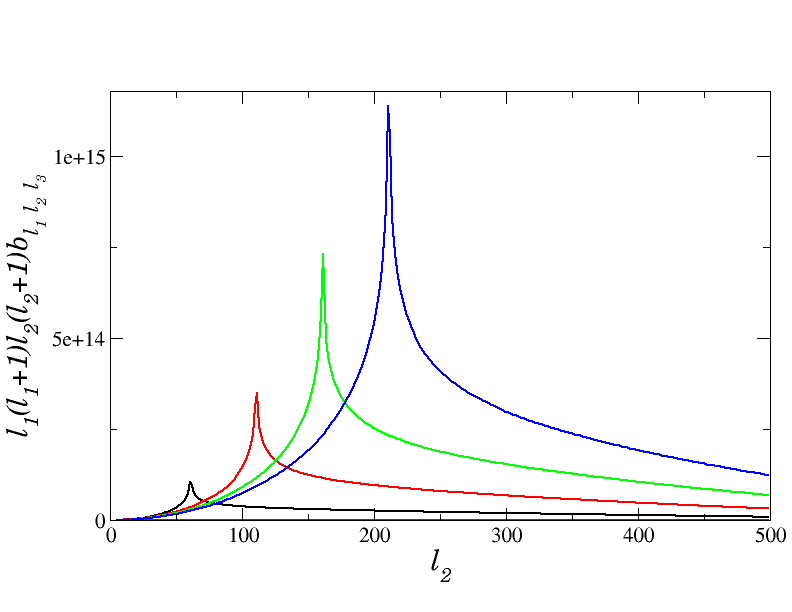}
        \caption{{\footnotesize Reduced bispectrum of compensated PMFs for $n=-3/2$.}} 
        \label{fig11a}
    \end{subfigure}
      \begin{subfigure}[b]{0.45\textwidth}
        \includegraphics[width=\textwidth]{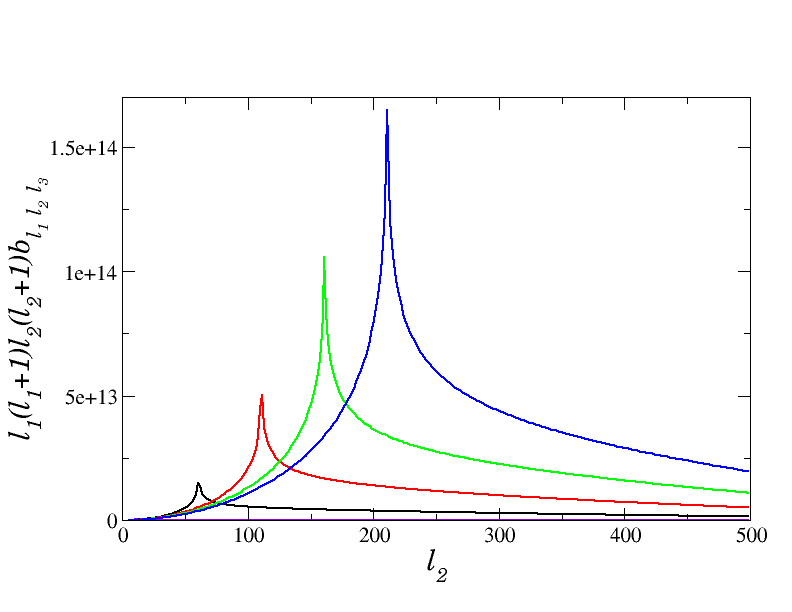}
        \caption{{\footnotesize  Reduced bispectrum  of compensated PMFs for $n=-1.9$.}}
        \label{fig11b}
    \end{subfigure}
    ~ 
 
    ~ 
    \begin{subfigure}[b]{0.45\textwidth}
        \includegraphics[width=\textwidth]{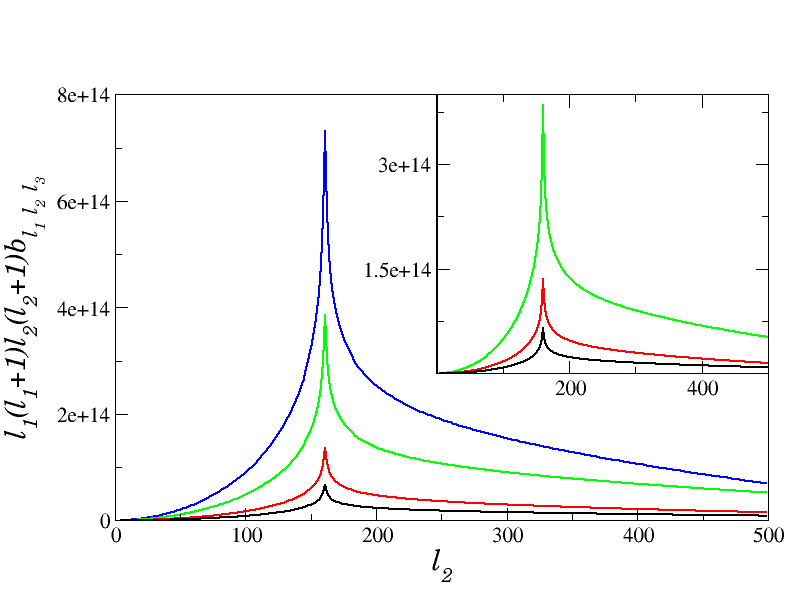}
        \caption{{\footnotesize Effects of infrared cut-off on the reduced bispectrum with $l_1=161$ for $n=-3/2$.}}
        \label{fig11c}
    \end{subfigure}
    \begin{subfigure}[b]{0.45\textwidth}
        \includegraphics[width=\textwidth]{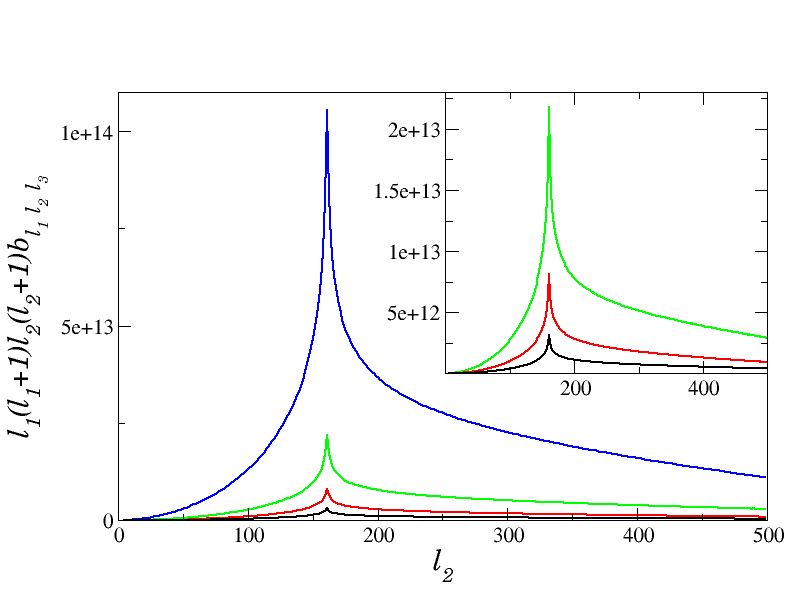}
        \caption{{\footnotesize  Effects of infrared cut-off on the reduced bispectrum with $l_1=161$ for $n=-1.9$.}}
        \label{fig11d}
    \end{subfigure}
    \caption{{\footnotesize Absolute values of  reduced bispectrum seeded by  compensated PMFs using the squeezed collinear configuration. The figures (a), (b) show the  total contribution of the compensated mode of the magnetic field for $n=-3/2$ and $n=-1.9$ respectively;  here the lines refers to different values of $l_1$, violet($l_1=11$), black($l_1=61$), red($l_1=110$), green($l_1=161$) and blue line($l_1=210$).  The last figures (c),(d) show the effects of an infrared cut-off on the reduced bispectrum for difference values of multipolar numbers $l_1$. Blue;  green; red;   and black lines refer to lower cut-off of $\alpha=0.001$, $\alpha=0.4$, $\alpha=0.6$, $\alpha=0.8$, respectively.
    The bispectrum is in units of $4\pi10^{16}A_P/(8\pi^2\rho_{\gamma,0})^3$.
      }}\label{fig7s2}
\end{figure}

\begin{figure}[h!]
    \centering
    \begin{subfigure}[b]{0.45\textwidth}
        \includegraphics[width=\textwidth]{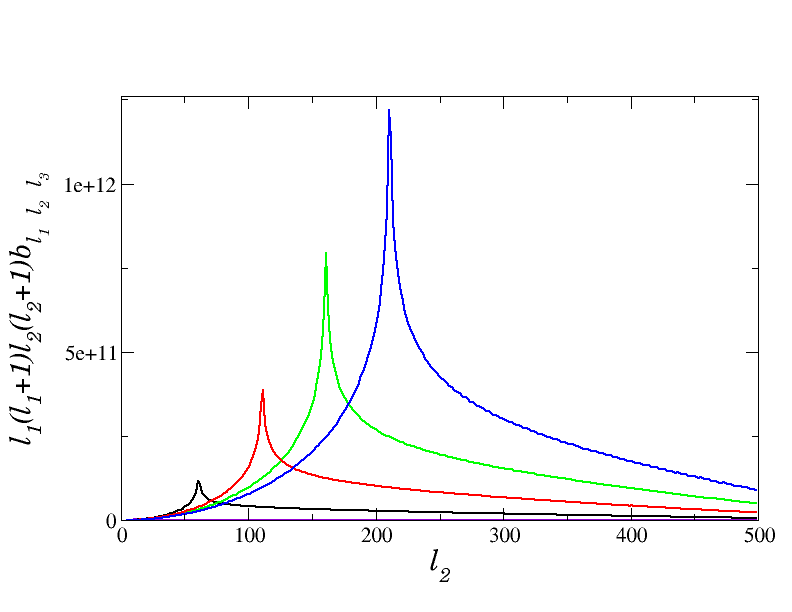}
        \caption{{\footnotesize Reduced bispectrum of compensated PMFs for $n=-5/2$.}} 
        \label{fig12a}
    \end{subfigure}
      \begin{subfigure}[b]{0.45\textwidth}
        \includegraphics[width=\textwidth]{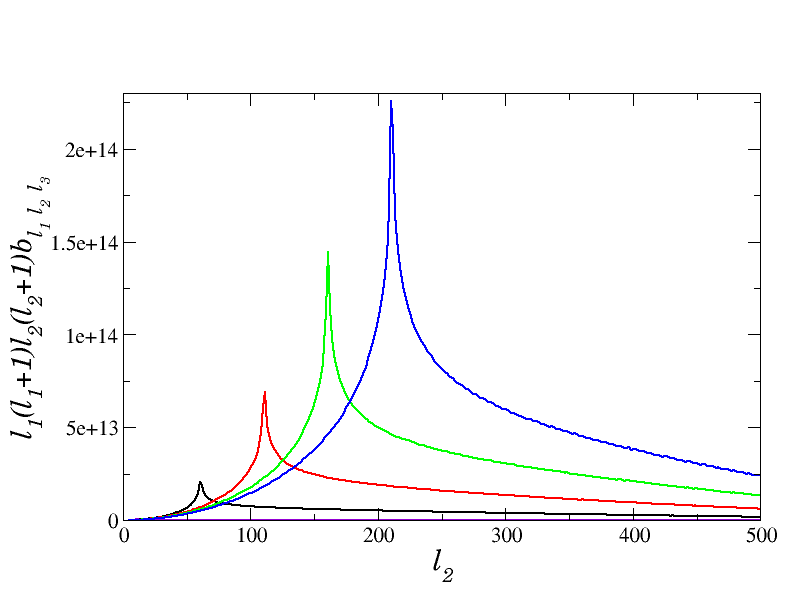}
        \caption{{\footnotesize  Reduced bispectrum  of compensated PMFs for $n=-3/2$.}}
        \label{fig12b}
    \end{subfigure}
    ~ 
    \begin{subfigure}[b]{0.45\textwidth}
        \includegraphics[width=\textwidth]{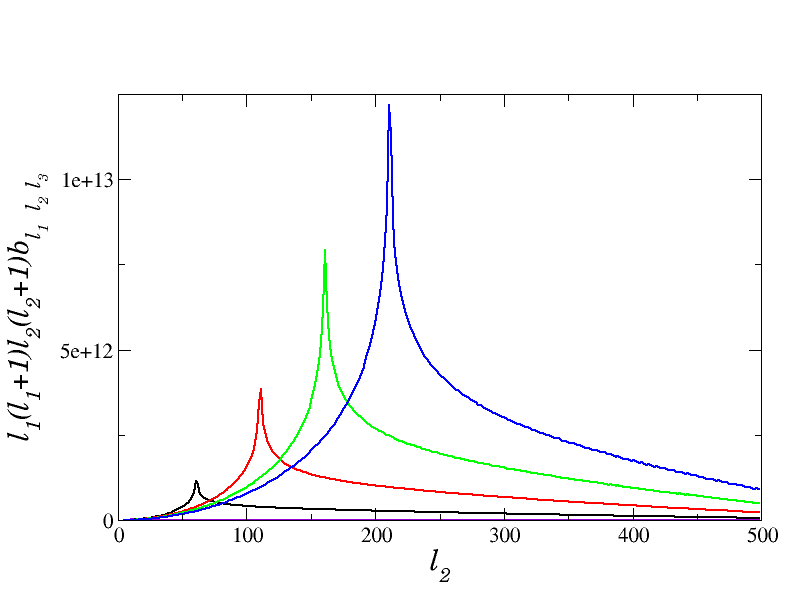}
        \caption{{\footnotesize \footnotesize Reduced bispectrum  of passive PMFs for $n=-5/2$.}}
        \label{fig12c}
    \end{subfigure}
       \begin{subfigure}[b]{0.45\textwidth}
        \includegraphics[width=\textwidth]{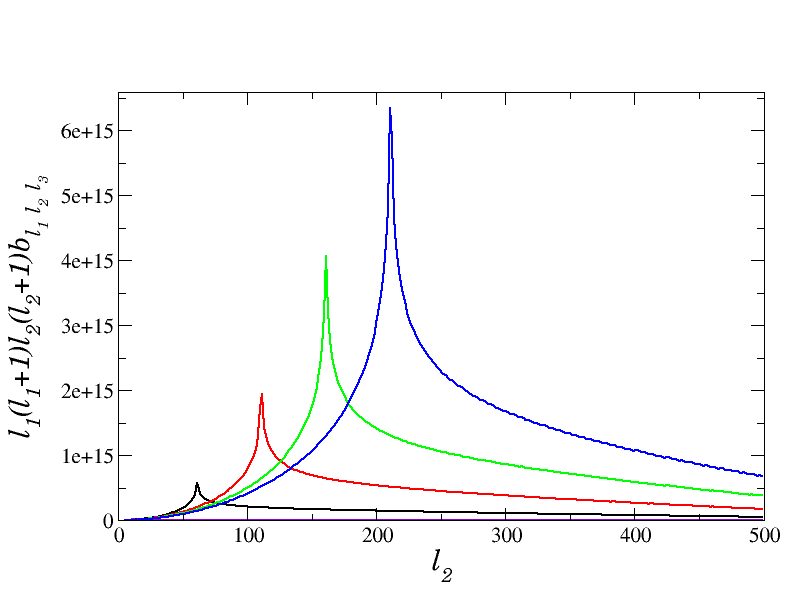}
        \caption{{\footnotesize  Reduced bispectrum  of passive PMFs for $n=-3/2$.}}
        \label{fig12d}
    \end{subfigure}
    ~ 
    \caption{{\footnotesize Absolute values of  reduced bispectrum seeded by  passive and compensated PMFs  under  p-independent approximation. The figures (a), (b) show the  total contribution of the compensated mode of the magnetic field for $n=-5/2$ and $n=-3/2$ respectively.  The figures (c), (d) show the  total contribution of the passive mode of the magnetic field for $n=-5/2$ and $n=-3/2$ respectively.  Here the lines refer to different values of $l_1$, violet($l_1=11$), black($l_1=61$), red($l_1=110$), green($l_1=161$) and blue line($l_1=210$).
      }}\label{fig7s3}
\end{figure}

In Figure (\ref{fig8}) we show the CMB reduced bispectrum generated by PMFs passive modes under collinear configuration. Meanwhile, figures (\ref{fig8a}) and (\ref{fig8b}) describe the signal for the non-helical and total contribution respectively. An interesting feature is that reduced bispectrum generated by helical contribution is totally negative, then, by using this unusual behavior we would have direct evidence of a helical component in the field.

Other important result of this paper is reported in figures (\ref{fig8c}) and (\ref{fig8d}). Here we show again the effect of an IR cut-off on the reduced bispectrum and we have found out that the biggest contribution of the bispectrum comes from an IR cut-off near $\alpha \sim 0.5$ instead of $\alpha = 0$. This peak might correspond to a type of dynamics in large scales and help us to determine the nature of PMFs (Since we are trying with causal fields,  this infrared cut-off  would correspond to  the maximum scale  in  which  magnetic fields may be generated at later times). Therefore the evidence of this cut-off in the bispectrum would reveal an interesting
signal from passive magnetic scalar mode. 
In addition, the change of the  reduced bispectrum for helical PMFs in presence of an IR cut-off is showed in figure (\ref{fighel}). Here we observe how the signal decreases when the IR cut-off increases for the compensated mode and how change the behavior for the passive case.
We want to remark on some  approximations used so far. Since we assume that effects of PMFs are important for small multipolar numbers, we write the transfer functions in terms of spherical Bessel function. Previous papers have worked without this approximation. For instance, \cite{32} computes the full radiation transfer function taking into account the effect of PMFs. The  full numerical integration of the bispectrum can be done via  second-order Einstein-Boltzmann codes like SONG\cite{tram} improving the estimation of the amplitude of PMFs.
Moreover, we must note that \cite{43} found a WMAP bound on non-helical passive mode for tensor temperature bispectrum of $B_{1Mpc}<3.1$nG and the Planck paper \cite{33} reported  $B_{1Mpc} < 2.8$nG all of them for scale-invariant fields. Actually, the tensor mode is dominant in the passive mode, and can give quite tighter constraint of the PMF amplitude than the scalar mode.
Thus, tensor mode contribution and a full transfer function determined by the presence of these fields will improve our results and therefore they will be interesting subjects of our future work.
\subsection{Non-causal Fields}
Now let us consider a red magnetic spectrum. Figure (\ref{fig7s1}) shows the reduced bispectrum for compensated mode with $n_H=n_B=n=-5/2$. Again, we plot the non-helical (\ref{fig10a})  and total (\ref{fig10b}) contribution of the bispectrum while plots (\ref{fig10c}) and (\ref{fig10d}) correspond to the change of the signal due to an IR cut-off. Since the PMFs bispectrum is almost determined by the poles in each $k$, the value of it peaked for $l_1=l_2=l_3$ as we can observe in figures (\ref{fig7s1}),(\ref{fig7s2}), (\ref{fig7s3}). Additionally, the figure (\ref{fig7s2}) shows signals for $n_H=n_B=n=-3/2$ and $n_H=n_B=n=-1.9$. As a matter of fact, some mechanisms of inflationary magnetogenesis with parity violating terms which lead to  helical magnetic field ($n=-1.9$), stand for the lower bound for the which the field can satisfy  the intensity of magnetic fields in the intergalactic medium; thereby the signal described in figure (\ref{fig7s2}) constraints  models for providing the seed for galactic magnetic fields \cite{18l2}. On the other hand, figure (\ref{fig7s3})  shows  the reduced bispectrum taking into account the p-independent approximation implemented in section \ref{pindep}. Due to complexity of the angular structure  on the PMF bispectrum for passive mode, the numerical computation for reduced bispectrum requires a great deal of time. To avoid this problem we can use the p-independent approximation by reducing the PMF bispectrum to an independent angular form as we 
studied above. From figures  (\ref{fig12a}) and (\ref{fig12b}) we observe an increase in the signal as we expected due to the lack of angular terms in the bispectrum. Finally, figures (\ref{fig12c}) and (\ref{fig12d}) show the total contribution of passive
modes. Note that the amplitude is larger than the compensated mode. This behavior have also been reported in \cite{25}. This result is interesting  because  an  estimation of $B_\lambda$ through a local-type primordial NG in curvature perturbation generates  constraints stronger than the compensated ones. Finally, if we compare the results reported in this section with the ones shown in above section for causal fields, we can observe that effect of $k_m$ is more significant in negative spectral indices, specially for nearly scale invariant scale fields. We then conclude that $k_m$  plays an important role in the study of these non-causal fields and this generates the possibility of determining some important clues in the mechanisms of magnetogenesis.      
\subsection{Estimation of the Magnetic Field Amplitude}
In fact,  it is possible to obtain a rough estimate  of $B_\lambda$ using the formula for the  primary reduced bispectrum found in   \cite{38} 
\begin{equation}
b_{lll}\sim l^{-4}\times2\times10^{-17}f_{NL},
\end{equation}
where  the non-Gaussianity (local) is fully specified by a single constant parameter $f_{NL}$. As we mentioned above, the k-dependence on the magnetic bispectrum is similar to the CMB bispectrum arising from the local type NG of curvature perturbations, therefore, by comparing the last equation with eq.(\ref{redbis1}),  allow us  to express a simple relation between $F_{NL}$ and $B_\lambda$ given by
\begin{equation}\label{eqng}
f_{NL}\propto \left( \frac{B_\lambda}{10^{-9}\text{G}}\right)^6.
\end{equation}
In Table \ref{tabla2} we present the  constant of proportionality of the last expression. Here we use $\frac{\tau_\nu}{\tau_B}=10^{17}$ which corresponds to the PMF generated at the grand unification energy scale(GUT) scale, $\left(\frac{B_\lambda^2}{8\pi\rho_{\gamma,0}}\right)\sim10^{-7}\left(\frac{B_\lambda}{10^{-9}\text{G}}\right)^2$, and $R_\gamma\sim0.6$. In order to constrain the smoothed amplitude of the magnetic field on a  scale of $1$Mpc ($B_1$), we will  use the $f_{NL}$ value reported by  Planck Collaboration \cite{pl2} of $f_{NL}<5.8$ at $68\%$ CL. The results of $B_1$ are shown in table \ref{tabla3}. Our results for compensated modes lead to  upper bounds on the PMF smoothed  amplitude which are  consistent with the 2015 Planck analysis \cite{33}, but for passive modes our results are  slightly tigher because  these were based on a rough estimation and may involve some uncertainties (except for the causal case where the bound coincides with  Planck analysis), however  notice that for passive modes the limits are almost 10 times more stringent than the compensated ones as was reported in \cite{25} for no-helical and scale invariant case, hence CMB-observation are sensitive to the magnetic induced modes. Since our results were  obtained under the  Sachs-Wolfe approximation, we expected a lower value of $B_1$ for causal fields with respect to the non causal fields, and therefore the blue spectra generated by these fields is strongly disfavoured by the CMB bispectrum.
On the other hand, we  constrain $B_1$ through the helical contribution and we observed an enhance of its  amplitude. We see this same effect  in the  two point correlation as was reported by  Planck analysis   $B_1<5.6$nG at 95\% CL \cite{33} for that contribution.    In the tables  also show the bounds when  a high value of the IR cut-off ($\alpha\sim 0.8$) is used. Since cut-off reduces the amplitude magnetic bispectrum signal, the  upper limit of $B_\lambda$ becomes  somewhat relaxed and therefore, we are able to  illustrate the impact $k_m$ can have on constraints on the PMF amplitude.
Although this effect becomes very small for a tiny value of $k_m$, the presence of this scale in  the  analysis of NG are complementary to the ones found by the two point correlation case and will provide new insight into the nature of primordial magnetic fields.

\begin{table}[htb]
\centering
\begin{tabular}{|l|l|l|l|l|l|l|l|}
\hline
& \multicolumn{2}{c|}{n$=-\frac{5}{2}$} & \multicolumn{1}{c|}{n$=-1.9$}& \multicolumn{2}{c|}{n$=-\frac{3}{2}$}& \multicolumn{2}{c|}{n$=2$} \\
\cline{2-8}
& Comp. & Passive & Comp  & Comp. & Passive & Comp & Passive \\
\hline \hline
Helical & 0.86(1.92) & 0.12 & 0.47(0.80)  & 0.35(0.49) & 0.008 & 0.021 & 0.0069\\ \cline{1-8}
N-Helical & 1.04(2.43) & 0.13  & 0.38  & 0.33 & 0.067 & 0.018 & 0.0050\\ \cline{1-8}
Total & 0.92(1.85) & 0.14 & 0.40(0.73) &  0.38(0.44) & 0.064 & 0.017(0.021) & 0.0051(0.006) \\ \cline{1-8}
\end{tabular}
\caption{Constant of proportionality of the eq.(\ref{eqng}) for different spectral indices and modes(compensated(comp) or passive) of the PMF without considering an IR cut-off. Parentheses are used to represent this value for $\frac{k_m}{k_D}\sim0.8$.}
\label{tabla2}
\end{table}

\begin{table}[htb]
\centering
\begin{tabular}{|l|l|l|l|l|l|l|l|}
\hline
& \multicolumn{2}{c|}{n$=-\frac{5}{2}$} & \multicolumn{1}{c|}{n$=-1.9$}& \multicolumn{2}{c|}{n$=-\frac{3}{2}$}& \multicolumn{2}{c|}{n$=2$} \\
\cline{2-8}
& Comp. & Passive & Comp  & Comp. & Passive & Comp & Passive \\
\hline \hline
Helical & 1.15(2.58) & 0.16 & 0.64(1.07)  & 0.47(0.66) & 0.098 & 0.029 & 0.0092\\ \cline{1-8}
N-Helical & 1.39(3.25) & 0.17  & 0.52  & 0.44 & 0.089 & 0.024 & 0.0068\\ \cline{1-8}
Total & 1.24(2.48) & 0.19 & 0.54(0.98) &  0.52(0.59) & 0.085 & 0.023(0.029) & 0.0068(0.0083) \\ \cline{1-8}
\end{tabular}
\caption{Bound on smoothed amplitude of the magnetic field on a  scale of $1$Mpc ($B_1$ in units of nG) for different spectral indices and modes(compensated(comp) or passive) of the PMF without considering an IR cut-off. Parentheses are used to represent $B_1$ for $\frac{k_m}{k_D}\sim0.8$.
Here we use $f_{NL}<5.8$ reported by  Planck Collaboration, 2016 \cite{33}.}
\label{tabla3}
\end{table}

\section{Discussion and Conclusions} \label{discu}
In this paper we investigate the effects of helical PMFs in the CMB reduced bispectrum.
One of the main motivations to introduce the helicity  comes from the fact that the these fields are good observables to probe parity-violation in the early stages of the Universe.
Furthermore, since  magnetic fields  depend quadratically on the field, it must induce non-Gaussian signals on CMB anisotropies at lower order instead of the standard inflationary mechanisms  where this signal appears only at high orders\cite{27}.
We started our work  deriving  the full  even and odd parts of the bispectrum which comes from the helical magnetic fields, thus extending the previous results reported in \cite{37}.
We obtained the full expression for the PMF bispectrum but, we did not consider  modes  that arise from odd intensity-intensity-intensity bispectrum. Although these signals are smaller than the even ones,  the evidence of the odd signals  would be a decisive  observable  to probe parity-violating processes in the early Universe\cite{28}.  We will provide more details of the parity-odd signals in a future paper.
Then, through the methodology used in \cite{25}, \cite{26}, we found  that
 PMFs bispectrum   peaks at  $k1 \sim k2$  under a squeezed configuration implying that statistical properties of the PMFs are similar to those of the local-type NG of curvature perturbations.
By calculating the bispectrum given by PMFs anisotropic stress, we observed that its amplitude is larger that the density one and also has a negative contribution for values less than $k_D$.
Through  numerical calculations of the  intensity-intensity-intensity reduced bispectra of the scalar modes,  we also studied the total contributions of the helical PMF bispectrum and the presence of an IR cut-off in the convolution integrals. Here we observe the same behavior seen in the power spectral case and the presence of negative contribution due to helical terms $A_BA_H^2$.
Nevertheless, in the computation in the passive modes for causal fields we observed an unusual behavior in the bispectrum. Indeed, in Figure  (\ref{fig8}) we have found  out that biggest contribution of the bispectrum comes from an IR cut-off near to $\alpha \sim 0.5$ instead of $\alpha = 0$ . Since $k_m$ is dependent on PMF generation model, this behavior might set strong limits on  PMF amplitude. Finally, we investigated the effects of $k_m$  on the reduced bispectrum for $n<0$. Due to the fact that the magnetic field intensity can be enhanced when we use passive modes, it is expected that those modes  determine a very strong constraints on the amplitude of the magnetic field on a given characteristic scale $\lambda$.  We verify this statement by using the  primary reduced bispectrum found in  \cite{38} and calculating $B_\lambda$.
Our results showed in tables  \ref{tabla2}, \ref{tabla3} reflect the fact that the corresponding bound on the mean amplitude of the  field  is  dependent on  strong values of the minimal cut-off and the helical contribution relaxing the constraints of $B_\lambda$. We also found that  for passive modes the limits are almost 10 times more stringent than the compensated ones for both helical and hon helical contribution,
this result was also reported in \cite{27} for non helical fields.
 However, we can observe that effect of $k_m$ is more significant in the magnetic bispectrum driven by  negative spectral indices. Hence, the presence of  $k_m$  plays an important role in the analysis of the signatures that these non-causal fields may leave in cosmological observations.\\
In conclusion, we have studied the  effects of helicity and a minimal cut-off on the constraints of the PMFs amplitude by computing the CMB reduced bispectrum induced on large angular scales by those fields.
Even though $k_m$ for causal modes would be important when this scale is larger
than the wavenumber of interest, for non-causal modes it is related  with the horizon scale  of the beginning of inflation \cite{tk,21}, and therefore the study of this cut-off on the bispectrum give us information about the PMF generation mechanisms.  
 
\appendix
\section{Scalar bispectra of the magnetic field}\label{A1}
Here we present the results   of the Section 2 for the even and odd contributions of the magnetic field bispectrum. The expressions for the magnetic bispectrum  reported in this paper were performed with the help of the  tensor computer algebra xAct package in Mathematica \cite{xact}.
\subsection*{Scalar correlation ($<\rho\rho\rho>$)}
For the energy density of PMF bispectrum ($<\rho\rho\rho>$) we have
\begin{equation}
F_{\rho \rho \rho}^{1}=\beta^2+\gamma^2+\mu^2-\beta\gamma\mu,
\end{equation}
\begin{equation}
F_{\rho \rho \rho}^{2}=\beta\gamma+\mu,
\end{equation}
\begin{equation}
F_{\rho \rho \rho}^{3}=\beta\mu+\gamma,
\end{equation}
\begin{equation}
F_{\rho \rho \rho}^{4}=\beta+\gamma \mu,
\end{equation}
for the even part, and using  definition of eq.(\ref{rrr}) we have for the odd part 
\begin{equation}
\mathcal{B}_{\rho_B\rho_B \rho_B}^{(A\,)\,ljk}= B_{\rho_B\rho_B\rho_B}^{(A)}\widehat{\mathbf{k1}-\mathbf{p}}^l\widehat{\mathbf{k2}+\mathbf{p}}^j\hat{\mathbf{p}}^k,
\end{equation}
 with
\begin{eqnarray}
B_{\rho_B\rho_B\rho_B}^{(A)}&=&\frac{8}{(2\pi)^3(4\pi)^3}\int d^3p   \left(P_H(\left|\mathbf{p}+\mathbf{k2}\right|)(P_H(p)P_H(\left| \mathbf{k1}-\mathbf{p}\right|)-P_B(p)P_B(\left| \mathbf{k1}-\mathbf{p}\right|)\beta) \nonumber  \right.\\
&+& \left. P_B(\left| \mathbf{p}+\mathbf{k2}\right|)(P_B(p)P_H(\left| \mathbf{k1}-\mathbf{p}\right|)\gamma -P_H(p)P_B(\left| \mathbf{k1}-\mathbf{p}\right|)\mu )  \right). 
\end{eqnarray}
\subsection*{Scalar cross-correlation($<\rho\rho\Pi>$)}
For the three-point cross-correlation   ($<\rho\rho\Pi>$) we obtain
\begin{eqnarray}
F_{\rho \rho \Pi^{(S)}}^{1}&=& 3(-1 +\alpha_q^2 +\beta_q^2+\gamma_q^2) \nonumber  \\
&+& \gamma^2+\beta^2 +\mu^2 -\beta\gamma\mu +3\beta\beta_q\gamma\gamma_q -3\alpha_q(\beta \beta_q+\gamma \gamma_q)-3\mu\beta_q\gamma_q, 
\end{eqnarray}
\begin{equation}
F_{\rho \rho \Pi^{(S)}}^{2}=2\beta\gamma-3\alpha_q\beta_q\gamma-3\alpha_q\beta\gamma_q-\mu+3\alpha_q^2\mu,
\end{equation}
\begin{equation}
F_{\rho \rho \Pi^{(S)}}^{3}=\gamma(-2+3\beta_q^2)+3\alpha_q\gamma_q+\beta\mu-3\alpha_q\beta_q\mu,
\end{equation}
\begin{equation}
F_{\rho \rho \Pi^{(S)}}^{4}=\beta(-2+3\gamma_q^2)+3\alpha_q\beta_q+\gamma \mu-3 \alpha_q \gamma_q \mu,
\end{equation}
for the even contribution, and with the definition of the three point correlation eq.(\ref{rrr}), the odd part is given by 
\begin{eqnarray}
\mathcal{B}_{\rho_B\rho_B\Pi_B^{(S)}}^{(A\,)\,ljk}&=& B_{\rho_B\rho_B\Pi_B^{(S)}}^{(A \,1)}  \widehat{\mathbf{k2}+\mathbf{p}}^l\widehat{\mathbf{k3}}^j\hat{\mathbf{p}}^k
+B_{\rho_B\rho_B\Pi_B^{(S)}}^{(A\,2)} \widehat{\mathbf{k1}-\mathbf{p}}^l\widehat{\mathbf{k3}}^j\hat{\mathbf{p}}^k \nonumber \\
&+& B_{\rho_B\rho_B\Pi_B^{(S)}}^{(A\,3)}\widehat{\mathbf{k1}-\mathbf{p}}^l\widehat{\mathbf{k2}+\mathbf{p}}^j\hat{\mathbf{k3}}^k 
+B_{\rho_B\rho_B\Pi_B^{(S)}}^{(A\,4)}\widehat{\mathbf{k1}-\mathbf{p}}^l\widehat{\mathbf{k2}+\mathbf{p}}^j\hat{\mathbf{p}}^k,
\end{eqnarray}
where
\begin{eqnarray}
B_{\rho_B\rho_B\Pi^{(S)}_B}^{(A\,1)}&=&\frac{-12}{(2\pi)^3(4\pi)^3}\int d^3p P_B(\left| \mathbf{k1}-\mathbf{p}\right|)\left( P_H(p)P_B(\left|\mathbf{p}+ \mathbf{k2}\right|)\gamma_q      \nonumber  \right.\\
&+& \left.  P_B(p)P_H(\left|\mathbf{p}+\mathbf{k2}\right|)(\alpha_q-\beta \beta_q) \right),
\end{eqnarray}
\begin{eqnarray}
B_{\rho_B\rho_B\Pi^{(S)}_B}^{(A\,2)}&=&\frac{12}{(2\pi)^3(4\pi)^3}\int d^3pP_B(\left| \mathbf{p}+ \mathbf{k2}\right|)\left(P_H(p)P_B(\left|\mathbf{k1}-\mathbf{p}\right|)\beta_q               \nonumber  \right.\\
&+& \left. P_B(p)P_H(\left|\mathbf{k1}-\mathbf{p}\right|)(\gamma \gamma_q -\alpha_q) \right),
\end{eqnarray}
\begin{eqnarray}
B_{\rho_B\rho_B\Pi^{(S)}_B}^{(A\,3)}&=&\frac{12}{(2\pi)^3(4\pi)^3}\int d^3p\left( P_H(\left| \mathbf{p}+ \mathbf{k2}\right|)P_H(p)P_H(\left|\mathbf{k1}-\mathbf{p}\right|)\alpha_q               \nonumber  \right.\\
&+& \left. P_B(p)(P_H(\left|\mathbf{k1}-\mathbf{p}\right|)P_B(\left|\mathbf{p}+\mathbf{k2}\right|)\gamma_q  -P_B(\left|\mathbf{k1}-\mathbf{p}\right|)P_H(\left|\mathbf{p}+\mathbf{k2}\right|)\beta_q )         \right),
\end{eqnarray}
\begin{eqnarray}
B_{\rho_B\rho_B\Pi^{(S)}_B}^{(A\,4)}&=&\frac{-4}{(2\pi)^3(4\pi)^3}\int d^3p\left( P_H(\left| \mathbf{p}+ \mathbf{k2}\right|)P_H(p)P_H(\left|\mathbf{k1}-\mathbf{p}\right|)               \nonumber  \right.\\
&-&  P_B(p)(P_B(\left|\mathbf{k1}-\mathbf{p}\right|)P_H(\left|\mathbf{p}+\mathbf{k2}\right|)\beta  -P_H(\left|\mathbf{k1}-\mathbf{p}\right|)P_B(\left|\mathbf{p}+\mathbf{k2}\right|)\gamma )     \nonumber \\
&+& \left.  P_B(\left|\mathbf{k1}-\mathbf{p}\right|)P_H(p)P_B(\left|\mathbf{p}+\mathbf{k2}\right|)(3\gamma_q\beta_q-\mu) \right).
\end{eqnarray}
\subsection*{Scalar cross-correlation ($<\rho\Pi\Pi>$)}
The  result  for the even contribution of  the three-point cross-correlation   ($<\rho\Pi\Pi>$) is given by
\begin{eqnarray}
F_{\rho \Pi^{(S)}\Pi^{(S)}}^{1}&=&-6+3(\alpha_p^2+\beta_p^2+\gamma_p^2)+ 3(\alpha_q^2+\beta_q^2+\gamma_q^2)+\beta^2+\gamma^2+\mu^2+3\beta_q(\beta\gamma\gamma_q+3\beta_p\gamma_p\gamma_q)  \nonumber \\
&-&  3\beta_p\gamma_p\mu       -\beta\gamma\mu        -3\alpha_q(\beta\beta_q+\gamma\gamma_q)-9\theta_{pq}(\beta_q\beta_p+\gamma_p\gamma_q-\theta_{pq}) -3\beta_q\gamma_q\mu \nonumber \\
&-& 3\alpha_p\beta(\beta_p+3\beta_q\gamma_p\gamma_q-3\beta_q\theta_{pq}-\gamma_p\mu)-3\alpha_p(\gamma\gamma_p-3\alpha_q\gamma_p\gamma_q+3\alpha_q\theta_{pq}),
\end{eqnarray}

\begin{eqnarray}
F_{\rho \Pi^{(S)}\Pi^{(S)}}^{2}&=&-3\alpha_q\beta_q\gamma+9\gamma_q\beta_q+\beta(2\gamma-3\alpha_q\gamma_q) +\mu(-7+3\alpha_q^2+9\theta_{pq}^2) +\alpha_p^2(6\mu-9\beta_q\gamma_q)           \nonumber \\
&-&9\theta_{pq}(\beta_q\gamma_p+\beta_p\gamma_q)+\alpha_p\beta_p(-6\gamma+9\alpha_q\gamma_q)+ 9\alpha_p\theta_{pq}(\beta_q\gamma-\alpha_q\mu)+6\beta_p\gamma_p,
\end{eqnarray}

\begin{eqnarray}\label{no3}
F_{\rho \Pi^{(S)}\Pi^{(S)}}^{3}&=& 9\alpha_q\gamma_p(\beta_p\beta_q-\theta_{pq}) -3\alpha_p(3\gamma_p(\beta_q^2-1)+3\gamma_q\theta_{pq}+\beta_p\mu-3\beta_q\theta_{pq}\mu)+6\alpha_q\gamma_q \nonumber \\
&-& 3\beta\beta_p\gamma_p+\gamma(-7+3\beta_p^2+6\beta_q^2-9\beta_p\beta_q\theta_{pq}+9\theta_{pq}^2)+2\beta\mu-6\alpha_q\beta_q\mu, 
\end{eqnarray}
\begin{eqnarray}
 F_{\rho \Pi^{(S)}\Pi^{(S)}}^{4}&=& 3\alpha_p\beta_p+3\alpha_q\beta_q-3\beta_p\gamma\gamma_p+9\alpha_q\beta_p(\gamma_p\gamma_q-\theta_{pq}) \nonumber \\
&+&  \beta (-5+3\gamma_p^2+3\gamma_q^2-9\gamma_p\gamma_q\theta_{pq}+9\theta_{pq}^2)+\gamma \mu-3\alpha_q\gamma_q \mu. 
\end{eqnarray}
Using eq.(\ref{rrr}),  the  odd contribution can be written as
\begin{eqnarray}
\mathcal{B}_{\rho_B\Pi_B^{(S)}\Pi_B^{(S)}}^{(A\,)\,ljk} &=&  B_{\rho_B\Pi_B^{(S)}\Pi_B^{(S)}}^{(A \,1)} \widehat{\mathbf{k2}+\mathbf{p}}^l\widehat{\mathbf{k3}}^j\widehat{\mathbf{p}}^k \nonumber \\
&+&  B_{\rho_B\Pi_B^{(S)}\Pi_B^{(S)}}^{(A\,2)} \widehat{\mathbf{k2}}^l\widehat{\mathbf{k3}}^j\hat{\mathbf{p}}^k+  B_{\rho_B\Pi_B^{(S)}\Pi_B^{(S)}}^{(A\,3)}\widehat{\mathbf{k2}}^l\widehat{\mathbf{k2}+\mathbf{p}}^j\hat{\mathbf{k3}}^k \nonumber \\
&+&  B_{\rho_B\Pi_B^{(S)}\Pi_B^{(S)}}^{(A\,4)}\widehat{\mathbf{k1}-\mathbf{p}}^l\widehat{\mathbf{k2}+\mathbf{p}}^j\hat{\mathbf{k3}}^k+ B_{\rho_B\Pi_B^{(S)}\Pi_B^{(S)}}^{(A\,5)}\widehat{\mathbf{k1}-\mathbf{p}}^l\widehat{\mathbf{k2}+\mathbf{p}}^j\hat{\mathbf{p}}^k       \nonumber \\
&+& 
 B_{\rho_B\Pi_B^{(S)}\Pi_B^{(S)}}^{(A\,6)}\widehat{\mathbf{k1}-\mathbf{p}}^l\widehat{\mathbf{k3}}^j\hat{\mathbf{p}}^k+ B_{\rho_B\Pi_B^{(S)}\Pi_B^{(S)}}^{(A\,7)}\widehat{\mathbf{k1}-\mathbf{p}}^l\widehat{\mathbf{k2}}^j\widehat{\mathbf{k2}+\mathbf{p}}^k     \nonumber \\
&+& 
 B_{\rho_B\Pi_B^{(S)}\Pi_B^{(S)}}^{(A\,8)}\widehat{\mathbf{k1}-\mathbf{p}}^l\widehat{\mathbf{k2}}^j\hat{\mathbf{k3}}^k+ B_{\rho_B\Pi_B^{(S)}\Pi_B^{(S)}}^{(A\,9)}\widehat{\mathbf{k1}-\mathbf{p}}^l\widehat{\mathbf{k2}}^j\hat{\mathbf{p}}^k\nonumber \\
&+&   B_{\rho_B\Pi_B^{(S)}\Pi_B^{(S)}}^{(A\,10)}\widehat{\mathbf{k2}}^l\widehat{\mathbf{k2}+\mathbf{p}}^j\widehat{\mathbf{p}}^k,
\end{eqnarray}
with
\begin{eqnarray}
B_{\rho_B\Pi^{(S)}_B\Pi^{(S)}_B}^{(A\,1)}&=&\frac{6}{(2\pi)^3(4\pi)^3}\int d^3pP_B(\left|\mathbf{k1}-\mathbf{p}\right|)\left(P_B(p)P_H(\left|\mathbf{p}+\mathbf{k2}\right|)(\alpha_q-\beta\beta_q)\nonumber  \right.\\
&+& \left. P_H(p)P_B(\left|\mathbf{p}+\mathbf{k2}\right|)\gamma_q  \right),             
\end{eqnarray}

\begin{equation}
B_{\rho_B\Pi^{(S)}_B\Pi^{(S)}_B}^{(A\,2)}=\frac{18}{(2\pi)^3(4\pi)^3}\int d^3pP_B(\left|\mathbf{k1}-\mathbf{p}\right|)P_H(p)P_B(\left|\mathbf{p}+\mathbf{k2}\right|)(-\gamma_p\gamma_q+\theta_{pq}),             
\end{equation}

\begin{eqnarray}
B_{\rho_B\Pi^{(S)}_B\Pi^{(S)}_B}^{(A\,3)}&=&\frac{18}{(2\pi)^3(4\pi)^3}\int d^3p P_H(\left|\mathbf{p}+\mathbf{k2}\right|)\left( P_H(p)P_H(\left|\mathbf{k1}-\mathbf{p}\right|)(\beta_p\alpha_q-\beta\theta_{pq}) \nonumber  \right.\\
&+& \left. P_B(\left|\mathbf{k1}-\mathbf{p}\right|) P_B(p)(-\alpha_p\alpha_q+\alpha_p\beta\beta_q-\beta_p\beta_q+\theta_{pq}) \right),             
\end{eqnarray}

\begin{eqnarray}
B_{\rho_B\Pi^{(S)}_B\Pi^{(S)}_B}^{(A\,4)}&=&\frac{-6}{(2\pi)^3(4\pi)^3}\int d^3p \left(P_H(\left|\mathbf{p}+\mathbf{k2}\right|) P_H(p)P_H(\left|\mathbf{k1}-\mathbf{p}\right|)\alpha_q \right. \nonumber\\
&+& \left. P_B(p)(P_H(\left|\mathbf{k1}-\mathbf{p}\right|)P_B(\left|\mathbf{p}+\mathbf{k2}\right|)\gamma_q-  P_B(\left|\mathbf{k1}-\mathbf{p}\right|)P_H(\left|\mathbf{p}+\mathbf{k2}\right|)\beta_q              ) \right),             
\end{eqnarray}

\begin{eqnarray}
B_{\rho_B\Pi^{(S)}_B\Pi^{(S)}_B}^{(A\,5)}&=&\frac{2}{(2\pi)^3(4\pi)^3}\int d^3p \left(P_B(p)\left( P_B(\left|\mathbf{p}+\mathbf{k2}\right|)P_H(\left|\mathbf{k1}-\mathbf{p}\right|)(\gamma-3\alpha_p\gamma_p) \right.\right. \nonumber \\
&-& \left. P_B(\left|\mathbf{k1}-\mathbf{p}\right|)P_H(\left|\mathbf{p}+\mathbf{k2}\right|)\beta \right) +P_H(p)\left(P_H(\left|\mathbf{k1}-\mathbf{p}\right|)P_H(\left|\mathbf{p}+\mathbf{k2}\right|) \right. \nonumber \\
&+&\left. \left. P_B(\left|\mathbf{k1}-\mathbf{p}\right|)P_B(\left|\mathbf{p}+\mathbf{k2}\right|)(3\beta_q\gamma_q-\mu) \right)\right),             
\end{eqnarray}

\begin{eqnarray}
B_{\rho_B\Pi^{(S)}_B\Pi^{(S)}_B}^{(A\,6)}&=&\frac{6}{(2\pi)^3(4\pi)^3}\int d^3p P_B(\left|\mathbf{p}+\mathbf{k2}\right|)\left( -P_H(p) P_B(\left|\mathbf{k1}-\mathbf{p}\right|)\beta_q\right. \nonumber \\
&+&\left. P_H(\left|\mathbf{k1}-\mathbf{p}\right|)P_B(p)(\alpha_q-\gamma\gamma_q+3\alpha_p\gamma_p\gamma_q-3\alpha_p\theta_{pq})\right),             
\end{eqnarray}

\begin{eqnarray}
B_{\rho_B\Pi^{(S)}_B\Pi^{(S)}_B}^{(A\,7)}&=&\frac{-6}{(2\pi)^3(4\pi)^3}\int d^3p P_B(p)\left( P_B(\left|\mathbf{p}+\mathbf{k2}\right|) P_H(\left|\mathbf{k1}-\mathbf{p}\right|)\gamma_p\right. \nonumber \\
&+&\left. P_B(\left|\mathbf{k1}-\mathbf{p}\right|)P_H(\left|\mathbf{p}+\mathbf{k2}\right|)(\alpha_p\beta-\beta_p)\right),             
\end{eqnarray}

\begin{equation}
B_{\rho_B\Pi^{(S)}_B\Pi^{(S)}_B}^{(A\,8)}=\frac{-18}{(2\pi)^3(4\pi)^3}\int d^3p P_B(p) P_B(\left|\mathbf{p}+\mathbf{k2}\right|) P_H(\left|\mathbf{k1}-\mathbf{p}\right|)(-\gamma_p\gamma_q+\theta_{pq}),            
\end{equation}

\begin{eqnarray}
B_{\rho_B\Pi^{(S)}_B\Pi^{(S)}_B}^{(A\,9)}&=&\frac{6}{(2\pi)^3(4\pi)^3}\int d^3p P_B(\left|\mathbf{p}+\mathbf{k2}\right|)\left(P_B(p)P_H(\left|\mathbf{k1}-\mathbf{p}\right|)\alpha_p\right. \nonumber \\
&+&\left. P_B(\left|\mathbf{k1}-\mathbf{p}\right|)P_H(p)(-\beta_p-3\beta_q\gamma_p\gamma_q+3\beta_q\theta_{pq}+\gamma_p\mu)\right).             
\end{eqnarray}

\begin{eqnarray}
B_{\rho_B\Pi^{(S)}_B\Pi^{(S)}_B}^{(A\,10)}&=&\frac{6}{(2\pi)^3(4\pi)^3}\int d^3p P_H(\left|\mathbf{p}+\mathbf{k2}\right|)P_B(p)P_B(\left|\mathbf{k1}-\mathbf{p}\right|)\alpha_p \nonumber \\
&+& P_H(p)\left(P_B(\left|\mathbf{k1}-\mathbf{p}\right|)P_B(\left|\mathbf{k2}+\mathbf{p}\right|)\beta_p-P_H(\left|\mathbf{k1}-\mathbf{p}\right|)P_H(\left|\mathbf{k2}+\mathbf{p}\right|)\gamma_p\right).             
\end{eqnarray}
 
\subsection*{Scalar cross-correlation($<\Pi\Pi\Pi>$)}
Finally,  the even contribution for the three-point cross-correlation of scalar anisotropic stress is written as 

\begin{eqnarray}
F_{\Pi^{(S)} \Pi^{(S)}\Pi^{(S)}}^{1}&=&-9+3(\alpha_k^2+\alpha_p^2+\alpha_q^2) -9\theta_{pq}(\beta_p\beta_q+\gamma_p\gamma_q-3\beta_k\beta_q\theta_{kp}+3\theta_{kp}\theta_{kq}) \nonumber\\
&+& 3(\beta_q^2+\beta_p^2+\beta_k^2)+\beta^2+\gamma^2+\mu^2-\beta\gamma\mu+3(\gamma_k^2+\gamma_p^2+\gamma_q^2)-3\alpha_q(\beta\beta_q+\gamma\gamma_q) \nonumber \\
&+& 3\beta_q(\beta\gamma\gamma_q+3\beta_k\gamma_k\gamma_q+3\beta_p\gamma_p\gamma_q)-9\theta_{kp}(\beta_k\beta_p+\gamma_k\gamma_p+3\beta_q\beta_k\gamma_p\gamma_q-\theta_{kp})          \nonumber \\
&-& 3\alpha_p(\gamma\gamma_p+3\alpha_q(-\gamma_p\gamma_q+\theta_{pq})+\beta(\beta_p+3\beta_q\gamma_p\gamma_q-3\beta_q\theta_{pq}-\gamma_p\mu) )+9\theta_{pq}^2    \nonumber \\
&+& 9\theta_{kq}^2-3 \alpha_k\left(\beta\beta_k-3\alpha_q\beta_k\beta_q+\gamma\gamma_k+3\beta_k\beta_q\gamma\gamma_q+3\alpha_q\theta_{kq}-3\gamma\gamma_q\theta_{kq}-\beta_k\gamma\mu \right. \nonumber \\
&-& \left.3  \alpha_p(\gamma_p\gamma_k-\theta_{kp}-3\gamma_p\gamma_q\theta_{kq}+3\theta_{kq}\theta_{pq}+\beta_k(\beta_p+3\beta_{q}\gamma_p\gamma_q-3\beta_q\theta_{pq}-\gamma_p\mu ))\right) \nonumber \\
&-& 3\mu(\beta_k\gamma_k+\gamma_p\beta_p+\beta_q\gamma_q-3\beta_k\gamma_p\theta_{kp}) -9\theta_{kq}(\beta_k\beta_q+\gamma_k\gamma_q-3\theta_{kp}\gamma_p\gamma_q),  
\end{eqnarray}
\begin{eqnarray}
F_{\Pi^{(S)} \Pi^{(S)}\Pi^{(S)}}^{2}&=& 3\beta_q(3\alpha_k\alpha_q\gamma_k-\alpha_q\gamma+6\gamma_q-3\alpha_k^2\gamma_q-3\gamma_k\theta_{kq}) +\alpha_p^2(-9\beta_q\gamma_q+6\mu)    \nonumber \\
&+& 6\beta_p(\gamma_p-3\gamma_k\theta_{kp})-9\gamma_q(3\beta_q\theta_{kp}^2+\beta_k\theta_{kq}-3\beta_p\theta_{kq}\theta_{kp}+\beta_p\theta_{pq})          \nonumber \\
&+& \beta(2\gamma-6\alpha_k\gamma_k-3\alpha_q\gamma_q+9\alpha_k\gamma_q\theta_{kq})+9\theta_{pq}(3\theta_{kp}\gamma_k\beta_q-\beta_q\gamma_p) +6\beta_k\gamma_k  \nonumber \\ 
&+& \mu(-13+6\alpha_k^2+3\alpha_q^2+18\theta_{kp}^2-9\alpha_k\alpha_q\theta_{kq} +9\theta_{kq}^2-27\theta_{kp}\theta_{kq}\theta_{pq}+9\theta_{pq}^2) \nonumber \\
&+& 3\alpha_p\left(\alpha_k\beta_q\gamma_q\theta_{kp}+\beta_p( -2\gamma+6\alpha_k\gamma_k+3\alpha_q\gamma_q-9\alpha_k\gamma_q\theta_{kq})+3\beta_q\gamma\theta_{pq} \right. \nonumber\\
&-& \left. 9\alpha_k\beta_q\gamma_k\theta_{pq}-6\alpha_k\theta_{kp}\mu-3\alpha_q\theta_{pq}\mu+9\alpha_k\theta_{kq}\theta_{pq}\mu\right),
\end{eqnarray}

\begin{eqnarray}
F_{\Pi^{(S)} \Pi^{(S)}\Pi^{(S)}}^{3}&=& 6\beta_p(\gamma\gamma_p-\alpha_p)-6\alpha_q(\beta_q+3\beta_p\gamma_p\gamma_q)-27\alpha_q\beta_k\theta_{kp}\theta_{pq}-2\mu(\gamma-3\alpha_q\gamma_q)   \nonumber \\
&+& 3\beta_k(\gamma\gamma_k-3\alpha_q\gamma_k\gamma_q+3\alpha_p\theta_{kp}-3\gamma\gamma_p\theta_{kp}+9\alpha_q\gamma_p\gamma_q\theta_{kp}+3\alpha_q\theta_{kq})+18\alpha_q\beta_p\theta_{pq}                  \nonumber \\
&-& \beta\left(-13 +3(\gamma_k^2+2\gamma_p^2+2\gamma_q^2)+9(\theta_{kp}^2+\theta_{kq}^2+2\theta_{pq}^2) -9\gamma_k(\gamma_p\theta_{kp}+\gamma_q\theta_{kq})                \right.          \nonumber \\
&+& \left. 9\gamma_p\gamma_q(3\theta_{kp}\theta_{kq}-2\theta_{pq})-27\theta_{kp}\theta_{kq}\theta_{pq}\right)+3\alpha_k\left(3\beta_q\theta_{kq} + \beta_p\left(-3\gamma_k\gamma_p+3\theta_{kp}  -9\theta_{kq}\theta_{pq}     \right.   \right. \nonumber \\
&+&\left.  \left. 9\gamma_p\gamma_q\theta_{kq}\right)+3\beta_k(-2+\gamma_p^2+\gamma_q^2-3\gamma_p\gamma_q\theta_{pq}+3\theta_{pq}^2)+ \mu(\gamma_k-3\gamma_q\theta_{kq}) \right),
\end{eqnarray}

\begin{eqnarray}
F_{\Pi^{(S)} \Pi^{(S)}\Pi^{(S)}}^{4}&=& -27\alpha_q\beta_k\beta_q\gamma_p\theta_{kp}-3\beta\beta_p\gamma_p  +\mu(2\beta-3\alpha_p\beta_p-6\alpha_q\beta_q+9\alpha_p\beta_q\theta_{pq})  \nonumber \\
&+& 6 \gamma_k(\alpha_k-\beta\beta_k+3\alpha_q\beta_k\beta_q)+9\alpha_p\beta_k\beta_p\gamma_k+9\gamma_p(2\alpha_p-\alpha_p\beta_k^2+\alpha_q\beta_p\beta_q-\alpha_p\beta_q^2)  \nonumber \\                                                        
&+& 6 \alpha_q\gamma_q  -9 \left(\alpha_k\gamma_p\theta_{kp}-\beta\beta_k\gamma_p\theta_{kp}+2\alpha_q\gamma_k\theta_{kq}-3\alpha_q\gamma_p\theta_{kp}\theta_{kq}+\alpha_q\gamma_p\theta_{pq} \right. \nonumber  \\
&+& \left.  \alpha_p \left(3\gamma_p\theta_{kq}^2 -3\beta_k\beta_q\gamma_p\theta_{kq}+\gamma_q\theta_{pq}+\gamma_k(\theta_{kp}+3\beta_k\beta_q\theta_{pq}-3\theta_{kq}\theta_{pq}) \right) \right)    \nonumber \\
&+& \gamma \left(-13+3(2\beta_k^2+\beta_p^2+2\beta_q^2) +9(\theta_{kp}^2+2\theta_{kq}^2+\theta_{pq}^2)-9\theta_{pq}(\beta_p\beta_q+3\theta_{kp}\theta_{kq}) \right. \nonumber \\
&-& \left. 9  \beta_k(\beta_p\theta_{kp}+2\beta_q\theta_{kq}-3\beta_q\theta_{kp}\theta_{pq})\right),
\end{eqnarray}
for the odd contribution we found
\begin{eqnarray}
B_{\Pi^{(S)}_B\Pi^{(S)}_B\Pi^{(S)}_B}^{(A\,1)}&=&\frac{3}{(2\pi)^3(4\pi)^3}\int d^3p\left( P_H(p) P_H(\left|\mathbf{k1}-\mathbf{p}\right|) P_H(\left|\mathbf{p}+\mathbf{k2}\right|) (-2 \alpha_q + 3 \alpha_k \theta_{kq}) \right.\nonumber \\
&-& \left. P_B(p)( P_B(\left|\mathbf{k1}-\mathbf{p}\right|) P_H(\left|\mathbf{p}+\mathbf{k2}\right|) \beta_q -  P_B(\left|\mathbf{p}+\mathbf{k2}\right|) P_H(\left|\mathbf{k1}-\mathbf{p}\right|) \gamma_q)\right),             
\end{eqnarray}

\begin{eqnarray}
B_{\Pi^{(S)}_B\Pi^{(S)}_B\Pi^{(S)}_B}^{(A\,2)}&=&\frac{1}{(2\pi)^3(4\pi)^3}\int d^3p \left(P_B(p) (P_B(\left|\mathbf{k1}-\mathbf{p}\right|) P_H(\left|\mathbf{p}+\mathbf{k2}\right|) (\beta - 3 \alpha_k \beta_k) \right.   \nonumber\\         
 &-&\left. P_B(\left|\mathbf{p}+\mathbf{k2}\right|) P_H(\left|\mathbf{k1}-\mathbf{p}\right|) (\gamma - 3 \alpha_p \gamma_p)\right)+ P_H(p)\left(2 P_H(\left|\mathbf{k1}-\mathbf{p}\right|) P_H(\left|\mathbf{p}+\mathbf{k2}\right|)
 \right.   \nonumber\\
&+&\left.\left. P_B(\left|\mathbf{k1}-\mathbf{p}\right|)P_B(\left|\mathbf{p}+\mathbf{k2}\right|) (-3 \beta_q \gamma_q + \mu)\right)\right)
\end{eqnarray}

\begin{eqnarray}
B_{\Pi^{(S)}_B\Pi^{(S)}_B\Pi^{(S)}_B}^{(A\,3)}&=&\frac{3}{(2\pi)^3(4\pi)^3}\int d^3p P_B(\left|\mathbf{p}+\mathbf{k2}\right|)\left( P_H(p)P_B(\left|\mathbf{k1}-\mathbf{p}\right|)\beta_q \nonumber  \right.\\
&+& \left. P_H(\left|\mathbf{k1}-\mathbf{p}\right|) P_B(p)(-\alpha_q +\gamma\gamma_q-3\alpha_p\gamma_p\gamma_q  +3\alpha_p\theta_{pq}) \right),             
\end{eqnarray}

\begin{eqnarray}
B_{\Pi^{(S)}_B\Pi^{(S)}_B\Pi^{(S)}_B}^{(A\,4)}&=&\frac{3}{(2\pi)^3(4\pi)^3}\int d^3p P_B(p)\left(P_B(\left|\mathbf{p}+\mathbf{k2}\right|) P_H(\left|\mathbf{k1}-\mathbf{p}\right|)\gamma_p \right. \nonumber\\
&+& \left.P_B(\left|\mathbf{k1}-\mathbf{p}\right|)P_H(\left|\mathbf{p}+\mathbf{k2}\right|)(\alpha_p\beta-3\alpha_k\alpha_p\beta_k-\beta_p+3\beta_k\theta_{kp})             \right),             
\end{eqnarray}

\begin{equation}
B_{\Pi^{(S)}_B\Pi^{(S)}_B\Pi^{(S)}_B}^{(A\,5)}=\frac{9}{(2\pi)^3(4\pi)^3}\int d^3p P_B(p) P_B(\left|\mathbf{p}+\mathbf{k2}\right|)P_H(\left|\mathbf{k1}-\mathbf{p}\right|)(-\gamma_p\gamma_q+\theta_{pq})
\end{equation}

\begin{eqnarray}
B_{\Pi^{(S)}_B\Pi^{(S)}_B\Pi^{(S)}_B}^{(A\,6)}&=&\frac{-3}{(2\pi)^3(4\pi)^3}\int d^3p  P_B(\left|\mathbf{p}+\mathbf{k2}\right|)\left( P_B(p) P_H(\left|\mathbf{k1}-\mathbf{p}\right|)\alpha_p \right.  \nonumber \\
&+&\left. P_B(\left|\mathbf{k1}-\mathbf{p}\right|)P_H(p)(-\beta_p-3\beta_q\gamma_p\gamma_q+3\beta_q\theta_{pq}+\gamma_p\mu)\right),             
\end{eqnarray}

\begin{eqnarray}
B_{\Pi^{(S)}_B\Pi^{(S)}_B\Pi^{(S)}_B}^{(A\,7)}&=&\frac{-3}{(2\pi)^3(4\pi)^3}\int d^3p P_B(\left|\mathbf{k1}-\mathbf{p}\right|)\left( P_B(\left|\mathbf{p}+\mathbf{k2}\right|)P_H(p) \gamma_q\right. \nonumber \\
&+&\left. P_B(p)P_H(\left|\mathbf{p}+\mathbf{k2}\right|)(\alpha_q-\beta\beta_q+3\alpha_k\beta_k\beta_q-3\alpha_k\theta_{kq} )\right),             
\end{eqnarray}

\begin{equation}
B_{\Pi^{(S)}_B\Pi^{(S)}_B\Pi^{(S)}_B}^{(A\,8)}=\frac{-9}{(2\pi)^3(4\pi)^3}\int d^3p P_H(p) P_B(\left|\mathbf{p}+\mathbf{k2}\right|) P_B(\left|\mathbf{k1}-\mathbf{p}\right|)(-\gamma_p\gamma_q+\theta_{pq}),            
\end{equation}

\begin{eqnarray}
B_{\Pi^{(S)}_B\Pi^{(S)}_B\Pi^{(S)}_B}^{(A\,9)}&=&\frac{-9}{(2\pi)^3(4\pi)^3}\int d^3p P_H(\left|\mathbf{p}+\mathbf{k2}\right|)\left( P_B(p)P_B(\left|\mathbf{k1}-\mathbf{p}\right|)\left(-\beta_p\beta_q+3\beta_k\beta_q\theta_{kp}  +\theta_{pq}                 \right.\right. \nonumber \\
&-& \left. 3\theta_{kp}\theta_{kq}+\alpha_p(-\alpha_q+\beta\beta_q-3\alpha_k\beta_k\beta_q+3\alpha_k\theta_{kq})\right) +P_H(\left|\mathbf{k1}-\mathbf{p}\right|)P_H(p) \left(-2\alpha_q\beta_p \right. \nonumber \\
&+& \left.\left.  3\alpha_q\beta_k\theta_{kp}+3\alpha_k\beta_p\theta_{kq}-3\beta\theta_{kp}\theta_{kq}+2\beta\theta_{pq}-3\alpha_k\beta_k\theta_{pq}\right)\right) 
\end{eqnarray}

\begin{eqnarray}
B_{\Pi^{(S)}_B\Pi^{(S)}_B\Pi^{(S)}_B}^{(A\,10)}&=&\frac{-3}{(2\pi)^3(4\pi)^3}\int d^3p \left( P_H(\left|\mathbf{p}+\mathbf{k2}\right|)P_H(p)P_H(\left|\mathbf{k1}-\mathbf{p}\right|)\alpha_k \right. \nonumber \\
&-& P_B(p) \left( P_B(\left|\mathbf{k1}-\mathbf{p}\right|) P_H(\left|\mathbf{p}+\mathbf{k2}\right|) \beta_k+   P_H(\left|\mathbf{k1}-\mathbf{p}\right|) P_B(\left|\mathbf{p}+\mathbf{k2}\right|)(\alpha_k\gamma  \right. \nonumber \\
&-&\left.\left. \gamma_k-3\alpha_k\alpha_p\gamma_p+3\gamma_p\theta_{kp})\right)\right),             
\end{eqnarray}

\begin{equation}
B_{\Pi^{(S)}_B\Pi^{(S)}_B\Pi^{(S)}_B}^{(A\,11)}=\frac{-9}{(2\pi)^3(4\pi)^3}\int d^3p P_B(p) P_B(\left|\mathbf{p}+\mathbf{k2}\right|)P_H(\left|\mathbf{k1}-\mathbf{p}\right|)  (-\alpha_k\alpha_p+\theta_{kp})      
\end{equation}

\begin{eqnarray}
B_{\Pi^{(S)}_B\Pi^{(S)}_B\Pi^{(S)}_B}^{(A\,12)}&=&\frac{-9}{(2\pi)^3(4\pi)^3}\int d^3p P_B(p) P_B(\left|\mathbf{p}+\mathbf{k2}\right|)P_H(\left|\mathbf{k1}-\mathbf{p}\right|)  \left(-\gamma_k\gamma_q+3\gamma_p\gamma_q\theta_{kp}       \right. \nonumber \\
&+&\left.  \theta_{kq}-3\theta_{kp}\theta_{pq}+\alpha_k(-\alpha_q+\gamma\gamma_q-3\alpha_p\gamma_p\gamma_q+3\alpha_p\theta_{pq})               \right),             
\end{eqnarray}

\begin{eqnarray}
B_{\Pi^{(S)}_B\Pi^{(S)}_B\Pi^{(S)}_B}^{(A\,13)}&=&\frac{-9}{(2\pi)^3(4\pi)^3}\int d^3p P_H(\left|\mathbf{p}+\mathbf{k2}\right|)\left( P_B(p)P_B(\left|\mathbf{k1}-\mathbf{p}\right|)(-\beta_k\beta_q+\theta_{kq})\right. \nonumber \\
&+&\left. P_H(\left|\mathbf{k1}-\mathbf{p}\right|)P_H(p)(-\alpha_q\beta_k+\beta\theta_{kq})\right),             
\end{eqnarray}

\begin{eqnarray}
B_{\Pi^{(S)}_B\Pi^{(S)}_B\Pi^{(S)}_B}^{(A\,14)}&=&\frac{-3}{(2\pi)^3(4\pi)^3}\int d^3p \left(- P_H(\left|\mathbf{p}+\mathbf{k2}\right|)P_B(p)P_B(\left|\mathbf{k1}-\mathbf{p}\right|)\alpha_k \right. \nonumber \\
&+& P_H(p) \left( P_H(\left|\mathbf{k1}-\mathbf{p}\right|) P_H(\left|\mathbf{p}+\mathbf{k2}\right|) \beta_k+   P_B(\left|\mathbf{k1}-\mathbf{p}\right|) P_B(\left|\mathbf{p}+\mathbf{k2}\right|)(-\gamma_k \right. \nonumber \\
&-&\left.\left. 3\beta_k\beta_q\gamma_q+3\gamma_q\theta_{kq}+\beta_k\mu)\right)\right),             
\end{eqnarray}

\begin{equation}
B_{\Pi^{(S)}_B\Pi^{(S)}_B\Pi^{(S)}_B}^{(A\,15)}=\frac{9}{(2\pi)^3(4\pi)^3}\int d^3p P_H(p) P_B(\left|\mathbf{p}+\mathbf{k2}\right|) P_B(\left|\mathbf{k1}-\mathbf{p}\right|)(-\beta_k\beta_q+\theta_{kq}),            
\end{equation}

\begin{eqnarray}
B_{\Pi^{(S)}_B\Pi^{(S)}_B\Pi^{(S)}_B}^{(A\,16)}&=&\frac{9}{(2\pi)^3(4\pi)^3}\int d^3p P_H(\left|\mathbf{p}+\mathbf{k2}\right|)\left( P_B(p)P_B(\left|\mathbf{k1}-\mathbf{p}\right|)(\alpha_k\alpha_p-\theta_{kp})\right. \nonumber \\
&+&\left. P_H(\left|\mathbf{k1}-\mathbf{p}\right|)P_H(p)(\alpha_k\beta_p-\beta\theta_{kp})\right),             
\end{eqnarray}

\begin{eqnarray}
B_{\Pi^{(S)}_B\Pi^{(S)}_B\Pi^{(S)}_B}^{(A\,17)}&=&\frac{3}{(2\pi)^3(4\pi)^3}\int d^3p P_H(p)P_B(\left|\mathbf{p}+\mathbf{k2}\right|)P_B(\left|\mathbf{k1}-\mathbf{p}\right|)\left(-\gamma_k\gamma_p+\theta_{kp} \right. \nonumber \\
&+&\left.  3\gamma_p\gamma_q\theta_{kq}-3\theta_{kq}\theta_{pq}+\beta_k(-\beta_p-3\beta_q\gamma_p\gamma_q+3\beta_q\theta_{pq}+\gamma_p\mu)                                      \right),             
\end{eqnarray}

\begin{eqnarray}
B_{\Pi^{(S)}_B\Pi^{(S)}_B\Pi^{(S)}_B}^{(A\,18)}&=&\frac{-3}{(2\pi)^3(4\pi)^3}\int d^3p P_B(p)P_H(\left|\mathbf{p}+\mathbf{k2}\right|)P_B(\left|\mathbf{k1}-\mathbf{p}\right|)\alpha_p \nonumber \\
&+&  P_H(p)\left(P_B(\left|\mathbf{p}+\mathbf{k2}\right|)P_B(\left|\mathbf{k1}-\mathbf{p}\right|)\gamma_p+P_H(\left|\mathbf{p}+\mathbf{k2}\right|)P_H(\left|\mathbf{k1}-\mathbf{p}\right|)(2\beta_p-3\beta_k\theta_{kp})                                     \right),      \nonumber       \\
\end{eqnarray}

\begin{eqnarray}
B_{\Pi^{(S)}_B\Pi^{(S)}_B\Pi^{(S)}_B}^{(A\,19)}&=&\frac{-3}{(2\pi)^3(4\pi)^3}\int d^3p P_B(\left|\mathbf{p}+\mathbf{k2}\right|)\left(P_H(\left|\mathbf{k1}-\mathbf{p}\right|)P_B(p)\alpha_k \right.\nonumber \\
&-& \left. P_H(p)P_B(\left|\mathbf{k1}-\mathbf{p}\right|)\beta_k                                     \right),      \nonumber       \\
\end{eqnarray}
where
\begin{eqnarray}
\mathcal{B}_{\Pi_B^{(S)}\Pi_B^{(S)}\Pi_B^{(S)}}^{(A\,)\,ljk} &=& B_{\Pi_B^{(S)}\Pi_B^{(S)}\Pi_B^{(S)}}^{(A \,1)}\widehat{\mathbf{k1}-\mathbf{p}}^l\widehat{\mathbf{k2}+\mathbf{p}}^j\hat{\mathbf{k3}}^k  + B_{\Pi_B^{(S)}\Pi_B^{(S)}\Pi_B^{(S)}}^{(A \,2)}\widehat{\mathbf{k1}-\mathbf{p}}^l\widehat{\mathbf{k2}+\mathbf{p}}^j\hat{\mathbf{p}}^k  \nonumber \\
&+&  B_{\Pi_B^{(S)}\Pi_B^{(S)}\Pi_B^{(S)}}^{(A\,3)} \widehat{\mathbf{k1}-\mathbf{p}}^l\widehat{\mathbf{k3}}^j\hat{\mathbf{p}}^k+  B_{\Pi_B^{(S)}\Pi_B^{(S)}\Pi_B^{(S)}}^{(A\,4)}\widehat{\mathbf{k1}-\mathbf{p}}^l\widehat{\mathbf{k2}}^j\hat{\mathbf{k2}+\mathbf{p}}^k \nonumber \\
&+&  B_{\Pi_B^{(S)}\Pi_B^{(S)}\Pi_B^{(S)}}^{(A\,5)}\widehat{\mathbf{k1}-\mathbf{p}}^l\widehat{\mathbf{k2}}^j\hat{\mathbf{k3}}^k+B_{\Pi_B^{(S)}\Pi_B^{(S)}\Pi_B^{(S)}}^{(A\,6)}\widehat{\mathbf{k1}-\mathbf{p}}^l\widehat{\mathbf{k2}}^j\hat{\mathbf{p}}^k       \nonumber \\
&+&  B_{\Pi_B^{(S)}\Pi_B^{(S)}\Pi_B^{(S)}}^{(A\,7)}\widehat{\mathbf{k2}+\mathbf{p}}^l\widehat{\mathbf{k3}}^j\hat{\mathbf{p}}^k
+  B_{\Pi_B^{(S)}\Pi_B^{(S)}\Pi_B^{(S)}}^{(A\,8)}\widehat{\mathbf{k2}}^l\widehat{\mathbf{k3}}^j\hat{\mathbf{p}}^k    \nonumber \\
&+&B_{\Pi_B^{(S)}\Pi_B^{(S)}\Pi_B^{(S)}}^{(A\,9)} \widehat{\mathbf{k2}}^l\widehat{\mathbf{k2}+\mathbf{p}}^j\hat{\mathbf{k3}}^k 
+ B_{\Pi_B^{(S)}\Pi_B^{(S)}\Pi_B^{(S)}}^{(A\,10)}\widehat{\mathbf{k1}}^l\widehat{\mathbf{k1}-\mathbf{p}}^j\hat{\mathbf{k2}+\mathbf{p}}^k \nonumber \\
&+& 
 B_{\Pi_B^{(S)}\Pi_B^{(S)}\Pi_B^{(S)}}^{(A\,11)}\widehat{\mathbf{k1}}^l\widehat{\mathbf{k1}-\mathbf{p}}^j\hat{\mathbf{k2}}^k+ B_{\Pi_B^{(S)}\Pi_B^{(S)}\Pi_B^{(S)}}^{(A\,12)}\widehat{\mathbf{k1}}^l\widehat{\mathbf{k1}-\mathbf{p}}^j\hat{\mathbf{k3}}^k \nonumber \\
&+& 
 B_{\Pi_B^{(S)}\Pi_B^{(S)}\Pi_B^{(S)}}^{(A\,13)}\widehat{\mathbf{k1}}^l\widehat{\mathbf{k2}+\mathbf{p}}^j\hat{\mathbf{k3}}^k+ B_{\Pi_B^{(S)}\Pi_B^{(S)}\Pi_B^{(S)}}^{(A\,14)}\widehat{\mathbf{k1}}^l\widehat{\mathbf{k2}+\mathbf{p}}^j\hat{\mathbf{p}}^k \nonumber \\
&+& 
 B_{\Pi_B^{(S)}\Pi_B^{(S)}\Pi_B^{(S)}}^{(A\,15)}\widehat{\mathbf{k1}}^l\widehat{\mathbf{k3}}^j\hat{\mathbf{p}}^k+ B_{\Pi_B^{(S)}\Pi_B^{(S)}\Pi_B^{(S)}}^{(A\,16)}\widehat{\mathbf{k1}}^l\widehat{\mathbf{k2}}^j\hat{\mathbf{k2}+\mathbf{p}}^k \nonumber \\
&+& 
 B_{\Pi_B^{(S)}\Pi_B^{(S)}\Pi_B^{(S)}}^{(A\,17)}\widehat{\mathbf{k1}}^l\widehat{\mathbf{k2}}^j\hat{\mathbf{p}}^k+ B_{\Pi_B^{(S)}\Pi_B^{(S)}\Pi_B^{(S)}}^{(A\,18)}\widehat{\mathbf{k2}}^l\widehat{\mathbf{k2}+\mathbf{p}}^j\hat{\mathbf{p}}^k+B_{\Pi_B^{(S)}\Pi_B^{(S)}\Pi_B^{(S)}}^{(A\,19)}\widehat{\mathbf{k1}}^l\widehat{\mathbf{k1}-\mathbf{p}}^j\hat{\mathbf{p}}^k . \nonumber\\\end{eqnarray}
Without  helical contributions of the field($A_H=0$), our results are in agreement with the ones found in \cite{37}, however there is an aditional factor of 3 in the eq.(\ref{no3}) in three terms.
\section{Integration domain}\label{apenb}
 The angular part of the integrals must be written in spherical coordinates $d^3p=2\pi p^2 dp d\alpha_k $, where $2\pi$ comes from of the integration of $\theta$. Since we consider an upper cut-off $k_D$ that corresponds to the damping scale at the spectrum, we must introduce the ($k1,k2$)-dependence on the angular integration domain. This implies that we should split the integral domain in different regions such that 
 \begin{equation}
 \left|\mathbf{k1}-\mathbf{p}\right| \leq k_D, \quad \left|\mathbf{k2}+\mathbf{p}\right| \leq k_D,
 \end{equation}
 obtaining  that region of the wave vectors where $0 <k1,k2< 2k_D$. Since we expect that most important contribution comes from  $\hat{\mathbf{k1}} \rightarrow -\hat{\mathbf{k2}}$ and
 using the above constraints we get the following integration domain in a squeezed configuration
 \begin{eqnarray}
 k_D>k2>0&&\nonumber \\
 &&k2>k1>0 \quad \int_{0}^{k_D-k2}dp\int_{-1}^{1}d\alpha_k+\int_{k_D-k2}^{k_D}dp\int_{\frac{k2^2+p^2-k_D^2}{2 k2 p}}^{1}d\alpha_k \nonumber\\
 &&k_D>k1>k2 \quad \int_{0}^{k_D-k1}dp\int_{-1}^{1}d\alpha_k+\int_{k_D-k1}^{k_D}dp\int_{\frac{k1^2+p^2-k_D^2}{2 k1 p}}^{1}d\alpha_k \nonumber\\
 &&2k_D>k1>k_D \quad \int^{k_D}_{k1-k_D}dp\int_{\frac{k1^2+p^2-k_D^2}{2 k1 p}}^{1}d\alpha_k \nonumber\\
 2k_D>k2>k_D&&\nonumber \\
  &&k2>k1>0 \quad \int^{k_D}_{k2-k_D}dp\int_{\frac{k2^2+p^2-k_D^2}{2 k2 p}}^{1}d\alpha_k \nonumber\\
  &&2k_D>k1>k2 \quad \int^{k_D}_{k1-k_D}dp\int_{\frac{k1^2+p^2-k_D^2}{2 k1 p}}^{1}d\alpha_k.
 \end{eqnarray}
The above integration domain was used to calculate the bispectrum for causal fields shown in figures (\ref{figparte1}) and (\ref{figparte2}). However, for the case of non-causal primordial magnetic fields (negative spectral indices) we can approximate the above result by selecting only regions where we can get the biggest contribution to the bispectrum (in fact, in \cite{22} they claimed that the biggest contribution comes from the  poles  of the integral). Then, we can work with the approximation made in \cite{31,25,27} where $k2<k1<k_D$ and  the angular part is neglected, finding that scheme of  integration is reduced to
 \begin{eqnarray} \label{sa}
 k_D>k2>0&&\nonumber \\
 &&k_D>k1>k2 \quad \int_{0}^{k_D}dp,
 \end{eqnarray}
and therefore the bispectrum can be approximated in following way:
The wave vector can be expressed in the basis defined in figure \ref{ref} as follows
\begin{eqnarray}
&&\hat{\mathbf{k1}}=\hat{\mathbf{e}}_z,\quad \hat{\mathbf{p}}=\sin\theta \cos\phi \hat{\mathbf{e}}_x+\sin\theta \sin\phi \hat{\mathbf{e}}_y+ \cos\theta \hat{\mathbf{e}}_z,\quad \nonumber \\
&&\hat{\mathbf{k2}}=-\sin\theta^{\prime}\hat{\mathbf{e}}_x-\cos\theta^{\prime}\hat{\mathbf{e}}_z,\quad \hat{\mathbf{k3}}=\sin\theta^{\prime\prime}\hat{\mathbf{e}}_x+\cos\theta^{\prime\prime}\hat{\mathbf{e}}_z,
\end{eqnarray}
being $\theta$, $\theta^{\prime}$, $\theta^{\prime\prime}$ the polar angle of $\mathbf{p}$,  $\mathbf{k2}$ and   $\mathbf{k3}$ respectively. With these formulas we can find the inner product between different wave vectors
\begin{eqnarray}
&&\hat{\mathbf{p}}\cdot \hat{\mathbf{k2}}=-\sin\theta\cos\phi\sin\theta^\prime-\cos\theta\cos\theta^\prime \nonumber \\
&&\hat{\mathbf{p}}\cdot \hat{\mathbf{k3}}=\sin\theta^{\prime\prime}\sin \theta \cos\phi+\cos\theta\cos\theta^{\prime\prime} \nonumber \\
&&\hat{\mathbf{k2}}\cdot \hat{\mathbf{k3}}=-\sin\theta^{\prime\prime}\sin \theta^\prime-\cos\theta^\prime\cos\theta^{\prime\prime}.
\end{eqnarray}
Thus, the expression (\ref{TEMbis1}) can be  written as
\begin{eqnarray}\label{apenb1}
\int p^n\left|\mathbf{k1}-\mathbf{p}\right|^n\left|\mathbf{k2}+\mathbf{p}\right|^n d^3p&\sim& 2\pi \int dp p^{n+2}\left(\left|k1-p \right|^n(p^2+k2^2-2pk2\cos\theta^\prime)^{n/2}  \right. \nonumber\\
&+&\left. \left|k2-p \right|^n(p^2+k1^2-2pk1\cos\theta^\prime)^{n/2}\right)\nonumber\\
&\sim&  \int dp p^{n+2}\left(k1^n\left|1-\frac{p}{k1}\right|^n k2^n(1+\left(\frac{p}{k2}\right)^2 -2\frac{p}{k2}\cos\theta^\prime)^{n/2}  \right. \nonumber\\
&+&\left. k2^n\left|1-\frac{p}{k2} \right|^nk1^n(1+\left(\frac{p}{k1}\right)^2-2\frac{p}{k1}\cos\theta^\prime)^{n/2}\right)\nonumber\\
&\sim&2\left(\frac{nk1^nk2^{2n+3}}{(n+3)(2n+3)}+\frac{nk1^{3n+3}}{(2n+3)(3n+3)}+\frac{k_D^{3n+3}}{(3n+3)}\right),
\end{eqnarray}
where in the last equality we have accounted eq.(\ref{sa}) and split into sub-ranges: $0<q<k2$, $k2<q<k1$ and $k1<q<k_D$. This result  was  derived analytically in \cite{31}.
\section{Integration domain for $\alpha\neq0$}\label{apenc}
We use the convolutions for the PMFs spectra with the parametrization for the
magnetic field  given in eqs. (\ref{PMFespectro1}), (\ref{powerPMF1}) and (\ref{powerPMF2}). Since $P_B\neq0$ and $P_H\neq0$ for $k_m<k1,k2< k_D$,
some conditions need to be taken into account: $ k_m<p<k_D$,  $ k_m<\left|\mathbf{k1}-\mathbf{p}\right|<k_D$ and  $ k_m<\left|\mathbf{k2} + \mathbf{p}\right|<k_D$.
The latter conditions introduce a k-dependence on the angular integration domain
and using the squeezed configuration ($\mathbf{k1}=-\mathbf{k2}\equiv\mathbf{k}$, $\mathbf{k3}\simeq0$),  the bispectrum is non zero only for $0<k<2k_D$. Such
constraints split the  integrals in differents parts as you can see in the appendix in \cite{21}. However, as claimed in \cite{23}, the p-integrals need a further splitting for odd $n_H,n_B$.
Here we will show only the result for $k_D>5k_m$ and $2k_m>k_D>k_m$.\\
For $k_D>5k_m$, we have:
\[
k_m>k>0
\]
\begin{equation}
\int_{k_m}^{k+k_m}d^3 p_{(p>k)} \int_{-1}^{\frac{k^2+p^{ 2}-k_m^2}{2kp}}d \gamma+\int_{k_m+k}^{k_D-k}d^3p_{(p>k)} \int_{-1}^{1}d \gamma+\int_{k_D-k}^{k_D}d^3p_{(p>k)} \int^{1}_{\frac{k^2+p^{ 2}-k_D^2}{2kp}}d \gamma
\end{equation}

\[
2k_m>k>k_m
\]
\begin{eqnarray}
\int_{k_m}^{k}d^3p_{(k>p)} \int_{-1}^{\frac{k^2+p^{ 2}-k_m^2}{2kp}}d \gamma&+&\int_{k}^{k_m+k}d^3p_{(p>k)} \int_{-1}^{\frac{k^2+p^{ 2}-k_m^2}{2kp}}d \gamma \nonumber\\
+\int_{k+k_m}^{k_D-k}d^3p_{(p>k)} \int_{-1}^{1}d \gamma&+&\int_{k_D-k}^{k_D}d^3p_{(p>k)} \int^{1}_{\frac{k^2+p^{ 2}-k_D^2}{2kp}}d \gamma
\end{eqnarray}

\[
\frac{k_D-k_m}{2}>k>2k_m
\]
\begin{eqnarray}
\int_{k_m}^{k-k_m}d^3 p_{(k>p)} \int_{-1}^{1}d \gamma&+&\int_{k-k_m}^{k}d^3p_{(k>p)} \int_{-1}^{\frac{k^2+p^{ 2}-k_m^2}{2kp}}d \gamma +\int_{k}^{k+k_m}d^3p_{(p>k)} \int_{-1}^{\frac{k^2+p^{ 2}-k_m^2}{2kp}}d \gamma \nonumber\\
+\int_{k_m+k}^{k_D-k}d^3p_{(p>k)} \int^{1}_{-1}d \gamma&+&\int_{k_D-k}^{k_D}d^3p_{(p>k)} \int^{1}_{\frac{k^2+p^{ 2}-k_D^2}{2kp}}d \gamma
\end{eqnarray}

\[
\frac{k_D}{2}>k>\frac{k_D-k_m}{2}
\]
\begin{eqnarray}
\int_{k_m}^{k-k_m}d^3 p_{(k>p)} \int_{-1}^{1}d \gamma&+&\int_{k+k_m}^{k_D}d^3p_{(p>k)} \int^{1}_{\frac{k^2+p^{ 2}-k_D^2}{2kp}}d \gamma +\int_{k}^{k_D-k}d^3p_{(p>k)} \int_{-1}^{\frac{k^2+p^{ 2}-k_m^2}{2kp}}d \gamma \nonumber\\
+\int_{k-k_m}^{k}d^3p_{(k>p)} \int_{-1}^{\frac{k^2+p^{ 2}-k_m^2}{2kp}}d \gamma&+&\int_{k_D-k}^{k+k_m}d^3p_{(p>k)} \int^{\frac{k^2+p^{ 2}-k_m^2}{2kp}}_{\frac{k^2+p^{ 2}-k_D^2}{2kp}}d \gamma
\end{eqnarray}

\[
\frac{k_D}{2}<k<\frac{k_D+k_m}{2}
\]
\begin{eqnarray}
\int_{k_m}^{k-k_m}d^3 p_{(k>p)} \int_{-1}^{1}d \gamma&+&\int_{k+k_m}^{k_D}d^3p_{(p>k)} \int^{1}_{\frac{k^2+p^{ 2}-k_D^2}{2kp}}d \gamma +\int_{k-k_m}^{k_D-k}d^3p_{(k>p)} \int_{-1}^{\frac{k^2+p^{ 2}-k_m^2}{2kp}}d \gamma \nonumber\\
+\int_{k_D-k}^{k}d^3p_{(k>p)} \int^{\frac{k^2+p^{ 2}-k_m^2}{2kp}}_{\frac{k^2+p^{ 2}-k_D^2}{2kp}}d \gamma&+&\int_{k}^{k+k_m}d^3p_{(p>k)} \int^{\frac{k^2+p^{ 2}-k_m^2}{2kp}}_{\frac{k^2+p^{ 2}-k_D^2}{2kp}}d \gamma
\end{eqnarray}
\[
k_D-k_m>k>\frac{k_D+k_m}{2}
\]
\begin{eqnarray}
\int_{k_m}^{k_D-k}d^3 p_{(k>p)} \int_{-1}^{1}d \gamma&+&\int_{k_D-k}^{k-k_m}d^3p_{(k>p)} \int^{1}_{\frac{k^2+p^{ 2}-k_D^2}{2kp}}d \gamma +\int_{k}^{k+k_m}d^3p_{(p>k)} \int^{\frac{k^2+p^{ 2}-k_m^2}{2kp}}_{\frac{k^2+p^{ 2}-k_D^2}{2kp}}d \gamma\nonumber\\
+\int_{k-k_m}^{k}d^3p_{(k>p)} \int^{\frac{k^2+p^{ 2}-k_m^2}{2kp}}_{\frac{k^2+p^{ 2}-k_D^2}{2kp}}d \gamma&+&\int_{k_m+k}^{k_D}d^3p_{(p>k)} \int^{1}_{\frac{k^2+p^{ 2}-k_D^2}{2kp}}d \gamma
\end{eqnarray}
\[
k_D>k>k_D-k_m
\]
\begin{eqnarray}
\int_{k_m}^{k-k_m}d^3 p_{(k>p)}\int^{1}_{\frac{k^2+p^{ 2}-k_D^2}{2kp}}d\gamma&+&\int_{k}^{k_D}d^3p_{(p>k)} \int^{\frac{k^2+p^{ 2}-k_m^2}{2kp}}_{\frac{k^2+p^{ 2}-k_D^2}{2kp}}d \gamma 
+\int_{k-k_m}^{k}d^3p_{(k>p)} \int^{\frac{k^2+p^{ 2}-k_m^2}{2kp}}_{\frac{k^2+p^{ 2}-k_D^2}{2kp}}d \gamma
\end{eqnarray}

\[
k_D+k_m>k>k_D
\]
\begin{eqnarray}
\int_{k_m}^{k-k_m}d^3p_{(k>p)} \int_{\frac{k^2+p^{2}-k_D^2}{2kp}}^{1}d\gamma&+&\int_{k-k_m}^{k_D}d^3p_{(k>p)} \int^{\frac{k^2+p^{ 2}-k_m^2}{2kp}}_{\frac{k^2+p^{2}-k_D^2}{2kp}}d \gamma
\end{eqnarray}

\[
2k_D>k>k_D+k_m
\]
\begin{equation}
\int_{k-k_D}^{k_D}d^3 p_{(k>p)} \int_{\frac{k^2+p^{2}-k_D^2}{2kp}}^{1}d \gamma.
\end{equation}


For the case where $2k_m>k_D>k_m$,  we have
\[
\frac{k_D-k_m}{2}>k>0
\]
\begin{equation}
\int_{k_m}^{k+k_m}d^3 p_{(p>k)} \int_{-1}^{\frac{k^2+p^{ 2}-k_m^2}{2kp}}d \gamma+\int_{k_m+k}^{k_D-k}d^3p_{(p>k)} \int_{-1}^{1}d \gamma+\int_{k_D-k}^{k_D}d^3p_{(p>k)} \int^{1}_{\frac{k^2+p^{ 2}-k_D^2}{2kp}}d \gamma
\end{equation}

\[
k_D-k_m>k>\frac{k_D-k_m}{2}
\]

\begin{equation}
\int_{k_m}^{k_D-k}d^3 p_{(p>k)} \int_{-1}^{\frac{k^2+p^{ 2}-k_m^2}{2kp}}d \gamma+\int_{k_D-k}^{k_m+k}d^3p_{(p>k)} \int_{\frac{k^2+p^{ 2}-k_D^2}{2kp}}^{\frac{k^2+p^{2}-k_m^2}{2kp}}d \gamma +\int_{k+k_m}^{k_D}d^3p_{(p>k)} \int^{1}_{\frac{k^2+p^{ 2}-k_D^2}{2kp}}d \gamma
\end{equation}

\[
k_m>k>k_D-k_m
\]

\begin{equation}
\int_{k_m}^{k_D}d^3p_{(p>k)} \int_{\frac{k^2+p^{ 2}-k_D^2}{2kp}}^{\frac{k^2+p^{ 2}-k_m^2}{2kp}}d \gamma
\end{equation}

\[
k_D>k>k_m
\]

\begin{equation}
\int_{k_m}^{k}d^3p_{(k>p)} \int_{\frac{k^2+p^{ 2}-k_D^2}{2kp}}^{\frac{k^2+p^{ 2}-k_m^2}{2kp}}d \gamma+\int_{k}^{k_D}d^3p_{(p>k)} \int_{\frac{k^2+p^{ 2}-k_D^2}{2kp}}^{\frac{k^2+p^{ 2}-k_m^2}{2kp}}d \gamma
\end{equation}

\[
2k_m>k>k_D
\]

\begin{equation}
\int_{k_m}^{k_D}d^3p_{(k>p)} \int_{\frac{k^2+p^{ 2}-k_D^2}{2kp}}^{\frac{k^2+p^{ 2}-k_m^2}{2kp}}d \gamma
\end{equation}

\[
k_m+k_D>k>2k_m
\]

\begin{equation}
\int_{k-k_m}^{k_D}d^3p_{(k>p)} \int_{\frac{k^2+p^{ 2}-k_D^2}{2kp}}^{\frac{k^2+p^{2}-k_m^2}{2kp}}d \gamma+\int_{k_m}^{k-k_m}d^3p_{(k>p)} \int^{1}_{\frac{k^2+p^{ 2}-k_D^2}{2kp}}d \gamma
\end{equation}
\[
2k_D>k>k_m+k_D
\]
\begin{equation}
\int_{k-k_D}^{k_D}d^3p_{(k>p)} \int^{1}_{\frac{k^2+p^{ 2}-k_D^2}{2kp}}d \gamma.
\end{equation}
The integration domain above generalizes the results obtained in \cite{21}. With this we can calculate the spectrum and bispectrum (under certain configurations) of PMFs for any value of the magnetic spectral index.



\end{document}